# Dynamic Interface Printing

Callum Vidler[1], Michael Halwes[1], Kirill Kolesnik[1], Philipp Segeritz[1,7], Matthew Mail[1], Anders J. Barlow[2], Emmanuelle M. Koehl[5], Anand Ramakrishnan[5,6], Daniel J. Scott[7], Daniel E. Heath[1,9], Kenneth B. Crozier[3,4,8], David J. Collins[1,9*]

Affiliations:

[1]Department of Biomedical Engineering, The University of Melbourne, Melbourne, Victoria, Australia

[2]Materials Characterisation and Fabrication Platform (MCFP), The University of Melbourne, Parkville, Victoria, 3010, Australia

[3]School of Physics, The University of Melbourne, Victoria 3010, Australia

[4]Department of Electrical and Electronic Engineering, The University of Melbourne, Victoria 3010, Australia

[5]Department of Plastic and Reconstructive Surgery, The Royal Melbourne Hospital, Victoria 3050, Australia

[6] The University of Melbourne, Melbourne Medical School, Department of Surgery, Royal Melbourne Hospital, Parkville VIC 3050

[7]The Florey and Department of Biochemistry and Pharmacology, The University of Melbourne, Parkville, Victoria, 3010, Australia

[8]Australian Research Council (ARC) Centre of Excellence for Transformative Meta-Optical Systems, The University of Melbourne, Victoria 3010, Australia

[9]The Graeme Clark Institute, The University of Melbourne, Parkville 3052, Victoria, Australia

*Corresponding author

E-mail: david.collins@unimelb.edu.au, vidlerc@student.unimelb.edu.au

**Abstract**

Additive manufacturing is an expanding multidisciplinary field encompassing applications including medical devices[1], aerospace components[2], microfabrication strategies[3,4], and artificial organs[5]. Among additive manufacturing approaches, light-based printing technologies, including two-photon polymerization[6], projection micro stereolithography[7,8], and volumetric printing[9–14], have garnered significant attention due to their speed, resolution and/or potential applications for biofabrication. In this study, we introduce dynamic interface printing (DIP), a new 3D printing approach that leverages an acoustically modulated, constrained air-liquid boundary to rapidly generate cm-scale three-dimensional structures within tens of seconds. Distinct from volumetric approaches, this process eliminates the need for intricate feedback systems, specialized chemistry, or complex optics while maintaining rapid printing speeds. We demonstrate the versatility of this technique across a broad array of materials and intricate geometries, including those that would be impossible to print via conventional layer-by-layer methods. In doing so, we demonstrate the rapid fabrication of complex structures in-situ, overprinting, structural parallelisation, and biofabrication utility. Moreover, we showcase that the formation of surface waves at this boundary enables enhanced mass transport, material flexibility, and permits three-dimensional particle patterning. We therefore anticipate that this approach will be invaluable for applications where high resolution, scalable throughput, and biocompatible printing is required.

**Main Text**

Rapid 3D printing, where whole parts are created on the scale of seconds to minutes rather than hours, is increasingly recognized as an enabling technology for a range of emerging bioprinting, prototyping, and manufacturing applications[15–18]. Conventional optical-based printing approaches, such as stereolithography, typically involve the application of light to cure materials one layer at a time, where such approaches have advantages in resolution and geometric fidelity. However, the printing rate is limited by the need to repeatedly reset the part position between layers to allow uncured resin to flow in, resulting in limitations on material stiffness and throughput.

Recently, volumetric printing approaches have been used to rapidly manufacture centimetre-scale constructs. In the case of computed axial lithography[9,12], a vial containing the photopolymer is rotated and a series of projections are exposed from azimuthal angles such that the cumulative intersection of light rays produces the desired object. As this process relies on the local depletion of oxygen within the volume to generate polymerization only in targeted regions, it accordingly is highly sensitive to polymerization dose and projection telecentricity. Though some effects can be partially corrected computationally[19], it introduces additional constraints for resins and bioinks utilising these systems. Other recent methods such as Xolography[10] or light sheet printing[20], are based on the use of a spiropyran photoswitch, or two-step absorption, wherein two wavelengths defined by a light sheet and an orthogonal projection simultaneously coincide to initiate polymerization. While such volumetric approaches permit rapid fabrication of free-floating isotropic structures, they are limited by the requirement of specialised optical systems or resin formulations. The need for transparent resins in particular limits the potential biofabrication applications of such volumetric printing processes, as it precludes many additives and limits the concentration of cells that can be suspended.

Although minimizing printing time is ideal for maintaining high cell viability[21], cell-induced optical scattering greatly limits the capacity for high-resolution cell-laden constructs. Alternatively, regulating the oxygen concentration at the printing interface in conventional bottom-up stereolithography processes can also enable high-speed printing. Continuous Liquid Interface Production (CLIP)[22,23], for instance, employs an oxygen-permeable membrane to prevent polymerization at the print boundary. In CLIP, however, the printed structure is progressively extracted from a liquid reservoir, posing difficulties for soft materials like hydrogels[18,19] due to structural instability during printing or the requisite manual handling in post-processing, and where the use of a solid-liquid boundary at the printing interface precludes the integration of capabilities such as overprinting.

In this work, we present a novel rapid 3D printing technique in which an object is generated at the boundary of an acoustically driven, constrained air-liquid interface, facilitating the rapid creation of arbitrary supportless structures without specialized chemistry or optical feedback systems. This approach is compatible with a range of materials including soft and biologically relevant hydrogels at speeds suitable for high-viability tissue engineering, scalable manufacturing, and rapid prototyping.

**Dynamic Interface Printing**

At its core, dynamic interface printing (DIP) comprises a hollow print head that is open at the bottom and sealed with a transparent glass window at the top. When the print head is submerged within a liquid prepolymer solution, air inside becomes trapped forming an air-liquid meniscus at the print head's tip. This meniscus becomes the print interface at which structures are polymerized by UV light transmitted through the glass window (**Fig. 1a, Supplementary Fig.1**). In our system, light is delivered using a 405 nm projection system with an in-plane resolution of 15.1 µm and adjustable irradiance levels ranging from zero to approximately 270 mW/cm$^2$. By controlling the air pressure inside the print head to adjust the meniscus' position and curvature, the print surface is brought co-planar with the focal plane. 2D slices of the desired object are then projected down through the print head onto the air-liquid interface, and the complete object is built up by continuously raising the entire print head relative to the print container while modifying the optical projections.

A distinguishing feature of Dynamic Interface Printing (DIP) is the ability to dynamically regulate the pressure within the print head, thereby controlling the shape and position of the meniscus during the printing process. This regulation can maintain the meniscus in a static state or modulate it acoustically across a range of amplitudes and frequencies to generate capillary-gravity waves at the print interface (**Fig. 1b**). The precise position of the meniscus at any given moment is determined by the superposition of the print head's vertical position, the static air pressure within the print head, and the driving amplitude and frequency of the acoustic modulation. This oscillatory actuation can be activated either continuously (**Fig. 1c(i)**) or transiently between the projection of 2D slices (**Fig. 1c(ii)**).

By localising the optics and acoustic modulation to the print head, DIP is inherently container-agnostic and as such does not specify constraints on the shape or optical properties of the printing vessel, as in volumetric printing approaches [9,10,14,17]. DIP therefore provides high fabrication rates by utilising a controllable meniscus as a print interface, which is enhanced by the formation of surface waves, enabling high resolution structures to be rapidly formed (**Fig. 1d-e, Movie 1**). This additional modality can be used to bolster fabrication rate, enhance material processing ranges, enable 3D particle patterning, in-situ structure formation and overprinting capabilities that are unique to DIP. With this approach, we demonstrate the fabrication of a wide variety of centimetre-scale objects in tens of seconds.

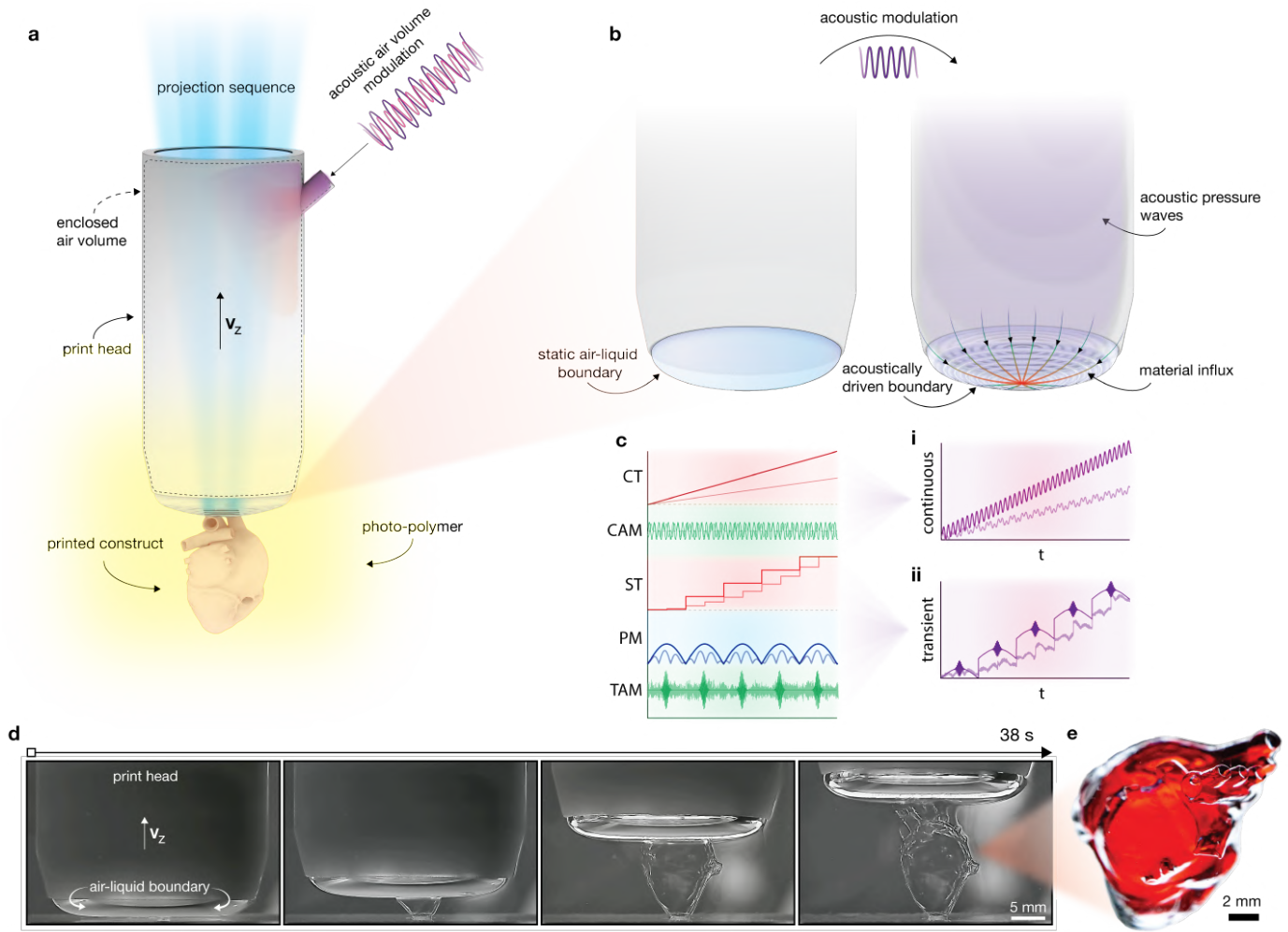

**Fig.1|Schematic illustration of dynamic interface printing**. **a**, An air-liquid boundary is formed at the base of a partially submerged print head, wherein the boundary acts as a print interface in which patterned projections are used to locally solidify the photopolymer. **b**, Acoustic manipulation of the internal print head air volume promotes enhanced material influx through capillary driven waves. **c**, (**i**) In continuous mode, air-liquid interface(s) global location depends on continuous translation of the print head (CT) and constant acoustic modulation (CAM). (**ii**) In transient mode, interface location depends on stepped translation (ST), internal pressure modulation (PM), and transient acoustic modulation (TAM). **d**, Timelapse photographs of the printing process for a heart geometry, demonstrating rapid fabrication of centimetre scale constructs in less than 40 seconds. **e**, Printed heart geometry as shown in (d), dyed red to improve visualisation.

**Convex Slicing**

As is also the case with other light-based printing techniques, a three-dimensional digital model of the desired geometry must first be translated into a series of images to be sequentially displayed by the projection system. However, in contrast to standard stereolithography configurations which employ a planar construction plane, DIP is characterized by a curved meniscus, necessitating a series of images that follow the profile of the interface and as such represent three-dimensional regions of the object. At the start of each print, the interface is first compressed against the base of the print volume to form a thin film region whose maximum extent is determined by the dimensions of the part. As the print advances, the compressed interface profile rises until its centre is tangent with the container's base. Beyond this transient region, the interface shape can be determined using the Young-Laplace equation, which relates the pressure difference sustained across the interface to its curvature. To predict the shape of the interface during initial compression and the subsequent transient region, we first solve the steady-state Young-Laplace solution using Bézier curves[25,26], which provides the approximate un-compressed interface profile. The transient contact shape is consequently approximated via volume equivalency under uniaxial deformation (**Fig.2a, Supplementary Fig.4a**). By utilizing an axisymmetric cylindrical print head, the three-dimensional shape of the convex interface can be readily computed from the two-dimensional Bézier solution by revolving the half-profile about the printhead's central axis (**Supplementary Fig.4b-c**). The convex projections that follow the interface shape for each point in time are therefore constructed by intersecting the series of Bézier surfaces with the input geometry discretised as a voxel array (**Fig.2b**). This slicing scheme therefore ensures that two-dimensional projections incident on the meniscus results in the correct three-dimensional mapping to the object domain for all intermediate interface locations (**Fig.2c, Supplementary Fig.5**).

**Dynamic Interface Printing Characterisation**

Beyond the importance of high throughput manufacturing, DIP provides unique utility for the creation of biological models due to its rapid printing rate and minimal shear along the air-liquid interface, in comparison to other light-based printing techniques and especially in contrast to extrusion-based bioprinting[27]. Here, we demonstrate the ability to create not only structures in hard acrylates such as 1,6-Hexanediol diacrylate (HDDA), but also common biologically relevant materials such as Poly(ethylene glycol) diacrylate (PEGDA) and Gelatin methacryloyl (GelMA). To demonstrate and characterize the ability to print across multiple material types, we evaluate a parameter space that quantifies the achievable print speed as a function of the optical power (**Fig. 2d**). Though the maximum print speed for a resin is a function of the photoinitiator concentration, monomer functionalization, material viscosity, and optical power, here we utilize common formulations for ease of comparison. The relatively low LAP photoinitiator concentrations ($\leq$ 0.3%) we use, for instance, are biocompatible[28]. Evidencing the high printing speeds achievable with DIP, translation rates greater than 700 µm/s are realized using a PEGDA hydrogel, with an optical dose of 270 mW/cm$^2$. Notably, lower optical power dosages (e.g. 30-50 mW/cm$^2$) which are more appropriate for biofabrication, still enable high-speed printing of centimetre-scale constructs in tens of seconds.

As the print interface is inherently curved, mapping a planar projection to its surface results in defocusing of the image near the boundary of the print head. The magnitude of defocusing is primarily contingent on material properties that impact the capillary length and the materials' contact angle with the print head. A theoretical constraint can be applied to determine the projected area fraction that can be used in cases where the uniformity of the projection resolution is critical. This is achieved by ensuring that for a given pixel, the in-plane resolution $P_{FP}$ does not exceed $\sqrt{2}P_{FP}$ when mapped to the interface. A theoretical analysis using Gaussian beam theory (**Supplementary Fig.6**) can then be used to predict the equivalent defocused pixel size (PSF) at a location $z$ (mm) above the focal plane. This can then be mapped to the local interface height determined by the Bézier solution for each material combination, resulting in the fractional area of the interface that maintains $P_{xy} \leq \sqrt{2}P_{FP}$ (**Fig. 2e**). For equal material properties, smaller print heads achieve a lower accurate pixel area fraction due to ratio of the capillary length to the print heads diameter. Conversely, for increasing print head size, the fraction of the diameter dominated by the capillary length decreases as the print head transitions towards a free surface. Similarly, materials with lower surface tensions and/or higher densities exhibit shorter capillary lengths, increasing thresholded pixel area fraction.

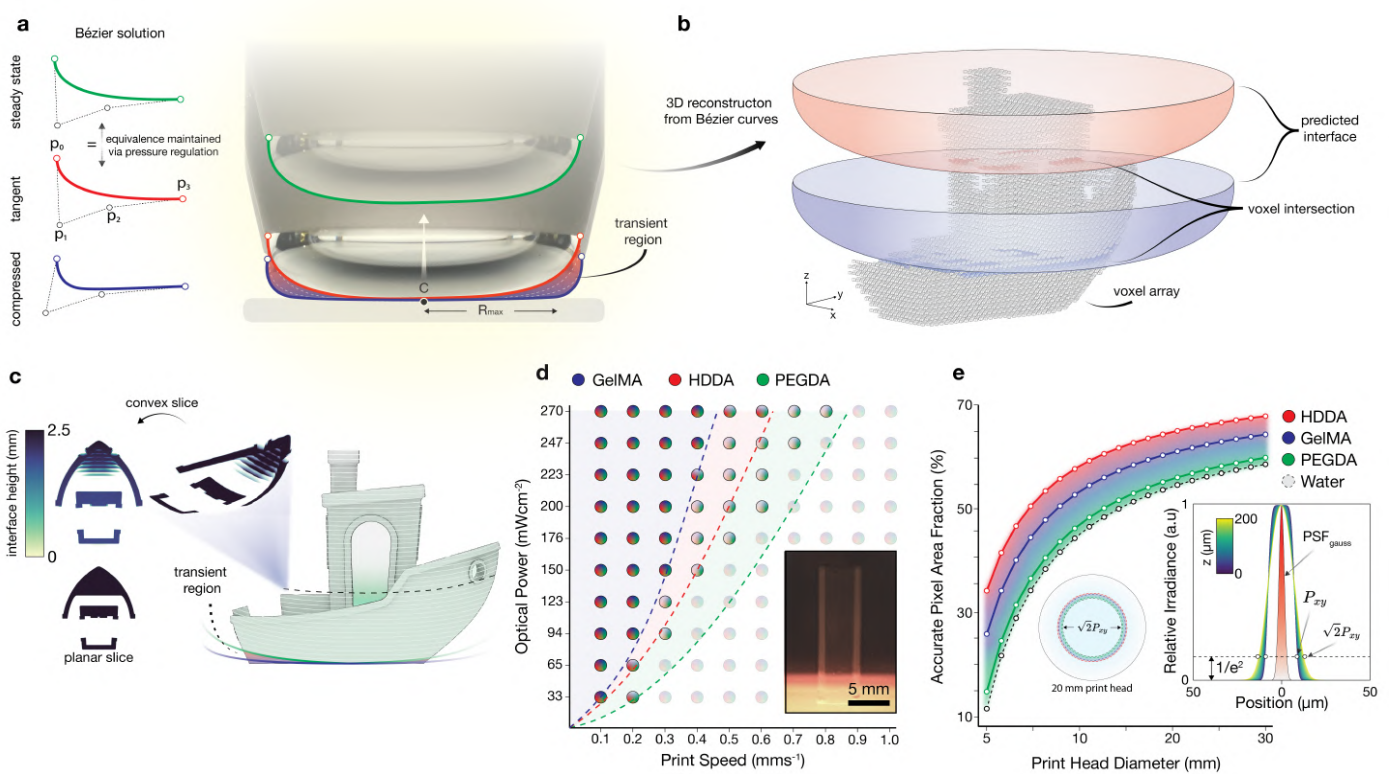

**Fig.2 | Dynamic interface printing system characterization. a**, Images of the air-liquid interface profile formed at the base of the print head under the compressed, tangent and steady state modes. Bézier curves are used to predict the shape of the interface during printing, corresponding to each of the interface modes. **b**, The convex slicing scheme is determined by first revolving the Bézier half profile about the central axis and computing the voxel-wise intersection. **c**, Convex optimised projections extend in three-dimensions and follow the boundary curvature under each interface mode. **d**, Print parameter space showing optical power, print speed pairs for GelMA (blue), HDDA (red) and PEGDA (green). Inset shows an example of the rectangular test structure used to assess the parameter space. **e**, Accurate pixel area fraction for increasing print head size for HDDA (red), GelMA (blue), PEGDA (green) and water (grey-dot). Inset shows the simulated deviation in the pixel size based on Gaussian beam theory for a range of z values and the overlayed variability in the area fraction for the 20 mm print, dependent on material formulation.

**Acoustic Modulation**

Using a constrained air-liquid boundary as a printing interface improves interface stability, enhancing resin influx, and results in a highly configurable system in which additional functionality can be multiplexed with light-based patterning. Importantly, this enables the ability to create capillary-gravity waves and sub-surface streaming via acoustic excitation (**Fig. 3b**), greatly enhancing material influx. To achieve this, we employ a novel method that modulates the enclosed printhead volume above the meniscus (**Fig. 3a, Supplementary Fig.1c-d**). This not only maintains the container-agnostic benefit of DIP, but largely eliminates the inherent coupling and modulation between the pressure field and the shape of the material container[37,38]. Moreover, air-coupled actuation in the print head generates variable amplitude capillary waves that enhance resin influx and that rapidly decay in time, enabling highly controlled excitation (**Supplementary Fig.7,22,25**).

Using an acoustic air-volume modulation approach, the frequency and amplitude of surface waves can be readily controlled during the printing process. To illustrate this, **Fig. 3c** shows the reflections from a ring of multi-coloured LEDs from a modulated surface for increasing driving frequency, resulting in azimuthally symmetric monochromatic modes at low frequency[34] and square symmetry at higher frequencies[39]. The transport capacity of these waves was evaluated using particle image velocimetry (PIV) and computational fluid dynamics (CFD) for various frequency and amplitude pairings (**Supplementary Fig.8-9,23-24, Movie 2**). Here low amplitude modulation, actuated at frequencies synchronised with the projection framerate (to minimise blurring), significantly augments mass-transport allowing for translational flow across the meniscus (**Fig. 3d**). Additionally, significant fluid velocities on the order of 10's of mm/s can be achieved along the interface and within the fluid bulk, allowing for the recirculation of resin material (**Fig.3e, Supplementary Fig.8,23-24, Movie 3**). Acoustic actuation therefore significantly increases the influx of material during printing, which is further amplified by meniscus curvature due to secondary streaming effects[36] (**Supplementary Fig.10**). To quantify the rate of material influx, highspeed photography was captured with and without acoustic excitation for various material viscosities and printing speeds (**Fig. 3f, Supplementary Fig.11**). The application of acoustic modulation here results in an extremely rapid decrease in the dry region below the interface compared to the case without, due to the rapid ingress of new material. By measuring the area of the instantaneous dry region during this process, we derived spatially averaged velocity magnitudes of ~ 16 – 40 mms$^{-1}$ across a range of printing rates. This heuristic is used within our slicing software in conjunction with the object's topology to predict and adapt the printing speed based on the fluidic path length (**Supplementary Fig.13**).

While other volumetric printing approaches[12,16,40,41] have shown unique utility for biofabrication by enabling the construction of low stiffness structures from cell-laden materials, these processes often rely on thermal gelation of the material (e.g. GelMA) before printing to prevent object and cellular sedimentation. Therefore, although the print time for these parts is short, the total time required to cool, print, and rewarm these structures extends into the tens of minutes, with workflow times essentially comparable with existing DLP processes. As an alternative, we show here not only that the supporting bath of DIP enables fabrication of soft biological structures without thermal gelation (**Supplementary Fig.14,27**) but also that acoustic modulation can significantly diminish sedimentation and enhance encapsulation efficiency through active mixing (**Fig. 3g, Supplementary Fig.15**). This greatly augments the processing of biological materials, all whilst maintaining physiologically relevant environmental conditions, processing parameters and cellular homogeneity.

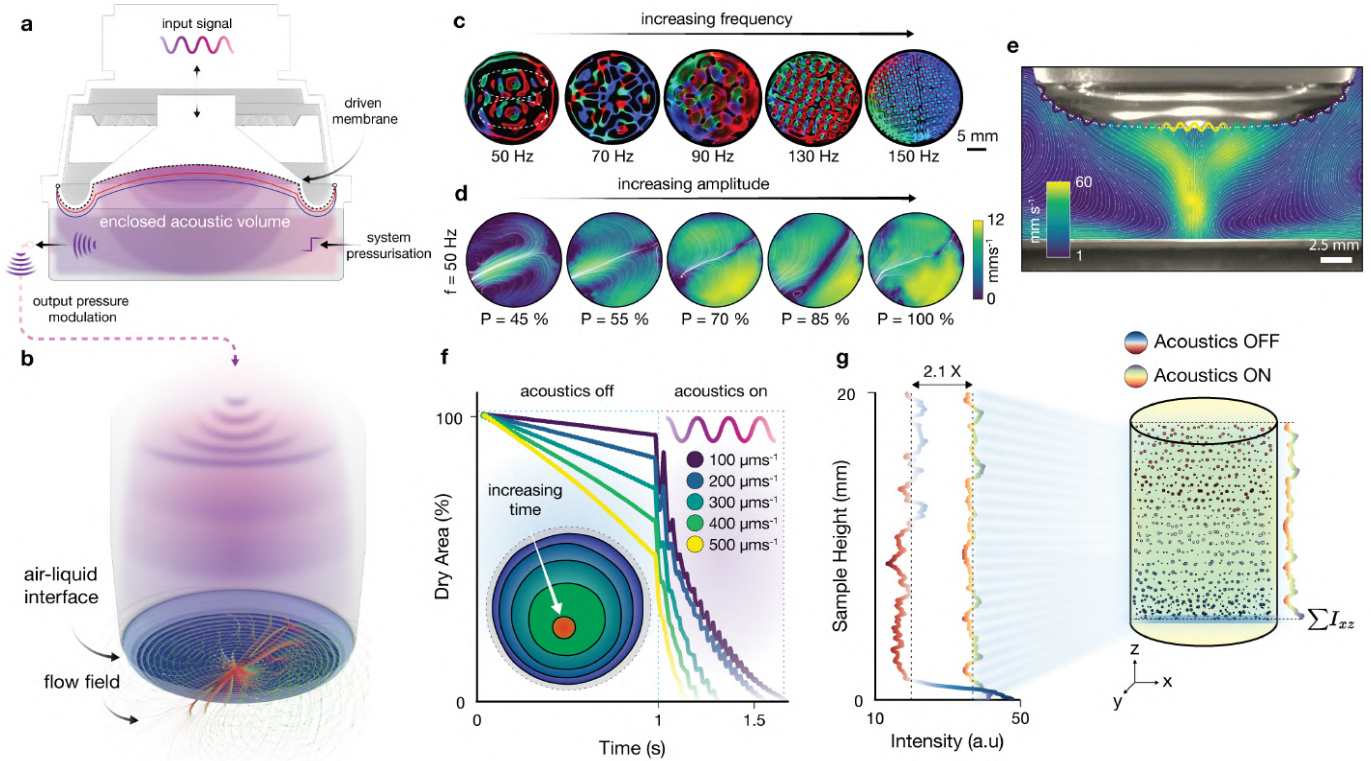

**Fig.3 | Acoustic modulation in dynamic interface printing**. **a**, Schematic illustration of the acoustic air volume modulation device. **b**, Illustration of dynamic interface printing under acoustic modulation, whereby capillary-gravity values formed on the free-surface of the print head result in flow fields that extend in three-dimensions. **c**, Multi-coloured light scattered from the air-liquid interface and imaged under acoustic excitation. **d**, Particle image velocimetry (PIV) normal to the interface at 50 Hz and increasing amplitude. **e**, Particle image velocimetry (PIV) perpendicular to the interface at 40 Hz and maximum amplitude, demonstrating the formation of high velocity jetting flows. **f**, Effect of acoustic actuation on the reduction in dry material area below the interface. Inset shows an example of the tracked dry boundary over time during the wetting process, circular-coloured regions (blue to red) and black contours indicate the impending material boundary as a function of increasing time. **g,** Effect of acoustic stimulation on cellular sedimentation, whereby encapsulation density (optical intensity) is plotted over the height of a circular pillar containing encapsulated 17 μm particles.

**Dynamic Interface Printing Capabilities**

Previous approaches, such as top-down SLA[42] that rely on unconstrained air-liquid interfaces suffer from mass transport and material film uniformity, which in many commercial systems necessitates the need for a wiping mechanism, limiting throughput. While minimising fluidic path length can reduce this (**Supplementary Fig.13,23-24**), it imposes significant geometric constraints on what can be printed. To investigate the advantage that DIP provides over established air-liquid modalities, the average fluid velocity during printing was computed across a range of structural sizes and acoustic parameters (**Supplementary Fig.22-24**). These results indicate an almost 4X improvement in fluid flow without acoustic excitation, which increases to approximately 10X with acoustic excitation (**Supplementary Fig.25**). To practically validate this improvement, we printed a series of fluidic manifolds containing multiple independent fluidic pathways (**Fig. 4a**). After printing, these manifolds were injected with a coloured silicone to visualise the channel topologies and imaged under uniform illumination (**Fig. 4b-c**). Next, to investigate the effect of print head size on accurate area fraction, we utilised a small 10 mm diameter print head to fabricate a series of micro-lattice structures and pillar arrays in HDDA. **Fig. 4d** highlights the effect of printing a structure whose edges exceed $\sqrt{2P_{xy}}$ without convex slicing, where structures at the periphery of the print field are distorted and poorly adhered due to local defocussing. Conversely, by ensuring that the object size does not exceed the accurate area fraction and with convex slicing enabled, an array of high uniformity pillars was formed with features ranging from ~ 30 - 100 µm (**Fig. 4e**).

DIP further enables the fabrication of structures from partially occluded or completely opaque materials. For instance, hydrogels containing high cell populations are often opaque in appearance due to the refractive index mismatch between their cellular constituents and bulk hydrogel prepolymer[43]. In the case of volumetric printing, light must travel unobstructed through the entire volume, and as such index matching between the material and cellular medium becomes crucial[40,44]. While there has been some promise to overcome this constraint computationally[45], in the case of DIP, light transmission is confined to a thin region at the air-liquid interface, minimizing the scattering and photo absorbing effects of suspended materials. To highlight this, we created a norbornene-functionalized sodium alginate, in which the opacity was augmented by pH until it completely occluded a USAF test target (**Fig. 4f**). Despite this material being completely opaque, a 10 mm tall alginate tricuspid valve was fabricated in 33 seconds containing 300 µm thick internal leaflets when imaged under micro-CT (µCT) (**Fig. 4g**). DIP further enables the control, modulation, and spatial distribution of the fabrication surface(s) through the design of the print head's geometry, for example utilizing a print head comprised of an array of individual surfaces, here demonstrated via the parallelized fabrication of the letters 'DIP' (**Fig. 4h, Movie 5**). The effective amplitude and frequency of a constrained surface(s) (and by extension the resulting flow profile) can be further modulated by careful design, sizing, and arrangement of holes located at the print head's tip.

While acoustic actuation can be used to augment material influx [46] in the thin film below the meniscus, standing waves at the liquid-air interface can be also used for suspended particle patterning. By momentarily stopping the printing process and raising the print head above the previous layer in the absence of optical exposures, standing waves can be formed at specific driving frequencies. These standing waves produce a hydrodynamic potential field at the top surface of the underlying structure, whereby particles within this region migrate from regions of high to low potential energy. The resulting interaction results in particle aggregation at nodal positions (**Supplementary Fig. 26**), and as such is contingent on the frequency and nodal locations of the standing wave[35,47]. While the use of nodal patterning is not new[48], with recent hydrodynamic analogues utilising surface waves to pattern cells and particles[29,30,32], these approaches are inherently constrained to two dimensions. In our case, however, by combining faraday-wave patterning with light-based printing, we can generate 3D particle arrangements (**Fig. 4i**). Here we demonstrate patterning variation via frequency and amplitude modulation, with further potential to explore the utilization of designed boundary topology[31,32]. In addition to the adaptability of the print head geometry, employing a constrained air-liquid interface offers several benefits beyond the integration of acoustic stimulation. Notably, this allows solid parts to pass through the printing interface, thereby enabling direct overprinting of multi-material or multi-component structures in-situ. To demonstrate this, we fabricated a ball-and-socket joint (**Fig. 4j**) in which a socket housing was initially printed, a 10 mm ball bearing was inserted, and finally the socket cap and rod were printed over or onto the ball, respectively.

Finally, to investigate our premise that DIP provides a unique advantage for soft materials, we seek to evaluate its potential as a biofabrication tool. To further contrast with previous printing approaches such as CLIP[22,23] and volumetric printing[9,14] which typically utilize a fixed printing area or impose constraints on the optical properties or shape of the resin container, DIP can use any material container shape in which the printing head can be inserted, including common laboratory consumables. Additionally, as the print head can move freely in 3D space, it is feasible to parallelize the system to sequentially create multiple structures (**Supplementary Fig.16a**) or even utilise varying material densities as fabrication platforms (**Supplementary Fig.17**). This permits not only the printing of multiple structures in the same resin bath, but also the sequential in-situ production of structures in multiple volumes (such as a well plate, **Movie 4**), each of which could contain various cell types, materials, or geometries. To assess the preliminary viability of this technique for generating cell-laden, biologically relevant constructs, a simplified kidney-shaped hydrogel structure was printed using HEK-293-F cells at a density of $7.2 \times 10^6$ cells mL$^{-1}$ directly in a 12-well plate (**Fig. 4k**). Fluorescence microscopy was employed to image the construct over 24 hours, demonstrating low process cytotoxicity and high cell viability (~93%) (**Fig. 4l**, **Supplementary Fig.18**).

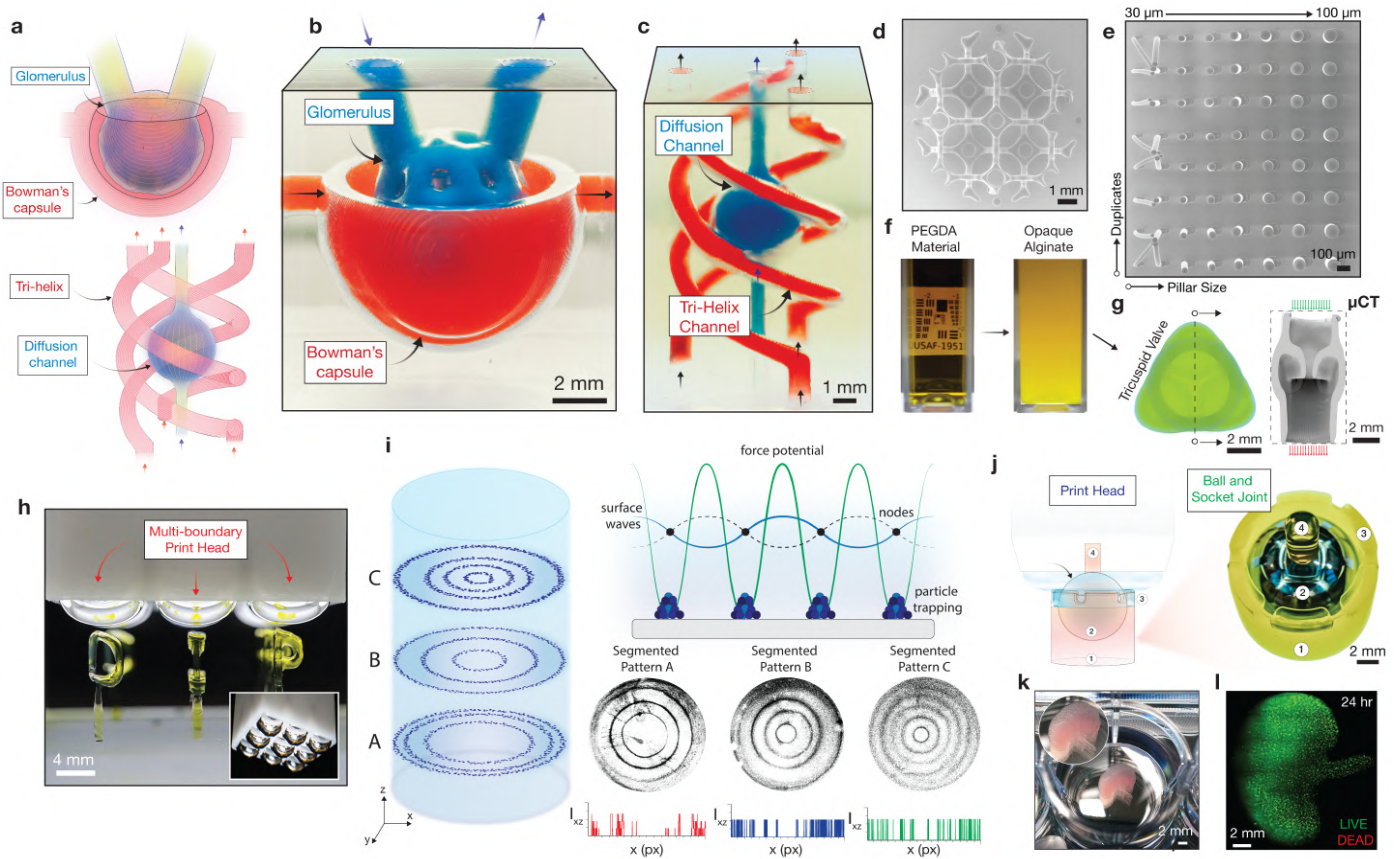

**Fig.4 | DIP capabilities. a,** Rendered illustration of the Bowman's capsule and tri-helix model. **b,** Printed Bowman's capsule model showing the Glomerulus and Capsule injected with red and blue dye. Print time was approximately 2 minutes. **c,** Tri-helix structure perfused with red and blue dye. **d,** Stitched top-down scanning electron microscope (SEM) image of a kelvin lattice printed with a 10 mm print head, scale bar 1 mm. Object FOV corresponds to approximately $1.5 \times \sqrt{2}P_{xy}$ diameter for HDDA. **e,** Stitched top-down helium ion microscopy (HIM) image of an equal height micro pillar array printed in HDDA, scale bar 200 µm. Distorted pillars are caused by surface tension effects during drying due to the high aspect ratio of the structures. **f,** Opacity comparison between the PEGDA (transparent) and Alginate (opaque) hydrogel materials when imaged against a standard USAF test target in a 10 mm cuvette. **g,** Top-down image of the tricuspid valve printed in an Alginate bioink and corresponding µCT cross-section, scale bar 2 mm. **h,** Print head with multiple independent air-liquid interfaces, used to create a 3x3 array of the letters 'DIP'. **i,** Three-dimensional patterning via standing surface waves. Suspended particles are trapped in nodal locations dependent on driving frequency, as shown in segmented patterns A-C. Corresponding image section intensity profile is shown below each patterned region. **j,** 4 Step overprinting process of a ball and socket joint, scale bar 2 mm. **k,** Simplified kidney model containing 7.2 million cells mL[-1] printed in situ in a 12-well plate. **l,** Stitched and deconvolved fluorescence image of (**j**) after 24 hours showing high cell viability maintained through the printing process.

**Discussion and Future Perspectives**

We have demonstrated volumetric fabrication rates on the order of $10^4$ mm$^3$/min, which exceeds that of other high speed printing processes including computed axial lithography (CAL)[9,14] and Xolography[10], without the need for specialized photochemistries, or optical feedback mechanisms. This is aided by surface-tension mediated printing interface stabilization and acoustically generated flow across the printing domain, permitting rapid material influx. This, coupled with the ability to multiplex print heads containing a greater number of individual interfaces, presents the opportunity to scale the throughput of this system arbitrarily, potentially permitting simultaneous fabrication across an entire multi-well plate. Moreover, unique to DIP is the ability to manipulate and excite a fluidic interface which enables direct fluid manipulation during the printing process, over-printing, configurable mass transport and three-dimensional patterning. It's foreseeable that additional modalities including acoustically driven transport systems[33] and multi-material sequencing could be incorporated into future design iterations, further extending the capabilities. Additionally, further patterning enhancement could be achieved by actuating the underlying structure[49] or configuring the print heads boundary topology to extend the available pattern complexity[32,33].

We envision that DIP offers significant advancements for biofabrication due to its ability to print high resolution constructs in soft, biologically relevant materials without necessitating thermal gelation or imposing optical characteristics of the underlying resin. The capability to spatially localise the print interface in three dimensions is also advantageous for biofabrication, enabling in-situ fabrication into multi-well plates, underscoring the future potential for automation. Additionally, our demonstration of acoustic excitation in three-dimensions enables simultaneous cellular patterning, crucial for cellular functionality in many tissue constructs[50,51]. Future DIP iterations could further extend to higher numerical aperture optics[52], wherein microscale structures could be created at high speeds without the cost associated with two-photon systems[53,54].

In summary, DIP represents a rapid and conceptually elegant printing approach that relies on the formation of a constrained and acoustically modulated air-liquid interface. Using DIP, we demonstrate a multifunctional, high-speed and high-throughput approach that has unique benefits for the fabrication of soft and biologically relevant materials. We therefore posit that DIP is best utilized where high-speed, high-resolution, and in situ fabrication of three-dimensional structures is required.

**References**

1. Bao, Y., Paunović, N. & Leroux, J. Challenges and Opportunities in 3D Printing of Biodegradable Medical Devices by Emerging Photopolymerization Techniques. Adv Funct Mater 32, 2109864 (2022).
2. Martinez, D. W., Espino, M. T., Cascolan, H. M., Crisostomo, J. L. & Dizon, J. R. C. A Comprehensive Review on the Application of 3D Printing in the Aerospace Industry. Key Eng Mater 913, 27–34 (2022).
3. Lee, K.-S., Kim, R. H., Yang, D.-Y. & Park, S. H. Advances in 3D nano/microfabrication using two-photon initiated polymerization. Prog Polym Sci 33, 631–681 (2008).
4. Zheng, X. et al. Design and optimization of a light-emitting diode projection micro-stereolithography three-dimensional manufacturing system. Review of Scientific Instruments 83, 125001 (2012).
5. Lee, A. et al. 3D bioprinting of collagen to rebuild components of the human heart. Science (1979) 365, 482–487 (2019).
6. Farsari, M. & Chichkov, B. N. Two-photon fabrication. Nat Photonics 3, 450–452 (2009).
7. Raman, R. et al. High-Resolution Projection Microstereolithography for Patterning of Neovasculature. Adv Healthc Mater 5, 610–619 (2016).
8. Ge, Q. et al. Projection micro stereolithography based 3D printing and its applications. International Journal of Extreme Manufacturing 2, 022004 (2020).
9. Kelly, B. E. et al. Volumetric additive manufacturing via tomographic reconstruction. Science (1979) 363, 1075–1079 (2019).
10. Regehly, M. et al. Xolography for linear volumetric 3D printing. Nature 588, 620–624 (2020).
11. Sanders, S. N. et al. Triplet fusion upconversion nanocapsules for volumetric 3D printing. Nature 604, 474–478 (2022).
12. Bernal, P. N. et al. Volumetric Bioprinting of Complex Living-Tissue Constructs within Seconds. Advanced Materials 31, 1904209 (2019).
13. Toombs, J. T. et al. Volumetric additive manufacturing of silica glass with microscale computed axial lithography. Science (1979) 376, 308–312 (2022).
14. Loterie, D., Delrot, P. & Moser, C. High-resolution tomographic volumetric additive manufacturing. Nat Commun 11, 852 (2020).
15. Dinc, N. U. et al. From 3D to 2D and back again. Nanophotonics 12, 777–793 (2023).
16. Gehlen, J., Qiu, W., Schädli, G. N., Müller, R. & Qin, X.-H. Tomographic volumetric bioprinting of heterocellular bone-like tissues in seconds. Acta Biomater 156, 49–60 (2023).
17. Rodríguez-Pombo, L. et al. Volumetric 3D printing for rapid production of medicines. Addit Manuf 52, 102673 (2022).
18. Wang, B. et al. Stiffness control in dual color tomographic volumetric 3D printing. Nat Commun 13, 367 (2022).
19. Orth, A., Sampson, K. L., Ting, K., Boisvert, J. & Paquet, C. Correcting ray distortion in tomographic additive manufacturing. Opt Express 29, 11037 (2021).
20. Hahn, V. et al. Light-sheet 3D microprinting via two-colour two-step absorption. Nat Photonics 16, 784–791 (2022).
21. Xu, H., Casillas, J., Krishnamoorthy, S. & Xu, C. Effects of Irgacure 2959 and lithium phenyl-2,4,6-trimethylbenzoylphosphinate on cell viability, physical properties, and microstructure in 3D bioprinting of vascular-like constructs. Biomedical Materials 15, 055021 (2020).
22. Tumbleston, J. R. et al. Continuous liquid interface production of 3D objects. Science (1979) 347, 1349–1352 (2015).
23. Lipkowitz, G. et al. Injection continuous liquid interface production of 3D objects. Sci Adv 8, (2022).
24. Zhang, F. et al. The recent development of vat photopolymerization: A review. Addit Manuf 48, 102423 (2021).
25. Lewis, K. & Matsuura, T. Bézier Curve Method to Compute Various Meniscus Shapes. ACS Omega 8, 15371–15383 (2023).
26. Lewis, K. & Matsuura, T. Calculation of the Meniscus Shape Formed under Gravitational Force by Solving the Young–Laplace Differential Equation Using the Bézier Curve Method. ACS Omega 7, 36510–36518 (2022).

27. Nair, K. et al. Characterization of cell viability during bioprinting processes. Biotechnol J 4, 1168–1177 (2009).
28. Nguyen, A. K., Goering, P. L., Reipa, V. & Narayan, R. J. Toxicity and photosensitizing assessment of gelatin methacryloyl-based hydrogels photoinitiated with lithium phenyl-2,4,6-trimethylbenzoylphosphinate in human primary renal proximal tubule epithelial cells. Biointerphases 14, (2019).
29. Hong, S.-H. et al. Surface waves control bacterial attachment and formation of biofilms in thin layers. Sci Adv 6, (2020).
30. Chen, P., Güven, S., Usta, O. B., Yarmush, M. L. & Demirci, U. Biotunable Acoustic Node Assembly of Organoids. Adv Healthc Mater 4, 1937–1943 (2015).
31. Liu, X. & Wang, X. Polygonal patterns of Faraday water waves analogous to collective excitations in Bose–Einstein condensates. Nat Phys (2023) doi:10.1038/s41567-023-02294-y.
32. Chen, P. et al. Microscale Assembly Directed by Liquid-Based Template. Advanced Materials 26, 5936–5941 (2014).
33. Guan, J. H., Magoon, C. W., Durey, M., Camassa, R. & Sáenz, P. J. Traveling Faraday waves. Phys Rev Fluids 8, 110501 (2023).
34. Zhang, S., Borthwick, A. G. L. & Lin, Z. Pattern evolution and modal decomposition of Faraday waves in a brimful cylinder. J Fluid Mech 974, A56 (2023).
35. Périnet, N., Gutiérrez, P., Urra, H., Mujica, N. & Gordillo, L. Streaming patterns in Faraday waves. J Fluid Mech 819, (2017).
36. Huang, Y., Wolfe, C. L. P., Zhang, J. & Zhong, J. Q. Streaming controlled by meniscus shape. J Fluid Mech 895, (2020).
37. Kolesnik, K. et al. Periodic Rayleigh streaming vortices and Eckart flow arising from traveling-wave-based diffractive acoustic fields. Phys Rev E 104, 045104 (2021).
38. Raymond, S. J. et al. A deep learning approach for designed diffraction-based acoustic patterning in microchannels. Sci Rep 10, 8745 (2020).
39. WESTRA, M.-T., BINKS, D. J. & VAN DE WATER, W. Patterns of Faraday waves. J Fluid Mech 496, S0022112003005895 (2003).
40. Bernal, P. N. et al. Volumetric Bioprinting of Organoids and Optically Tuned Hydrogels to Build Liver-Like Metabolic Biofactories. Advanced Materials 34, 2110054 (2022).
41. Größbacher, G. et al. Volumetric Printing across Melt Electrowritten Scaffolds Fabricates Multi-Material Living Constructs with Tunable Architecture and Mechanics. Advanced Materials (2023) doi:10.1002/adma.202300756.
42. Schmidleithner, C. & Kalaskar, D. M. Stereolithography. in 3D Printing (InTech, 2018). doi:10.5772/intechopen.78147.
43. Belay, B. et al. Optical projection tomography as a quantitative tool for analysis of cell morphology and density in 3D hydrogels. Sci Rep 11, 6538 (2021).
44. Madrid-Wolff, J., Boniface, A., Loterie, D., Delrot, P. & Moser, C. Controlling Light in Scattering Materials for Volumetric Additive Manufacturing. Advanced Science 9, 2105144 (2022).
45. Madrid-Wolff, J., Boniface, A., Loterie, D., Delrot, P. & Moser, C. Controlling Light in Scattering Materials for Volumetric Additive Manufacturing. Advanced Science 9, (2022).
46. Hsiao, K. et al. Single-digit-micrometer-resolution continuous liquid interface production. Sci Adv 8, (2022).
47. Wright, P. H. & Saylor, J. R. Patterning of particulate films using Faraday waves. Review of Scientific Instruments 74, 4063–4070 (2003).
48. Chladni, E. F. F. Entdeckungen Uber Die Theorie Des Klanges. (Readex Microprint, 1967).
49. Harley, W. S., Kolesnik, K., Xu, M., Heath, D. E. & Collins, D. J. 3D Acoustofluidics via Sub-Wavelength Micro-Resonators. Adv Funct Mater 33, (2023).
50. Di Marzio, N. et al. Sound-based assembly of a microcapillary network in a saturn-like tumor model for drug testing. Mater Today Bio 16, 100357 (2022).
51. Armstrong, J. P. K. et al. Engineering Anisotropic Muscle Tissue using Acoustic Cell Patterning. Advanced Materials 30, (2018).
52. Vidler, C., Crozier, K. & Collins, D. Ultra-resolution scalable microprinting. Microsyst Nanoeng 9, 67 (2023).
53. Jing, X., Fu, H., Yu, B., Sun, M. & Wang, L. Two-photon polymerization for 3D biomedical scaffolds: Overview and updates. Front Bioeng Biotechnol 10, (2022).

54. Geng, Q., Wang, D., Chen, P. & Chen, S.-C. Ultrafast multi-focus 3-D nano-fabrication based on two-photon polymerization. Nat Commun 10, 2179 (2019).
55. Zhu, M. et al. Gelatin methacryloyl and its hydrogels with an exceptional degree of controllability and batch-to-batch consistency. Sci Rep 9, 6863 (2019).
56. Ooi, H. W. et al. Thiol–Ene Alginate Hydrogels as Versatile Bioinks for Bioprinting. Biomacromolecules 19, 3390–3400 (2018).
57. Brainard, D. H. The Psychophysics Toolbox. Spat Vis 10, 433–436 (1997).
58. Behroodi, E., Latifi, H. & Najafi, F. A compact LED-based projection microstereolithography for producing 3D microstructures. Sci Rep 9, 19692 (2019).

# Methods

**3D Printer Assembly**

System components were mounted on a pair of orthogonal optical breadboards to facilitate the alignment of the vertical and horizonal components of the system. Patterned cross-sections of the object were generated using a high-power projection module (LRS-WQ, Visitech) with a resolution of 2560 x 1600 pixels and pixel size of 15.1 µm. The projection module was mounted to a linear stage (MOX-02-100, Optics Focus), which was affixed to the vertically oriented optical breadboard. Direct control of the dynamic interface was performed via another linear stage (MOX-02-50, Optics Focus) that controlled the displacement of a 50 mL syringe connected to the print head via a silicone hose. An additional pair of linear stages (MOX-02-100, Optics Focus) was used to position the cuvette/well plate below the print head in two dimensions for sequential or multi-step printing (**Supplementary Fig.2-3**). System control was executed via a custom MATLAB graphical user interface (GUI) that enabled the management of motorized linear stages through RS232, control of the acoustic modulation device, control of the projection module parameters, and the transmission of cross-sectional images via HDMI.

**Print Head**

In this study, the print head was tailored to various dimensions, contingent upon the desired dimensions of the resin container. For almost all configurations, we utilized axis-symmetric cylindrical print heads with the benefit of simplifying the computation of the interface shape, although other (arbitrarily shaped) print head boundary contours were feasible as demonstrated. In general print heads ranging from 30 mm to 5 mm were primary utilised. The objects extent in the $x, y$ direction was limited by the projector's total field of view at the focal plane. The object height was limited by the length of the print head, which in our case was equal to the projection focal length. For our setup, total submergible print head length was approximately 70 mm. Of note, much taller structures are conceptually feasible by submerging the projection and illumination optics, or by increasing the working distance of the projection optics. The print head was fabricated using a commercial 3D printing system (Form 3+, Formlabs), with a threaded insert to quickly interchange print heads. Additionally, a glass window was clamped between a gasket to form an air-tight enclosed volume, whilst facilitating the transmission of light down its centre (**Supplementary Fig.1a-b**). An internal channel was also added to enable gas delivery into the print head cavity via the syringe system and acoustic modulation device. This port was used to either maintain or modulate the shape of the air-liquid interface during printing.

**Acoustic Modulation Device**

Acoustic modulation of the air-liquid interface was achieved via direct volume manipulation of the air-volume contained within the print head. Conceptually, the approach was straightforward and consisted of a 3" 15W speaker driver (Techbrands, AS3034) affixed to an enclosed 3D printed manifold containing an inlet and outlet port (**Fig. 3a**, **Supplementary Fig.1c-d**). The speaker was driven by a commercially available amplifier (Adafruit, MAX9744) using the supplied auxiliary port, with specified waveforms sent by the MATLAB GUI. Frequency ranges of 5 - 500 Hz were used, with fixed or transient frequency switching depending on the structure. By specifying a waveform for each degree of freedom, it was straightforward to synchronise the acoustic modulation with the remainder of the motion and pressure control (**Fig. 1c**). The acoustic modulation device operated as an in-line control unit, such that the inlet port was connected to the syringe system and the outlet port was connected to the print head. This facilitated pressurisation of the enclosed system and modulation about the pressurised set-point.

**Material composition and preparation**

**PEGDA materials:** various PEGDA materials were utilised in this study ranging from 10% w/v to 100% w/v. Each formulation followed the same protocol whereby the required weight fraction of PEGDA $M_n$ 700 (#455008, Sigma) dissolved into 40g of 40°C deionized water (excluding 100% w/v) and thoroughly mixed for 10 minutes. Subsequently, 0.1% w/w of Tartrazine (#T0388, Sigma) and 0.25% w/w of Lithium phenyl-2,4,6-trimethylbenzoylphosphinate (LAP) (#900889, Sigma) were added to the mixture and stirred until complete dissolution. Materials were then stored in light-safe falcon tubes until required.

**HDDA material:** A solution of 500 mg of Phenylbis(2,4,6-trimethylbenzoyl) phosphine oxide (511447, Sigma) and 50 g of 1,6-Hexanediol diacrylate (#246816, Sigma) was prepared by warming the mixture to 40°C and stirring for 30 minutes. To control the resolution in the z-direction, the photo-absorber Sudan I (#103624, Sigma) was added in various quantities ranging from 0 – 0.04% w/w. Materials were then stored in light-safe falcon tubes until required.

**GelMA material:** GelMA was synthesized following a previously reported protocol (Ref [55]), yielding a degree of substitution of 93% (confirmed by NMR). Next, a 10% w/v GelMA solution was prepared by dissolving 1 g of GelMA in 10 mL of cell culture media (Freestyle 293 Expression Medium, Thermofisher) preheated to 37°C. After complete dissolution of GelMA, 100 mg of Tartrazine and 25 mg of LAP were added to the solution, which was maintained at 37°C until complete dissolution. The mixture was sterilized by passing it through a 0.22 µm sterile filter within a biosafety cabinet and subsequently stored in refrigerated light-safe falcon tubes until required.

**Alginate material:** Norbornene-functionalized sodium alginate (AN) was synthesized based on a previously reported protocol[56]. In short, 10 g of sodium alginate were dissolved in 500 ml of 0.1 M 2-(N-Morpholino) ethane-sulfonic acid buffer (#145224-94-8 Research Organics) and fixed to pH 5.0; 9.67 g of 1-Ethyl-3-(3-dimethylaminopropyl) carbodiimide•HCl, 2.90 g of N-hydroxysuccinimide, and 3.11 ml of 5-norbornene-2-methylamine were added. The pH was fixed at 7.5 with 1 M NaOH, and the reaction was carried out at room temperature for 20 hours. The mixture was dialyzed against water for 5 days prior to lyophilisation. The degree of norbornene functionalization was determined to be 16.2% by $^1$H NMR. A 7% w/v AN solution was prepared by dissolving 1 g of AN in 14.29 ml of phosphate buffered saline (PBS) solution. Next, 200 mg of Tartrazine, 20 mg of LAP, and 122.7 µl of 2,2'-(ethylenedioxy)diethanethiol were dissolved in 5.59 ml of PBS, and added to the AN solution and mixed until homogenous. The pH was adjusted with 1 M NaOH until the solution was visibly opaque.

**UDMA support material:** A solution of 50 mg of Phenylbis(2,4,6-trimethylbenzoyl) phosphine oxide (511447, Sigma) and 5 g of Diurethane dimethacrylate (#436909, Sigma) was prepared by warming the mixture to 45ºC and stirring for 30 minutes. To remove trapped air-bubbles the mixture was then transferred to a light-safe falcon tube and centrifuged at 4000 rpm for 10 minutes to remove residual air bubbles. This material was then used as the base support for the free-floating print test.

### Cell printing

Human embryonic kidney (HEK) 293-F cells (Freestyle 293-F, Thermo Fisher) were used to determine the preliminary viability of the DIP 3D printing system. Unlike other volumetric printing methods, high cell densities can be printed easily without requiring the refractive index between the cells and the printing medium to be matched. In this work, a cell solution with 7.2 million cells/mL was used for both the model of the kidney and the cell-viability measurements. To determine the cell viability, a thin 500 µm wall was printed to minimize the effect of cell death via insufficient media diffusion and imaged using a LIVE/DEAD viability/toxicity kit (L3224, Invitrogen). Three wall structures were printed, and measurements were taken after 24 hours to determine the preliminary viability of the technique. Cell viability was determined in three locations for each sample, with the total viability being an average of all collection points (**Supplementary Fig.18**).

To create the cell-loaded bio-ink, the GelMA solution was warmed to 37ºC followed by the resuspension of cells into the solution. The solution was passed through a cell strainer (#0877123, Thermo Fisher) and stored in a water bath at 37ºC while not in use. The printing process involved pipetting approximately 3 mL of the GelMA ink into a single well of a 12-well plate and lowered the print head into the well. Before each print, the motorized syringe was used to resuspend the cells by sequentially applying positive and negative pressure (analogous to pipetting the liquid up and down) to prevent cell settling prior to printing. To reduce the likelihood of the printed object detaching from the bottom of the well-plate, a slower print velocity of 150 µm/s was used, resulting in an object creation time of approximately 30 seconds.

### Data pre-processing, printing, and post-processing

3D design models of the Bowman's capsule, tri-helix structure and kelvin cell were created using nTopology, nTop. Tricuspid, heart, and buckyball models were downloaded from Thingiverse.com. For each geometry, the STL file was extracted and sliced using Chitubox into a stack of PNG images. As the framerate of the HDMI signal was limited to 120 fps, we commonly utilised projection frame rates that matched the acoustic driving frequencies to minimise interface motion blurring. The object was therefore discretised, into a voxel array according to the desired linear print speed and frame rate. The layer height ($L_h$) was determined as $L_h = \frac{V_z}{f}$, where $V_z$ was the linear print speed and $f$ was the acoustic excitation frequency which matched the projection frequency. Once discretised, the image stack was transformed by the convex slicing algorithm producing a secondary convex-optimized image stack, with the sequence being sent to the projector via a HDMI signal using Psychtoolbox-3[57]. The print sequence started by moving the print head to a defined distance above the print surface (or high-density material), where the interface was automatically generated by displacing the syringe dependent on the selected print head. The MATLAB GUI was operated by first sending a signal to turn on the LED module and subsequently controlling the air-liquid interface location by modulating the pressure, acoustic driving and translation location. The optical power of the projection module was automatically set dependent on the selected print velocity using the parameter space matrix. For prints made with HDDA, the printed structures were removed from print volume and washed with isopropyl alcohol. For soft structures made from PEGDA and GelMA, the excess material was gently removed using a pipette (recycled) and the structures were resuspended in deionized water or cell culture media to wash away remaining un-polymerized material. The structures were then gently fluidically detached from the bottom of the material container and stored under deionized water or cell culture media.

**Convex slicing algorithm**

The developed convex slicing algorithm aims to correct for geometrical discretisation differences between a traditionally flat construction surface and the curved surface utilised in this work. A full explanation of the convex slicing process is outlined in the Supplementary Materials; however, the main components will be briefly restated here. Firstly, the general shape of the interface is determined by the Young-Laplace equation which describes the Laplace pressure difference ($\Delta p$) sustained across a gas-liquid boundary dependent on the materials surface tension ($\gamma$) and surface normal ($\hat{n}$).

$$\Delta p = -\gamma \nabla \cdot \hat{n}$$

In this study we utilised axisymmetric print containers such that $\hat{n}$ can be easily found by substituting the general expressions for principal curvatures. By normalising by the capillary length, $l = \sqrt{\frac{\gamma}{\rho g}}$, where $\gamma$ is the materials surface tension, $\rho$ is the materials density and, $g$ is gravity. The radial and vertical coordinates of the interface non-depersonalised to $x = \frac{r}{l}$ and $y = \frac{z}{l}$, resulting in the partial differential equation for the interface shape is given by:

$$\frac{y''}{(1+(y')^2)^{\frac{3}{2}}} + \frac{y'}{x(1+(y')^2)^{\frac{1}{2}}} - y = 0$$

This equation can be readily solved using numerical integration with appropriate boundary conditions (**Supplementary Materials: S4**), however utilizing this method would require numerical integration for not only the steady-state case, but for each intermediate region during interface compression. We instead opted to solve the PDE using a cubic Bézier curve approximation for the steady-state case and geometrically deform the Bézier control points under an equality constraint (**Supplementary Materials: S7**). This is significantly computationally faster given the large number of intermediate surfaces within the transient region. To produce each surface, the boundaries half profile was revolved about the print heads z-axis, to form a three-dimensional surface (**Supplementary Materials: S5**). This therefore produced a sequence of surfaces starting at the compressed state and transitioning to the tangent state, followed by the steady-state interface profile for the remainder of the model's height. The corresponding convex projection(s) were determined by Euclidian distance minimisation between the cartesian voxel grid and the surface arrays (**Supplementary Materials: S6**). Reconstruction accuracy was validated by 'replaying' the projections over an empty voxel array and computing the Jaccard index between reconstructed voxel array and the input voxel array (**Supplementary Materials: S8-9**).

**Optical Modelling:**

To determine our theoretical optical model (**Supplementary Materials: S10**), we employed a similar approach to Behroodi *et al*.[58] , that models the in-plane resolution as the spatial convolution of the point spread function ($PSF(x,y)$) and the micro-mirror spatial arrangement ($f(x,y)$), given by the following:

$$f(x,y) * PSF(x,y) = \int_{\tau_1=-\infty}^{\infty} \int_{\tau_2=-\infty}^{\infty} f(\tau_1, \tau_2) \cdot PSF(x-\tau_1, y-\tau_2) d\tau_1 d\tau_2$$

This was then used to determine the effective delivered energy and depth of cure across the meniscus by decomposing a planar incident ray into reflective and transmissive components $\eta(n_1, n_2, \hat{u}, \hat{n})$ and modelling the energy intensity along the transmissive vector $\gamma_z$ as a material dependent Beer-Lambert decay (**Supplementary Materials: S11**), given by:

$$\mathcal{H}(\boldsymbol{\gamma_x}, \boldsymbol{\gamma_y}, \boldsymbol{\gamma_z}) = \eta(n_1, n_2, \hat{u}, \hat{n}) \cdot \widehat{\boldsymbol{E}} \; e^{\frac{-\gamma_z}{\varepsilon_d[D]+\varepsilon_i[S]}}$$

As the interface is curved, the effective resolution is spatially dependent on the local meniscus height away from the focal plane. Therefore, for each pixel in the projected image, we mapped its local coordinate to corresponding coordinate on the meniscus surface. This was used to theoretically predict the accurate area fraction for a given print head size, shape and material properties, and its practical impact on resolution (**Supplementary Materials: S12**).

**Acoustically Driven Flow**

**Analytical solution:** To understand the formation of acoustically driven capillary-gravity waves, we utilised many established analytical approaches that describe the induced velocity and secondary streaming effects[36] created by the meniscus (**Supplementary Materials: S13-15**). This analysis therefore establishes velocity scaling laws for capillary-gravity waves dependent on the dominance of capillary of gravity driven effects. The dispersion relation for capillary waves, which relates the wave frequency ($\omega$) to the wavenumber ($k$), is first given by:

$$\omega^2 = \frac{\gamma}{\rho} k^3 + gk$$

We therefore show (**Supplementary Materials: S14**) that $\left(\frac{\lambda}{l_{cap}}\right)^2$ provides a unitless quantity that relates the dominance of surface tension and acoustic parameters on flow magnitude, where the flow velocity scales with:

$$U \propto \frac{h_0^2 \rho g \phi}{\lambda \mu}, \text{for } \left(\frac{\lambda}{l_{cap}}\right) > 1$$

$$U \propto \frac{h_0^2 \gamma \phi}{\lambda^3 \mu}, \text{for } \left(\frac{\lambda}{l_{cap}}\right) < 1$$

**Supplementary Fig.19** shows the effect of material and acoustic parameters on velocity scaling.

**Experimental investigation:** Particle image velocimetry was employed to understand the three-dimensional flow field produced below the air-liquid boundary under acoustic excitation. A high-speed camera (Kron Technologies, Chronos 1.4 Camera) was used to capture footage of 20-50 μm PMMA particles during excitation normal and orthogonal to the air-liquid boundary. Particle tracing and velocity reconstruction was performed on the captured video sequences using PIVLab for MATLAB, the exact parameters and methodology used can be found in (**Supplementary Materials: S18**). The velocity profiles for top-down close to a boundary, side on close to a boundary and side on above a boundary are shown in (**Fig.3c-e, Supplementary Fig.8-9**).

**Interface re-stabilisation:** To determine transient interface re-stabilisation in bulk flow, high-speed photography under a uniform backlight was captured at 5000 fps (**Supplementary Fig.7**). Mencius edge tracking was achieved by a custom MATLAB script that segmented and detected the boundary edge for each image sequence. The boundary displacement was calculated at the centroid location and plotted over all frames. The termination of acoustic stimulation was software triggered, with the stability criterion set at $(1/e^2)$ of the starting amplitude.

**Image analysis of material influx rate:** Material influx rate under acoustic excitation was determined by filling a glass cuvette with materials doped with black dye to prevent light transmission. The cuvette was placed on top of a red backlight, such that when the air-liquid boundary is formed against the base of the cuvette light transmission is observed by a CCD (**Supplementary Materials: S19**). Material influx rate was measured by raising the air-liquid boundary with and without acoustic excitation and tracking the influx of dyed material which occluded the backlight transmission (**Fig. 3f**, **Supplementary Fig.11).**

## Print Parameter Space

To identify the ideal parameter space for dynamic interface printing, a range of print speed and optical dose combinations were tested using three materials—PEGDA, GelMA, and HDDA. For each combination, a 5 x 5 x 15 mm rectangular structure was printed, with successful outcomes being defined by the presence of a sharply delineated structure and smooth surface finish. Structures that did not meet these requirements, either by only partially resolving or producing no structure, were removed from the parameter map. Generally, the print speed parameter space is not only constrained by the optical dosage, but rather by the rate at which new material can 'wet' the interface. Inadequate wetting typically causes the interface to fluidically 'pin' to the underlying structure as the polymer solidifies quickly just below the interface, resulting in a local region of no material.

## Microscopy

**MicroCT (μCT):** μCT images were acquired using a Phoenix Nanotom M scanner (Waygate Technologies, voxel size = 10 μm³, 90 kV tube voltage, 200μA tube current, 8 min scan time). For hydrogel samples, the structures were briefly dried with tissue paper and mounted into a falcon tube for imaging. For hard materials such as HDDA, the structures were placed on a plastic cap to provide good contrast between the printed structure and the supporting medium. An STL surface mesh was extracted and imported into Keyshot 11 (Keyshot, Luxion) to render the final microCT representation.

**Florescence microscopy:** Florescence microscopy images were performed using a Zeiss Axio Observer Z1 (Zeiss, Germany) using either a 4X or 10X objective. For large constructs that were greater than the objective's field of view, the images were stitched within the Zeiss Zen software to create a large format image. Once the fluorescence images were acquired, cell counting was performed on each live/dead image pair using a custom MATLAB script.

**Scanning electron microscopy (SEM):** SEM images were acquired on a FlexSEM 1000 (Hitachi High Technologies, Japan). Printed structures on glass slides were mounted directly to the microscope stage with no further sample preparation. The samples did not have a conductive coating applied. The FlexSEM was operated in variable-pressure mode at 50 Pa, and images were acquired with a 15 keV beam using the ultra-variable detector (UVD). To cover the field of view needed for the large structures, the working distance was typically 40-50 mm, and multiple images were collected in a tiled manner and stitched together in post-processing.

**Helium ion microscopy images (HIM):** HIM images were acquired using the Zeiss NanoFab using the helium source. During imaging the flood-gun was used to actively neutralize the surface removing the need for a conductive coating. All structures were imaged using an accelerating voltage of 30kV, a beam current of between 1-2pA and a field of view of 1100 µm. Structures were printed directly onto a silanized glass slide and were mounted to the stage using the integrated mounting clips. To facilitate capturing structures larger than the field of view, multiple images were taken and later stitched using ImageJ/Fiji.

**Acknowledgments**
We thank R. Dagastine for his continuous support of the project and K. Brenner for her help with manuscript proof reading. We would also like to acknowledge that this work was performed in part at the Materials Characterisation and Fabrication Platform (MCFP) at the University of Melbourne and the Victorian Node of the Australian National Fabrication Facility (ANFF).

Funding: Dr David Collins is the recipient of a Discovery Early Career Researcher Award from the Australian Research Council (DECRA, DE200100909; Discovery Project, DP230102550), and funding from the National Health and Medical Research Council (Ideas, APP2003446). Associate Professor Daniel Heath is the recipient of a Future Fellowship from the Australian Research Council (FT190100280). Kenneth Crozier acknowledges funding from the Australian Research Council Centre of Excellence for Transformative Meta-Optical Systems (TMOS, Project No. CE200100010). We would also like to acknowledge that this project was partially supported by the Royal Melbourne Hospital and University of Melbourne Innovation Acceleration Program (IAP).

**Author contributions:**
Conceptualization: C.V and D.C. Methodology C.V and D.C. Investigation: C.V, M.H, K.K, P.S, M.M, A.B Visualization: C.V, D.C. Funding acquisition: C.V, D.C, E.K, A.R, D.S, D.H, K.C. Project administration: C.V, D.C. Writing – original draft: C.V. Review and editing: C.V, M.H, K.K, P.S, M.M, A.B, E.K, A.R, D.S, D.H, K.C, D.C.

**Competing interests:**

C.V, M.H and D.C have submitted a provisional patent (Application Number: 2023901976) held/submitted by The University of Melbourne that covers dynamic interface printing of three-dimensional structures. The authors certify that they have no affiliations with or involvement in any organization or entity with any financial interest or nonfinancial interest in the subject matter or materials discussed in this manuscript.

**Data and materials availability:**

The data that support the findings of this study are available within the paper and Supplementary Information. Additional supporting data generated during the present study such as software and geometry files are available from the corresponding authors upon reasonable request.

Supplementary Materials for
# Dynamic Interface Printing

Callum Vidler[1], Michael Halwes[1], Kirill Kolesnik[1], Philipp Segeritz[1,7], Matthew Mail[1], Anders J. Barlow[2], Emmanuelle M. Koehl[5], Anand Ramakrishnan[5,6], Daniel J. Scott[7], Daniel E. Heath[1,9], Kenneth B. Crozier[3,4,8], David J. Collins[1,9*]

Affiliations:

[1]Department of Biomedical Engineering, The University of Melbourne, Melbourne, Victoria, Australia

[2]Materials Characterisation and Fabrication Platform (MCFP), The University of Melbourne, Parkville, Victoria, 3010, Australia

[3]School of Physics, The University of Melbourne, Victoria 3010, Australia

[4]Department of Electrical and Electronic Engineering, The University of Melbourne, Victoria 3010, Australia

[5]Department of Plastic and Reconstructive Surgery, The Royal Melbourne Hospital, Victoria 3050, Australia

[6] The University of Melbourne, Melbourne Medical School, Department of Surgery, Royal Melbourne Hospital, Parkville VIC 3050

[7]The Florey and Department of Biochemistry and Pharmacology, The University of Melbourne, Parkville, Victoria, 3010, Australia

[8]Australian Research Council (ARC) Centre of Excellence for Transformative Meta-Optical Systems, The University of Melbourne, Victoria 3010, Australia

[9]The Graeme Clark Institute, The University of Melbourne, Parkville 3052, Victoria, Australia

*Corresponding author
E-mail: david.collins@unimelb.edu.au

**This PDF File Includes:**

    Materials and Methods S1 to S22

    Supplementary Figs. 1 to 28

    Table 1

    Captions for Movies 1 to 5

    References

# Materials and Methods

## S1: DIP printing system mechanical design

**Version 1**

System components were mounted on a pair of orthogonal optical breadboards to facilitate the alignment of the vertical and horizonal components of the system (**Supplementary Fig.2a).** Patterned cross-sections of the object were generated using a high-power projection module (LRS-WQ, Visitech) with a resolution of 2560 x 1600 pixels and pixel size of 15.1 µm. The projection module was mounted to a linear stage (MOX-02-100, Optics Focus), which was affixed to the vertically oriented optical breadboard. Direct control of the dynamic interface was performed via another linear stage (MOX-02-50, Optics Focus) that controlled the displacement of a 50 mL syringe connected to the print head via a silicone hose. An additional pair of linear stages (MOX-02-100, Optics Focus) was used to position the cuvette/well plate below the print head in two dimensions for sequential or multi-step printing. Stage motion control was achieved using a commercially available 3D printer control board (BIGTREETECH, SKR 3) and a custom designed DB9 breakout board to interface with the motion stages.

Orthogonal video of the printing process was captured using a 4K CCD camera (AmScope, HD408) with a 16 mm lens (Raspberry Pi, RPI-16MM-LENS).

**Version 2**

To incorporate in-situ imaging, a second revision of the system was developed, allowing the probe and associated optical hardware to remain stationary while the printing vessel moved relative to the fixed probe (**Supplementary Fig. 3).** The primary mechanical modifications included the integration of a custom-made coreXY translation system and a NEMA 23 ball-screw linear stage for the z-axis to accommodate the increased vertical payload of the XY gantry. Additionally, a blue reflective dichroic mirror (#35-519, Edmund Optics) and a 50:50 beam splitter (#43-359, Edmund Optics) were added to facilitate in-situ imaging of the interface and structures during fabrication. Illumination for the system can be provided coaxially via an expanded fiber optic light source or through collimated back illumination. In the latter case, this was achieved with a custom-manufactured collimated backlight featuring an integrated heating element to maintain physiologically relevant temperatures during fabrication of cell-laden materials or materials that undergo thermal gelation (e.g. GelMA).

**Acoustic Modulation**

An inline acoustic modulation device was placed between the syringe and the print head which facilitated direct volume manipulation of the air-liquid interface (**Supplementary Fig.2b**). This modulation device consisted of a 3" 15W speaker driver (Techbrands, AS3034) affixed to an enclosed 3D printed manifold containing an inlet and outlet port, such that the speaker cone forms a single side of the enclosed manifold (**Supplementary Fig.1c-d**). The speaker was driven using a 20W audio amplifier (Adafruit, ADA1752) enabling the direct control of the frequency and amplitude of the air-liquid interface via a 3.5 mm auxiliary cable.

**S2: Print head design**

The print head was tailored to various dimensions, contingent upon the desired dimensions of the resin container. We used axisymmetric cylindrical print heads to simplify the computation of the interface shape, although other (arbitrarily shaped) print head boundary contours were feasible. Print head sizes ranged from D = 25 mm to D = 5 mm. The object's size in the $x,y$ direction was limited by the projector's total field of view at the focal plane and the object height was limited by the length of the print head, which in our case was equal to the projection focal length. For our setup, the total submergible print head length was approximately 70 mm. Much taller structures would be feasible by submerging the projection and illumination optics, or by demagnifying the projection optics in order to increase the working distance. The print head consisted of 6 parts, which when combined created an enclosed air-volume with a glass window at its top to enable light transmission down its centre **(Supplementary Fig.1a-b)**. Additionally, a pneumatic channel located on the side of the print head enables direct pressurization and acoustic excitation of the air-liquid boundary. All components were 3D printed using a commercial resin 3D printer (Formlabs, Form3+).

**S3: Software control**

A custom MATLAB GUI was used to control the DIP printing system which enabled the voxelization of STL geometries, pre-processing of image arrays using the convex slicing algorithm, motion control via G-code over a serial data connection, video capture, video transmission to the projection module over HDMI, projection module control, and acoustic modulation. Printing of structures was performed by first generating a waveform for each degree of freedom of the interface, whereby the global position of the interface was dependent on the summation of all waveforms. This approach allowed us to create highly complex motion control as shown in **Fig.1c.**

**S4: Air-liquid interface modelling**

In DIP, the shape of the interface can be approximately described by the Young-Laplace equation which relates the interface curvature to the differential pressure sustained across the boundary. In general, this can be written as the following:

$$\Delta p = -\gamma \nabla \cdot \hat{n}, \tag{1}$$

where $\Delta p$ denotes the Laplace pressure, $\gamma$ is the surface tension and $\hat{n}$ is the vector normal to the surface. Following the approach from Butt *et al.*[1], the shape of the non-dimensionalized interface can be found by substituting the general expressions for the principal curvatures[2] of an axisymmetric surface as shown below:

$$p_i - \rho g z = \gamma \left( \frac{r''}{(1+r'^2)^{\frac{3}{2}}} - \frac{1}{r(1+r'^2)^{\frac{1}{2}}} \right), z \in (0, h). \tag{2}$$

The coordinate origin is taken as the contact point of the meniscus edge with the print head, with the positive z-axis being directed downward along the print head's central axis and the r-axis parallel to the print-head's diameter. The superscript prime denotes the derivative with respect to z, and $h$ denotes the maximum height of the meniscus given by:

$$\frac{h}{R_o} = 2(Bo)^{-\frac{1}{2}} \sin\left(\frac{\theta_Y}{2}\right), \tag{3}$$

where $R_o$ is the radius of the theoretical spherical meniscus with volume $V_o = \frac{4\pi}{3} R_o^3$, $Bo$ denotes the bond number and $\theta_Y$ is the contact angle. Therefore, the shape of the meniscus can be determined via numerical integration of the above non-linear second order ODE. The integration starts at $z = 0$ to the point $z = h$, with the initial radius and radial slope equal to the print head radius and contact angle, respectively. Additionally, the solution is constrained such that the volume of the meniscus must match the total volume of air injected into the print head. To solve this, we chose to frame the Young-Laplace equation as an initial value problem, using an implementation of the shooting method in MATLAB[2]. The solution for this problem was defined with initial values that satisfy the following boundary constraints:

$$\mathbf{M} = \begin{Bmatrix} r(0) - R \\ r'(0) - \cot(\theta_Y) \\ \pi \int_0^h r^2 dz - V_o \end{Bmatrix} = 0. \tag{4}$$

This therefore converts the above boundary value problem into a root finding solution which aims to ensure that the boundary conditions $\mathbf{M} = 0$. A comparison between the curvature of the interface determined numerically and experimentally is shown in Supplementary Fig.3a, wherein the Young-Laplace model accurately predicts the interface curvature for an increasing internal pressure state $p_i$.

It is worth noting that the shape of the interface depends on the quantity $p_i - \rho g z$, where $p_i$ denotes the pressure within the print head. As the print head is withdrawn from the bath, the value of $\rho g z$ decreases linearly and therefore the value of $p_i$ must also change linearly to maintain the same interface shape.

## S5: Convex interface formation

As the DIP approach relies on the pressurization of a print container to produce an air-liquid meniscus, the profile of this boundary and consequently the cured region is non-planar. Traditional slicing schemes[3] assume that the projected geometry is parallel to the construction plane and as such would result in reconstructed artefacts in the case of DIP. To correct for this in the case of an axisymmetric print head, the three-dimensional surface can be reconstructed by revolving the Young-Laplace predicted surface about the z-axis (**Supplementary Fig.4b**). Let the discrete interface profile, $Z(r)$, be the solution to the boundary value problem, with the parametric expression of the reconstructed 3D surface given by:

$$x(\theta, r) = r \cos \theta, \qquad (5)$$

$$y(\theta, r) = r \sin \theta, \qquad (6)$$

$$z(\theta, r) = Z(r). \qquad (7)$$

**S6: Voxel intersection**

Unlike standard DLP printing, the projected images required for DIP arise from the intersection of a convex surface with the voxelized representation of the target geometry, resulting in a non-planar slicing scheme. The voxels which lie on this surface can be determined via a distance minimization of the surface to the voxel in the array. Let a point on the surface of the interface be defined by $S_p = (x_p, y_p, z_p)$ such that it satisfies the above parametric relationship, and the voxel representation of the desired model is given by $[V_i, V_j, V_k]$, where $i, j, k$ represent the dimensions of the voxel matrix whose $i, j$ dimensions define the number of corresponding pixels in-plane and the maximum size of $k$ is determined by the discretization of the object as a function of the sliced layer height. Additionally, $[V_i, V_j, V_k]$ represents a binary 3D matrix where the presence of geometry is defined by a '1' and the absence of geometry is defined by a '0'. Whether a voxel is located on the surface of the interface ($S_p$) is determined by the minimization of the Euclidian distance between that point and the closest voxel. If $n$ voxels are present within the same Euclidian distance, then the resultant value is averaged over $n$ samples in the following way:

$$\check{V} = \frac{1}{n}\sum_{n=1}^{n} \arg \min_{i,j,k}, V \qquad (8)$$

where $\check{V}$ represents the voxel coordinate and value which satisfies the above relationship. This approach is repeated for each location on the interface in three-dimensional space, where the desired image sent to the projection module for each layer is given by:

$$I_{ij}^k = \|\check{V}\|, \qquad (9)$$

where $\|\check{V}\|$ denotes the voxel value and the superscript $k$ denotes the projection in the sequence. This relationship represents the equivalent projection of the three-dimensional voxel array onto the interface surface in two dimensions. It's worth noting that as the voxel array has been reduced to two dimensions via the projection, therefore we have lost some information about its original position. To preserve this, we also store the absolute original z-location of the pixels for each projection, which becomes useful later for reconstruction:

$$IZ_{ij}^k = \check{V}_k, \qquad (10)$$

where $IZ$ is the matrix containing the absolute z-locations and $\check{V}_k$ denotes the global z-coordinate of the pixel stored in $I_{ij}^k$. As the print head moves up in the positive z-direction, the voxels intersected by the interface changes. To determine this intersection, the interface profile is translated in the z-direction corresponding to the discretization of the voxel array (layer height = $L_h$) in the z-direction:

$$z(\theta, r) = Z(r) + L_h. \tag{11}$$

## S7: Determining the intermediate interface shape

To ensure the printed object remains adhered to the print container, the meniscus must be flattened against the bottom surface such that the maximum extent of the printed part, $R_{max}$, is contained within the flat region of the meniscus. To predict the extent to which the print head must be lowered to create said flat region, a Bézier curve method for approximating the meniscus shape is used as previously described[4,5]. Briefly, a MATLAB script minimizes the error of the Young-Laplace equation for a meniscus in a cylindrical capillary at a print head position corresponding to an undeformed meniscus, which is defined via the control points of a Bézier curve. The print head position is then moved downwards, and the free control points are moved radially outwards in a stepwise fashion until the height of the meniscus at $R_{max}$ is approximately 0, while preserving the meniscus volume **(Supplementary Fig.4a)**. To solve for the initial meniscus shape, a dimensionless form of the Young-Laplace equation is used:

$$\frac{\frac{d^2y}{dx^2}}{\left(1+\left(\frac{dy}{dx}\right)^2\right)^{\frac{3}{2}}} + \frac{\frac{dy}{dx}}{x\left(1+\left(\frac{dy}{dx}\right)^2\right)^{\frac{1}{2}}} - y = 0, \tag{12}$$

where $x = \frac{r}{l}$, $y = \frac{h}{l}$, and $l = \sqrt{\frac{\sigma}{\rho g}}$ is the capillary length. The above equation is solved using the boundary conditions $\frac{dy}{dx} = 0$ at $x = 0$ and $\frac{dy}{dx} = \cot \theta_Y$ at $x = x_3$, where $x_3$ is the dimensionless form of the print head radius and $\theta_Y$ is the contact angle.

To solve the above equation, a cubic Bézier curve is defined along with its first and second derivatives:

$$B(t) = (1-t)^3 P_0 + 3(1-t)^2 t P_1 + 3(1-t) t^2 P_2 + t^3 P_3, \tag{13}$$

$$B'(t) = 3(1-t)^2 (P_1 - P_0) + 6(1-t)t(P_2 - P_1) + 3t^2 (P_3 - P_2), \tag{14}$$

$$B''(t) = 6(1-t)(P_2 - 2P_1 + P_0) + 6t(P_3 - 2P_2 + P_1), \tag{15}$$

where $t$ is a parametric variable and $P_0, P_1 \ldots P_n$ represent the control points of the Bézier curve. After defining the x- and y-coordinates of the control points, the kron function (Kronecker tensor product) in MATLAB is used to calculate the x- and y-coordinates of the meniscus curve as well as the first and second derivatives. Then, $\frac{dy}{dx}$ and $\frac{d^2y}{dx^2}$ are calculated using the following equations:

$$\frac{dy}{dx} = \frac{\frac{dy}{dt}}{\frac{dx}{dt}}, \tag{16}$$

$$\frac{d^2y}{dx^2} = \frac{\left(\frac{dx}{dt}\right)\left(\frac{d^2y}{dt^2}\right) - \left(\frac{dy}{dt}\right)\left(\frac{d^2x}{dt^2}\right)}{\left(\frac{dx}{dt}\right)^3}. \tag{17}$$

Afterwards, a loss term is defined corresponding to the sum of the dimensionless Young-Laplace equation divided by the sum of the y-coordinates of the meniscus curve. The undeformed meniscus shape is then calculated using the MATLAB function fmincon. Once this "steady-state" solution is found, the flattened meniscus shape is calculated by moving the control point $P_3$ (corresponding to the edge of the print head) down. Control points $P_1$ and $P_2$ are then moved radially outward and the meniscus shape is solved again, preserving the arc length of the meniscus so that meniscus volume is maintained. This is repeated until the meniscus height at $R_{max}$ is close enough to 0 as evaluated by a user-defined threshold.

Aside from remapping the 3D cartesian geometry in order to slice the volume with a curved print interface, the meniscus shape presents an additional challenge for the beginning layers of the print. If we define the first layer of the print where the air/liquid interface was lowered into the container using the "steady state" meniscus shape, any geometry of the desired object radially beyond the contact point C (**Supplementary Fig. 5a**) could not be cured correctly. The height of the air/liquid interface at all points between $r = 0$ and $r = R_{max}$ (the maximum extent of the object's base) would be too tall, and this would lead to (1) the print not adhering to the container bottom and (2) the print not correctly matching the desired geometry. Therefore, the air/liquid interface must be lowered past this initial contact point, deforming the interface to form a (pseudo) flat region.

The number of interpolation steps used is dependent on the discretization of the volume ($V_k$) and the layer height $L_h$. One challenge with applying such an approach for predicting the slicing profile within this interpolated region is that the solution is comprised of a series of overlapping surfaces wherein a voxel can be incident with more than one surface. To combat this, two approaches can be used to ensure that a single voxel and subsequently the projections that make up that voxel never result in a value that exceeds the target original value. One way to do this could be to reduce the greyscale value $I_{ij}^k$, such that its cumulative dosage never exceeds $\|\breve{V}\|$. However, if the number of interpolation steps exceeds 255, the solution lacks sufficient degrees of freedom. This is further complicated by the fact that the curing propagation rate $R_p$ scales non-linearly with intensity[6]:

$$R_p = \frac{k_p}{k_t^{1/2}}[M](\phi I_a)^{1/2}, \qquad (18)$$

where $k_p$ and $k_t$ represent the propagation and termination parameters of the material, $[M]$ is the initial monomer concentration, $\phi$ is the quantum yield and $I_a$ is the absorbed intensity, which in our case is proportional to the greyscale value described in $I_{ij}^k$. A more convenient approach is to exclude those voxels that have already been written by the interpolated surface from the voxel array. This approach solves both the degree of freedom constraint imposed by an 8-bit image and the non-linear intensity threshold which governs curing onset. Therefore, the available voxels for each loop can be written as the following, where $k$ represents the interpolation step over the domain $(0, z')$, with:

$$[V_i, V_j, V_k]^{k+1} = [V_i, V_j, V_k]^k - \|\widetilde{V_{ijk}}\|.$$

### S8: Geometry reconstruction from projections

To validate that the convex slicing algorithm results in the same input geometry, the generated projections are traced back through an empty target volume (**Supplementary Fig.5d**). For each image in the projection sequence of size $k$, the voxel value is determined by the quantity $I_{ij}^k$, whose global position in the matrix is determined by $IZ_{ij}^k$. Let $V'$ be an empty target volume with dimensions equal to the input geometry $[V_i', V_j', V_k'] = 0$. Therefore, any element in $V'$ is given by the coordinate transform of the projection sequence such that:

$$V'_{ijk} = [I_{ij}^k, IZ_{ij}^k], \quad \|V'_{ijk}\| = \|I_{ij}^k\|. \tag{19}$$

**S9: Computing the reconstruction error ($\delta$)**

The similarity between the reconstructed volume and the input volume was computed using the Jaccard Index. For each 2D orthogonal component plane of the volume, the Jaccard Index[7] was summed over the volume and averaged over each component axis, whereby:

$$J(V',V) = \frac{1}{3}\left(\frac{1}{i}\sum_{i=1}^{i}\frac{V'^i_{jk}\cap V^i_{jk}}{V'^i_{jk}\cup V^i_{jk}} + \frac{1}{j}\sum_{j=1}^{j}\frac{V'^j_{ik}\cap V^j_{ik}}{V'^j_{ik}\cup V^j_{ik}} + \frac{1}{k}\sum_{k=1}^{k}\frac{V'^k_{ij}\cap V^k_{ij}}{V'^k_{ij}\cup V^k_{ij}}\right) = \delta, \quad \delta \in [0,1]. \tag{20}$$

The reconstruction error is thus quantified by a single value $\delta \in [0,1]$, which determines the similarity between the target domain and the computed domain, where a value of $\delta = 1$ denotes a perfect match. A low value of $\delta$ normally denotes an interface step size mismatch between the voxel representation $V$ and the layer slice height $L_h$. Therefore, the slicing step size was reduced until the Jaccard Index exceeded a threshold value $\bar{\delta}$ which in our case was set $\bar{\delta} > 0.9$. A process flow diagram of this slicing scheme can be found in **Supplementary Fig.5e**.

**S10: Theoretical model of optical resolution**

In the case of DIP, print resolution is determined by exposure energy density, magnification, spatial distribution of the projection optics, and the photo-polymer response, which depends on photo-initiator concentration, monomer concentration and photo-absorber concentration. To quantify the theoretical resolution of the imaging system, we employed a similar approach outlined by Behroodi et al.[8] which predicts the final energy distribution at the projection plane as the superposition of the point-spread functions of all pixels reflected from the DMD surface via spatial convolution:

$$f(x,y) * PSF(x,y) = \int_{\tau_1=-\infty}^{\infty}\int_{\tau_2=-\infty}^{\infty} f(\tau_1,\tau_2) \cdot PSF(x-\tau_1, y-\tau_2)d\tau_1 d\tau_2, \tag{21}$$

where $f(x,y)$ denotes the spatial function of the micromirror cross section at the projection plane. For a single pixel on the DMD, the spatial function is determined by:

$$f(x,y) = \begin{cases} \frac{g_s}{255} & -m\frac{d_x}{2} < x < m\frac{d_x}{2}, -m\frac{d_y}{2} < y < \frac{d_y}{2} \\ 0 & x < -m\frac{d_x}{2}, m\frac{d_x}{2} < x, y < -m\frac{d_y}{2}, m\frac{d_y}{2} < y \end{cases}, \tag{22}$$

where $d_x$ and $d_y$ denote the dimensions of the micromirror, $m$ is the magnification of the projection optics and $g_s$ denotes the greyscale value $[0, 255]$. The spatial convolution equation determines the equivalent Gaussian distribution function ($\omega_0$) at the focal plane of the projection optics. The diameter of the point on the focal plane can subsequently be modelled by the Gaussian distribution, where the UV intensity of a point source at a given plane is defined by:

$$I(x,y,z) = \frac{2P}{\pi[\omega(z)]^2} e^{\frac{-2(x^2+y^2)}{[\omega(z)]^2}}, \tag{23}$$

where $I(x,y,z)$ is the optical intensity of the projected light with units (J/cm²s), $P$ is the total power of the UV light determined in units of (J/s) and $\omega(z)$ denotes the Gaussian radius at a location $z$, whose half-width is $1/e^2$ of the Gaussian maximum intensity $I_{max}$. In the case of DIP for an axisymmetric print head, the curvature of the meniscus imposes a radially symmetric change in the Gaussian radius dependent on the meniscus height $Z(r)$. As $Z(r)$ was initially determined from an origin located at the print head, an equivalent profile $X_m(z)$ will be implemented such that $X_m(0) = 0$ and $X_m(z) = \max Z(r)$, this ensures that the maximum meniscus deformation is coincident with the image plane when $z = 0$, **Supplementary Fig.6a**. The Gaussian beam width at a location on the meniscus is therefore given by:

$$\omega(z) = \omega_o \sqrt{1 + \left(\frac{X_m(z)}{Z_R}\right)^2}, Z_R = \frac{\pi n \omega_0^2}{\lambda}, \tag{24}$$

where $\omega_0$ denotes the beam waist, $n$ is the refractive index and $\lambda$ is the wavelength. Therefore, the exposure energy per unit area for a finite time $t$ and meniscus surface is:

$$E(x, y, z, t) = I(x, y, z)t.$$

### S11: Energy density across the meniscus

Light entering the resin through the surface of the meniscus is either absorbed or scattered by the material. These two effects determine the fraction of energy deposited into the material and therefore the polymerization thickness and penetration depth. From Beer-Lambert, the energy per unit area within the resin surface is governed by an exponential reduction in intensity based on material parameters[9]

$$E(x,y,z,z',t) = E(x,y,z,t)\, e^{\frac{-z'}{D_p}}, \ D_p = \frac{1}{\varepsilon_d[D] + \varepsilon_i[S]}, \tag{25}$$

where $D_p$ is the penetration depth at which the intensity falls to $1/e^2$ of the surface intensity, $\varepsilon_d$ and $\varepsilon_i$ are the molar absorption coefficients of the photo-initiator and photoabsorber, respectively, $D$ and $S$ are the concentrations of the photo-initiator and photo-absorberand $z'$ defines the coordinate system into the material which is not necessarily aligned with the optical axes. In the case of DIP, the curvature of the meniscus decomposes the incoming light at the location $(x, y, z)$ into scattered and transmitted components. At any location $z$ on the meniscus surface, the incident angle of the incoming ray of light can be defined by $\alpha_0$ relative to the print head axis. In our case $\alpha_0 \approx 0$, however, it could be feasible that $\alpha_0 \neq 0$ for some print head and optical configurations. Similarly, the angle of the meniscus $\alpha_m$ and its normal $\widehat{\alpha_m}$ at location $(x, y, z)$ relative to the print head axis can be determined by:

$$\alpha_m = \tan^{-1}\left(\frac{dX_m(z)}{dz}\right), \ \widehat{\alpha_m} = \pi - \cot^{-1}\left(\frac{dX_m(z)}{dz}\right), \tag{26}$$

where $X_m(z)$ denotes the radial position of the meniscus as a function of the meniscus height $z$. The definition of this ray in three-dimensions coincident with a point on the surface of the meniscus $(x, y, z)$, can be approximately described by the Gaussian-beam theory[10]. Under the assumption of an axisymmetric print head, the normal vector at this point in-two dimensions $\hat{n}$ and associated light ray $\hat{u}$, is given by:

$$\widehat{\boldsymbol{u}} = \langle \cos\alpha_0, \sin\alpha_0 \rangle, \tag{27}$$

$$\widehat{\boldsymbol{n}} = \langle \cos\widehat{\alpha_m}, \sin\widehat{\alpha_m} \rangle. \tag{28}$$

From Snell's Law, the angle of the outgoing ray relative to the surface normal is given by the relative change in the refractive index of the material and the incident angle $\theta_1$. As the angle between the incoming ray $\widehat{\boldsymbol{u}}$ and surface normal $\widehat{\boldsymbol{n}}$ is given by the dot product $\cos\theta_1 = \widehat{\boldsymbol{u}} \cdot \widehat{\boldsymbol{n}}$, the outgoing angle $\theta_2$ relative to the surface normal can be simplified to the following:

$$\theta_2 = \sin^{-1}\left(\frac{n_1}{n_2}\sqrt{1-(\widehat{\boldsymbol{u}}\cdot\widehat{\boldsymbol{n}})^2}\right). \tag{29}$$

The direction of this ray represents the new coordinate system $\widehat{\boldsymbol{\gamma}}$ for the Beer-Lambert solution where the rotation of the ray with respect to the local coordinate system vector $\widehat{\boldsymbol{z}} = \langle 1, 0 \rangle$ is given by:

$$\alpha_{\widehat{\gamma}} = \pi - \cot^{-1}\left(\frac{dX_m(z)}{dz}\right) + \sin^{-1}\left(\frac{n_1}{n_2}\sqrt{1-(\widehat{\boldsymbol{u}}\cdot\widehat{\boldsymbol{n}})^2}\right), \tag{30}$$

$$\alpha_{\widehat{\gamma}} = \pi - \cos^{-1}\widehat{\boldsymbol{z}}\cdot\widehat{\boldsymbol{n}} + \sin^{-1}\left(\frac{n_1}{n_2}\sqrt{1-(\widehat{\boldsymbol{u}}\cdot\widehat{\boldsymbol{n}})^2}\right), \tag{31}$$

$$\widehat{\boldsymbol{\gamma}} = \langle \cos\alpha_{\widehat{\gamma}}, \sin\alpha_{\widehat{\gamma}} \rangle, \tag{32}$$

where the new coordinate axes for the Beer-Lambert solution are given by:

$$\langle \boldsymbol{\gamma}_x, \boldsymbol{\gamma}_y, \boldsymbol{\gamma}_z \rangle = \langle \widehat{\boldsymbol{\gamma}_\perp}, \widehat{\boldsymbol{\gamma}_\perp} \times \widehat{\boldsymbol{\gamma}}, \widehat{\boldsymbol{\gamma}} \rangle. \tag{33}$$

Furthermore, the proportion of light transmitted into the material is dependent on the incident angle $\theta_1$ and the energy per unit area $E(x,y,z,t)$. Using the Fresnel equations[11], the transmission coefficients for parallel and perpendicular polarizations are proportional to the cosine of the incoming and outgoing rays, such that:

$$T_\perp = \frac{2n_1\cos\theta_1}{n_1\cos\theta_1 + n_2\cos\theta_2}, \tag{34}$$

$$T_\| = \frac{2n_1\cos\theta_1}{n_2\cos\theta_1 + n_1\cos\theta_2}. \tag{35}$$

Under the assumption that the incoming light is not polarized, the transmission coefficient $T$ is given by the average of the $S$ and $P$ polarization states. As the Fresnel coefficients represent amplitudes, the transmitted intensity $\eta$ at the ray-meniscus intersection is proportional to the square of the amplitude:

$$T = \frac{T_\perp + T_\|}{2}, \tag{36}$$

$$\eta = T^2. \tag{37}$$

The energy per unit area $E$ is dependent on the position of the incoming ray $\widehat{\boldsymbol{u}}$, its coordinate system is such that $\widehat{\boldsymbol{z}}\cdot\widehat{\boldsymbol{u}} \approx 1$ and $\widehat{\boldsymbol{x}}\cdot\widehat{\boldsymbol{u}} \approx 0$. In general, under the assumption that the beam's half-width is not sufficiently large and where the gradient at the surface $\left[\frac{\partial z}{\partial x}, \frac{\partial z}{\partial y}\right]$ ensures that $\Delta z_{-\omega_0}^{\omega_0} \approx 0$, then the vector $\widehat{\boldsymbol{E}}$ can be shown to be:

$$\widehat{\boldsymbol{E}} \approx \widehat{\boldsymbol{u}} \cdot E(x,y,z). \tag{38}$$

Similarly, if $\widehat{\boldsymbol{n}}$ defines the normal vector at the meniscus, then the transmitted component $\mathcal{H}(\boldsymbol{\gamma}_x, \boldsymbol{\gamma}_y, \boldsymbol{\gamma}_z)$ with coordinates relative to the transmissive beam, is given by

$$\mathcal{H}(\boldsymbol{\gamma_x}, \boldsymbol{\gamma_y}, \boldsymbol{\gamma_z}) = \eta(n_1, n_2, \widehat{\boldsymbol{u}}, \widehat{\boldsymbol{n}}) \cdot \widehat{\boldsymbol{E}} \; e^{\frac{-\gamma_z}{\varepsilon_d[D] + \varepsilon_i[S]}}.$$

## S12: Modelling the effective resolution, meniscus working region and interfacial focal map.

The effective printing resolution can be formed by convolving the point spread function over a single pixel, wherein the theoretical point spread function in three-dimensions can be simulated by defocusing in the Fourier domain by a modulated Gaussian function, where:

$$\widehat{\boldsymbol{h}}(\omega, z) = e^{-\sigma(z)^2 \omega^2 \cdot \frac{\sin \zeta(\omega,z)}{\zeta(\omega,z)}}, \tag{39}$$

$$\zeta(\omega, z) = \frac{\zeta \cdot \omega(1-\omega)}{K(z_i - z)}. \tag{40}$$

This approach results in the 3D representation of the point spread function as shown in **Supplementary Fig.6b**. Therefore, the effective pixel size was approximated over a $z$-range of 5 mm towards the projection lens. By assuming that the allowable pixel size can only deviate by $\sqrt{2}P_{xy}$, where $P_{xy}$ denotes the in-plane pixel size, the effective meniscus region can be plotted for varying print head sizes and material surface tensions, **Supplementary Fig.6c**. It's worth noting that the entirety of the meniscus can be used as demonstrated by the convex slicing algorithm, however, the spatial resolution is dependent on the curvature and its relative location away from the optical axis. To investigate the theoretical defocusing of a projected image in the plane, the standard USAF test pattern was convolved at $z = 0$ mm, $z = 3$ mm, and $z = 5$ mm, **Supplementary Fig.6d**.

To investigate the effective reduction in resolution across the meniscus, a checkerboard pattern was computed at numerous regions between 0 – 5 mm. By comparing the relative height of the interface at the location $(x, y, z)$ to the corresponding pixel in the defocused checkboard array, the effective defocused pixel representation was generated across the entirety of the interface as shown in **Supplementary Fig.6e**.

## S13: Derivation and modelling of capillary waves on the air-liquid boundary:

During printing, a thin fluid volume of uncured material is created between the meniscus and the previously cured region, the height of which depends on the oxygen inhibition zone of the material[12,13] and the $z$-translation of the meniscus. The Navier-Stokes continuity and momentum equations for incompressible Newtonian fluids can be written as:

$$\rho \left( \frac{\partial \boldsymbol{u}}{\partial t} + (\boldsymbol{u} \cdot \nabla) \boldsymbol{u} \right) = -\nabla \boldsymbol{T} + \rho \boldsymbol{g} \tag{41}$$

$$\rho \nabla \cdot \boldsymbol{u} = 0, \tag{42}$$

where $\rho$ is the fluid density, $\boldsymbol{u}$ denotes the velocity, $\boldsymbol{T} = -p\boldsymbol{I} + \mu(\nabla \boldsymbol{u} + (\nabla \boldsymbol{u})^T)$ is the stress tensor, $p$ is the pressure, and $\mu$ is the dynamic viscosity, $\boldsymbol{g}$ is the standard gravity. From lubrication theory[14–16], a thin fluid volume or film can be described under the assumption that the fluid depth is much shallower than the fluid's extent. Under this assumption, the full Navier-Stokes equations can be simplified. As the velocity at small length-scales is small compared to the viscous forces, the inertial terms on the left hand-side are negligible as they are proportional to $u^2/L$. This approximation reduces the momentum equation to $0 = -\nabla p + \rho \boldsymbol{g} + \mu \nabla^2 \boldsymbol{u}$. Furthermore, as the thickness of the film is assumed to be small, the velocity perpendicular to the plane is negligible. For the case of an axisymmetric print head, the coordinate system is such that $\boldsymbol{x}$ denotes the vector in the radial direction, $\boldsymbol{z}$ denotes the vector parallel to the optical axis, and $\boldsymbol{y}$ denotes the vector perpendicular to $\boldsymbol{xz}$, resulting in the following reduced momentum equations:

$$-\frac{dp}{dx} + \mu \frac{\partial^2 u}{\partial z^2} = 0, \tag{43}$$

$$-\frac{dp}{dy} + \mu \frac{\partial^2 v}{\partial z^2} = 0, \tag{44}$$

$$-\frac{dp}{dz} + \rho g = 0. \tag{45}$$

For a thin-film under incompressible flow, the increase in volumetric flow rate in the positive x-direction is given by:

$$\Delta Q = \left[\int_0^{h(x)} u \, dz\right]_x^{x+\Delta x}. \tag{46}$$

where $h(x)$ is the meniscus height. This must be balanced by the decrease in volumetric flow in the z-direction, $-\frac{\partial h}{\partial t} dx$. By assuming a non-slip condition at the previously printed interface and finite shear across the boundary, the horizontal velocity profile is given by:

$$u = \frac{1}{\mu} \frac{\partial p}{\partial x} \left(\frac{z^2}{2} - hz\right) + \frac{\tau}{\mu} z. \tag{47}$$

Equating the volumetric flow rate yields,

$$\frac{\partial Q}{\partial x} = \frac{\partial}{\partial x} \int_0^{h(x)} u \, dz = -\frac{\partial h}{\partial t}, \tag{48}$$

$$\frac{\partial}{\partial x} \int_0^{h(x)} \left(\frac{1}{\mu} \frac{\partial p}{\partial x} \left(\frac{z^2}{2} - hz\right) + \frac{\tau}{\mu} z\right) dz = -\frac{\partial h}{\partial t}, \tag{49}$$

$$\frac{\partial}{\partial x} \left(\frac{1}{\mu} \frac{\partial p}{\partial x} \left(\frac{-h^3}{3}\right) + \frac{\tau}{\mu} \frac{h^2}{2}\right) = -\frac{\partial h}{\partial t}, \tag{50}$$

$$\tau h \frac{\partial h}{\partial x} + \frac{h^2}{2} \frac{\partial \tau}{\partial x} - \frac{1}{3} \frac{\partial}{\partial x} \left(h^3 \frac{\partial p}{\partial x}\right) = -\mu \frac{\partial h}{\partial t}. \tag{51}$$

In our case, the volume of air within the print head is acoustically driven, causing a variation of pressure and shear force along the surface of the meniscus which is proportional to the local air pressure and velocity. The pressure in a sound wave is given by $P = P_0 + A\sin(\omega t - Kx)$, where the velocity is in-phase with the pressure $v = B\sin(\omega t - Kx)$, where $A$ and $B$ depend on factors such as the compressibility of the air and the amplitude of the wave. The pressure within the fluid can now be described by the following:

$$p - p_{atm} = \rho g h - \gamma \frac{d^2 h}{dx^2} + A\sin(\omega t - Kx) + f(z). \tag{52}$$

Inserting this into the above equation for the shear force and applying the assumption of small perturbations in the interface height $h = h_0 + \varepsilon$, the equation can be linearized as shown.

$$B\sin(\omega t - Kx)h\frac{\partial h}{\partial x} - \frac{h^2}{2}BK\cos(\omega t - Kx) - \frac{1}{3}\frac{\partial}{\partial x}\left(h^3\left(\rho g\frac{\partial h}{\partial x} - \gamma\frac{d^3 h}{dx^3} - KA\cos(\omega t - Kx)\right)\right) = -\mu\frac{\partial h}{\partial t}, \quad (53)$$

$$B\sin(\omega t - Kx)h_0\frac{\partial \varepsilon}{\partial x} - \frac{h_0^3}{3}\left(\rho g\frac{\partial^2 \varepsilon}{\partial x^2} - \gamma\frac{d^4 \varepsilon}{dx^4}\right) + \mu\frac{\partial \varepsilon}{\partial t} = \frac{h_0^2}{2}BK\cos(\omega t - Kx) - \frac{h_0^3}{3}K^2 A\sin(\omega t - Kx). \quad (54)$$

If B is negligible, then the meniscus would only be affected by the pressure from the acoustic driver, and not the shear from the driver. Under this assumption, the above equation can be solved analytically and simplified to the following:

$$\frac{h_0^3}{3}\left(\rho g\frac{\partial^2 \varepsilon}{\partial x^2} - \gamma\frac{d^4 \varepsilon}{dx^4}\right) - \mu\frac{\partial \varepsilon}{\partial t} = \frac{h_0^3}{3}K^2 A\sin(\omega t - Kx). \quad (55)$$

The homogenous solution to the above equation is of the form $\varepsilon = Ce^{ikx+st}$, where the constant $s$ describes the relaxation of the disturbance given by $s = \frac{-h_0^3}{3\mu}(\rho g k^2 + \gamma k^4)$. The particular solution should also be of the form $F\sin(\omega t - Kx) + G\cos(\omega t - Kx)$. Therefore, substituting this in gives the particular solution of,

$$\varepsilon_p = A_p(3\mu\omega\cos(\omega t - Kx) - h_0^3 K^2(\rho g + \gamma)\sin(\omega t - Kx)), \quad (56)$$

$$A_p = \frac{A h_0^3 K^2}{K^4 h_0^6(\rho g + \gamma)^2 + 9\omega^2\mu^2}. \quad (57)$$

The time dependent solution to a small perturbation in the interface height $\varepsilon$, can be determined by adding together the homogeneous and particular solutions resulting in an equation that describes the full solution of meniscus perturbation under acoustic sound waves,

$$\varepsilon = e^{-\frac{h_0^3}{3\mu}(\rho g k^2 + \gamma k^4)t}(C_1\cos kx + C_2\sin kx) + A_p(3\mu\omega\cos(\omega t - Kx) - h_0^3 K^2(\rho g + \gamma)\sin(\omega t - Kx)). \quad (58)$$

This equation describes the time dependent solution to the interface height as a function of the acoustically driven perturbations of the interface, whereby the produced waves are added to waves caused by previous perturbations that decay exponentially in time. It's worth noting that this solution can only accurately be applied for low frequency acoustic perturbations (like those used in this work), as the inertial terms in the momentum equation have been assumed to be negligible. Under high frequency acoustic driving, the fluid element would need to change position much faster than what is captured by this model, in this instance the components $\frac{\partial \boldsymbol{u}}{\partial t}$ and $\boldsymbol{u}\cdot\nabla\boldsymbol{u}$ would be non-zero. Inserting this back into the equation for the velocity in the x-direction, yields:

$$u = \frac{\rho g}{\mu}\frac{\partial \varepsilon}{\partial x}\left(\frac{z^2}{2} - (h_0 + \varepsilon)z\right), \quad (59)$$

$$\frac{\partial \varepsilon}{\partial x} = e^{-\frac{h_0^3}{3\mu}(\rho g k^2 + \gamma k^4)t}(-kC_1\sin kx + kC_2\cos kx) + A_p(h_0^3 K^3(\rho g + \gamma)\cos(Kx - \omega t) - 3K\mu\omega\sin(Kx - \omega t)), \quad (60)$$

$$u = \frac{\rho g}{\mu}\left(e^{-\frac{h_0^3}{3\mu}(\rho g k^2 + \gamma k^4)t}(-kC_1\sin kx + kC_2\cos kx) + A_p(h_0^3 K^3(\rho g + \gamma)\cos(Kx - \omega t) - 3K\mu\omega\sin(Kx - \omega t))\right) \times \left(\frac{z^2}{2} - \left(h_o + \left(e^{-\frac{h_0^3}{3\mu}(\rho g k^2 + \gamma k^4)t}(C_1\cos kx + C_2\sin kx) + A_p(3\mu\omega\cos(\omega t - Kx) - h_0^3 K^2(\rho g + \gamma)\sin(\omega t - Kx))\right)\right)\right). \quad (61)$$

In the above equation, the constants $C_1$ and $C_2$ describe the amplitudes of previous evolutions of the wave, $k$ is the wave number, $K$ is the dimensionless wavenumber which is normalized against the capillary length $\frac{K}{L_{cap}}$, $\omega$ is the driving frequency, $\gamma$ is the surface tension, $\rho$ is the density and $A_p$ is the driving amplitude. Note that this approximate model of the interface dynamics does not take into account the streaming effects generated by the meniscus curvature nor the squeeze flow due to the translation of the air-liquid interface in the z-direction, **Supplementary Fig.19a-c**. The result from the image based de-wetting analysis shown **Supplementary Fig.11** indicates that the influx rate of material under acoustic stimulation is approximately an order of magnitude higher than under lubrication-driven flow. Therefore, we can approximate the material influx rate based on acoustics alone as a conservative estimate. Furthermore, to account for the curvature of the interface, the constant fluid depth $h_0$ is replaced by $h(x)$ which describes the height of the meniscus as a function of the print head's diameter, **Supplementary Fig.19d-e.**

### S14: Scaling laws for acoustically driven flows

To understand how fluid transport is affected by different material parameters, we can assume the interface height in an idealized case is given by $h(x,t) = h_0 + \phi(t)\cos(qx)$, where $q$ denotes the wavenumber and $\phi(t)$ describes the amplitude of the wave. Under this approximation, the pressure within the fluid is given by the atmospheric pressure, gravitational effects and capillary effects. For lubrication theory to hold, the relative height change of the interface is small compared to its depth, therefore the curvature of the interface can be approximated by $\kappa = d^2h/dx^2$. The pressure difference between the crest and the trough scales with $\gamma\phi/\lambda^2$ in the capillary case and with $\rho g \phi$ in the gravity-driven case. We can therefore describe the ratio of gravitational to capillary effects as $\lambda^2 \rho g / \gamma$ or $(\lambda/l_{cap})^2$, whereby $(\lambda/l_{cap})^2 > 1$ denotes a system where gravity dominates and $(\lambda/l_{cap})^2 < 1$ denotes a system where capillary effects dominate[17]. Using this relationship, in conjunction with the generalized equation for velocity in the x-direction, we can derive a scaling relationship for the velocity $U$ dependent on material parameters. By using a length scale of $\lambda$ in the x-direction and $h_0$ in the z-direction, it can be shown that:

$$U \propto \frac{h_0^2 \rho g \phi}{\lambda \mu}, \text{for } \left(\frac{\lambda}{l_{cap}}\right)^2 \gg 1, \tag{62}$$

$$U \propto \frac{h_0^2 \gamma \phi}{\lambda^3 \mu}, \text{for } \left(\frac{\lambda}{l_{cap}}\right)^2 \ll 1, \tag{63}$$

where the wavelength $\lambda$ is found by solving the dispersion relation for capillary waves, which relates the wave frequency ($\omega$) to the wavenumber ($k$) and is given by:

$$\omega^2 = \frac{\gamma}{\rho}k^3 + gk. \tag{64}$$

An example of these scaling laws can be shown for different material processing parameters in **Supplementary Fig.19f-g**.

## S15: Effect of interface curvature on streaming flows

When acoustically driving the meniscus profile under low amplitude oscillations for increasing internal pressure states, we observed a dramatic increase in fluid flow below the interface when compared to a pseudo-level interface, **Supplementary Fig.10**. We postulate that this enhanced fluid transport is driven by two key factors related to the meniscus curvature and an increase in available mass volume. Firstly, following the analysis by Prinet *et al*. and Zhong *et. al*[18,19], the streaming velocity below the meniscus can be written as a function of the parallel streaming velocity,

$$\boldsymbol{u}_s|_{z=\zeta_0} = -\frac{3}{4\omega}\hat{u}_\parallel^2 \partial_\tau \hat{u}_\parallel \hat{\boldsymbol{t}}, \tag{65}$$

where $\hat{u}_\parallel$ denotes the primary flow parallel to the meniscus, $\hat{\boldsymbol{t}}$ denotes the tangential vector to the meniscus and $\zeta_0$ describes the surface elevation of the static meniscus, whose solution is given by the Young-Laplace equation. The value for the tangential velocity at the meniscus surface $\hat{u}_\parallel$ is obtained through linear interpolation of the meniscus wave along the meniscus at $z = \zeta_0(x)$:

$$\hat{u}_\parallel = -\cos \omega t \sum_n^\infty A_n k_n \frac{1+k_n\zeta_0}{[1+(\partial_x\zeta_0)^2]^{1/2}} (\sin k_n x - \partial_x\zeta_0 \cos k_n x), \tag{66}$$

where $k_n$ describes the eigen-wavenumber $k_n = n\pi/\lambda$, the coefficient $A_n = \frac{2\omega a_0}{\omega^2 - \omega_n^2} \frac{(-1)^n + s}{(m^2 + k_n^2)\lambda}$, $a_0$ is the forcing acceleration, $s$ is the interface symmetry and $m$ is the mode number in the y-direction. Substituting in the equation for $\hat{u}_\parallel$, the meniscus streaming velocity can be written as:

$$\boldsymbol{u}_s|_{z=\zeta_0} = \left\{-\frac{3}{4\omega}\sum_{n,m} A_n A_m (k_n + \zeta_0 k_n^2)(k_m + \zeta_0 k_m^2)\left[k_m - \frac{\rho g_0}{\sigma}(\zeta_0 - z*)\right]\sin(k_n x)\cos(k_m x) + O(\partial_x\zeta_0)\right\}\hat{\boldsymbol{i}} + O(\partial_x\zeta_0)\hat{\boldsymbol{j}}, \tag{67}$$

where $\hat{\boldsymbol{i}}$ and $\hat{\boldsymbol{j}}$ denote the horizontal and vertical unit vectors. Therefore, $\boldsymbol{u}_s$ is intrinsically dependent on the interface profile $\zeta_0$, which produces a periodic streaming profile anchored about the two nodal positions in the case where the meniscus profile is symmetric. The resulting streaming magnitude is therefore dependent on the Fourier spectrum of the static meniscus profile as highlighted by Zhong *et. al*, wherein the flow profile is shaped by the entire wave spectrum rather than a single monochromatic wave. Therefore, for a given frequency and amplitude, the curvature defines a velocity excitation mode, the magnitude of which depends on the shape of the interface. In addition to the meniscus curvature, we also hypothesize that for an increasing meniscus profile beyond the extent of the print head the available material influx scales with:

$$Q \propto \int_0^\kappa \zeta_0(x)\, dx, \tag{68}$$

where $\kappa$ denotes the maximum extent of the interface from the print head edge. The value of $Q$ will increase up until the point that $\partial_x\zeta_0|_{z=\kappa} = \cot\theta_Y$. Beyond this point the interface keeps increasing in volume laterally, however, the value of $\kappa$ does not substantially increase. We hypothesize that this lateral volume expansion is due to manufacturing inaccuracies in the print head, which result in lateral translation and rotation of the contact line producing a slightly wider and obtuse meniscus than the diameter of the print head. This overpressurization results in three key changes that impact the efficacy of the interface to induce streaming. Firstly, as the volume of the meniscus increases, we obtain an effective amplitude reduction due to the increase in internal volume. Furthermore, the contact line rotation produces a positive bulged region, which limits material influx due to flow separation and destructive flow interactions. Finally, the 'stiffness' of the bubble in the *x*-direction decreases in comparison to the *z*-direction, resulting in both vertical and horizontal driving modes, reducing the overall efficiency.

## S16: Interface wetting model

The rate at which you can print in DIP is primarily dependent on two key processing parameters, the responsiveness of the material to light and the rate at which new material can enter the printing interface. For the former, the polymerization kinetics are driven by the intensity of light, monomer concentration, oxygen inhibition region, photo absorber concentration and photo-initiator concentration[20]. For the latter, the rate of material influx is driven primarily by the velocity of the interface in the z-direction and the frequency and amplitude of acoustic driving. However, an important criterion to meet is to ensure that, independent of the part geometry, the interface is completely saturated with new material. To predict this infill time for a given geometry we employ a computational approach based on the distance transform of the voxel array[21], where the presence of geometry is defined as a '1' and the absence of geometry is defined by '0'. We can, therefore, treat the '0' regions in the voxel array as resin sources, which define the fluidic path length. For each voxel in the array the distance between a white pixel $w_{i,j,k}$ and the closest source $s_{i,j,k}$ is given by,

$$D_{i,j,k} = \arg\min_{i,j,k} \sqrt{(w_i - s_i)^2 + (w_j - s_j)^2 + (w_k - s_k)^2}. \tag{69}$$

Therefore, the time to until the voxel $w_{i,j,k}$ is completely filled with new material is approximately

$$t = \beta \frac{D_{i,j,k}}{\bar{u}}. \tag{70}$$

where $t$ represents the infill time, $D_{i,j,k}$ is the magnitude of the distance between the voxel $w_{i,j,k}$ and the closest source $s_{i,j,k}$ and $\beta$ represents a correction factor which depends on the geometry and volume of the available source material. Additionally, two constraints are applied to the solution $D_{i,j,k}$ which depend on the object geometry and interface shape. The first is that for a given voxel $w_{i,j,k}$, the search region for the closest source point cannot exceed $k$, as $k$ defines the printing surface. Source regions greater than $k$ contain no material as they exist above the air-liquid meniscus. Secondly, a source point is only valid if the vector between the source and the voxel $\vec{V}_{s \to w}$, does not intersect the geometry, **Supplementary Fig.13a**. This is to ensure that a minimum solution is not found which is blocked by neighbouring geometry. This approach is quite similar to voxel ray tracing, which is often used in computer graphics for modelling light transport[22,23]. For example, let $\vec{V}_{s \to w}$ be the vector formed between $w_{i,j,k}$ and $s_{i,j,k}$, whose distance is given by $D_{i,j,k}$. The origin of the vector $O(O_i, O_j, O_k)$ be located at the source $s_{i,j,k}$, with a direction vector $D(D_i, D_j, D_k)$, therefore any point along the vector its position is given by.

$$P = O + tD. \tag{71}$$

Let the bounding box of the voxel array, be given by $V_{min} = (V_{min,i}, V_{min,j}, V_{min,k})$ and $V_{max} = (V_{max,xi}, V_{max,j}, V_{max,k})$. To determine if the ray intersects the voxel, we compute the intersection points with the planes defining the voxel's surfaces. For each voxel face we compute the entrance and exit points of the ray:

$$t_{enter,i} = \frac{V_{min,i} - O_i}{D_i}, t_{exit,j} = \frac{V_{min,i} - O_i}{D_i}, \tag{72}$$

$$t_{enter,j} = \frac{V_{min,j} - O_j}{D_j}, t_{exit,j} = \frac{V_{min,j} - O_j}{D_j}, \tag{73}$$

$$t_{enter,k} = \frac{V_{min,k} - O_k}{D_k}, t_{exit,j} = \frac{V_{min,k} - O_k}{D_k}. \tag{74}$$

The ray intersects the voxel if and only if the intervals ($t_{enter}, t_{exit}$) for each axis overlap. The intersection occurs if the maximum value among all $t_{enter}$ values is less than or equal to the minimum value among all $t_{exit}$ values. If an intersection is found, its position can be calculated by using the value of $t$ over the interval in which the intersection occurred.

**S17: Print speed prediction using interface wetting model**

Using the interface wetting model, the fluidic path length and wetting time can be determined for representative slice planes, as shown in **Supplementary Fig. 13b-c**. By repeating this approach for all object planes (**Supplementary Fig. 13e**) and taking the maximum value for each plane, the fluidic path length $D_z$, interface wetting $t_z$ time and vertical print velocity $V_z$ (independent of curing kinetics) can be generated over the entire object (**Supplementary Fig. 13f**). Therefore, two independent solutions for an object's print time can be created. Firstly, a conservative approach can be applied wherein the print speed is dependent on the minimum $V_z$ value over the entire object. Alternatively, the print speed can be dynamically increased or decreased in a geometrically dependent way based on the local $V_z$ of that layer. A comparison between the cumulative print time of these two approaches can be shown in **Supplementary Fig. 13g**.

**S18: Finite element analysis of the printing resin influx**

To analyze the effects of the curved interface and acoustic actuation on printing speed, we model printing material inflow using finite element analysis (FEA) software COMSOL Multiphysics 6.1. Resin ingress across printing structure was investigated for two competing printing schemes: the DIP (**Supplementary Fig.20a-b**) and the classic top-down stereolithography printing approach (**Supplementary Fig.20c-d**). We employ axial symmetry of the problem by utilizing 2D axisymmetric modeling domains to drastically reduce computational effort. The print head and printed structure are treated as impermeable solids and excluded from the modeling.

The Laminar Flow module is used to model the pressure and velocity field in the printing material (PEGDA) and air subdomains. Assuming incompressible Newtonian fluids, this module utilize the Navier-Stokes equations. A non-slip boundary condition is used on all the outer walls of the domains except the free surface and the meniscus. For initial conditions, velocity components are zero and a zero reference pressure is induced at the top boundary. The air properties were set at a density of 1.204 kg/m³ and a viscosity of 18.1 µPas. The PEGDA density was 1012 kg/m³. Viscosity and surface tension data for PEGDA can be found in **Table 1**.

PEGDA-air interface and PEGDA free surface are simulated with the Moving Mesh module. The velocity and the normal stress boundary condition on the PEGDA-air interface are set as following:

$$\boldsymbol{u_1} = \boldsymbol{u_2}. \tag{75}$$

$$(\boldsymbol{T_1} - \boldsymbol{T_2})\hat{\boldsymbol{n}} = \gamma(\nabla_s\hat{\boldsymbol{n}})\hat{\boldsymbol{n}}. \tag{76}$$

where indices 1 and 2 denote the PEGDA and the air phases respectively, $\hat{n}$ is the unit normal, outward from the PEGDA domain, and $\nabla_s = (I - \hat{n}\hat{n}^T)\nabla$ is the surface gradient operator. The Moving Mesh interface enables spatial displacement of the corresponding domain boundaries in response to the fluid motion. It utilizes the arbitrary Lagrangian-Eulerian (ALE) formulation where the mesh grid mapping to the material domain enables solving a deforming Lagrangian-type systems [COMSOL Multiphysics Reference Manual, Version 6.1]. The Navier-Stokes equations are solved within a moving frame, fully coupled with the mesh equations. The mesh velocity normal component thus matches the normal fluid velocity on the boundary. In the case of a free surface, the expressions can be simplified accordingly.

A preliminary study is conducted to establish the shape of the PEGDA-air meniscus. To do it we use a domain which has no printed structure, and the meniscus equilibrium shape is evaluated by running a time-dependent study with stationary boundary conditions (zero boundary displacement). In the subsequent analysis, the shape of the formed meniscus defines the profile of the printed structure.

To model the transient fluid ingress during printing, a dynamic study is conducted. As an initial state, the interface is considered compressed against the printed structure, forming a uniform 50 µm thick layer of fluid. This was chosen to improve initial computational stability of the solution, especially under acoustic excitation. Selected boundaries are translated downward (along the z-axis) as shown in **Supplementary Fig. 20a,c,** to replicate the displacement of the print head during the printing process. While the print head is translated upwards in the experiment, the modeling set-up is inverted and the resin container with the printed structure is displaced down instead. The displacement is performed with a delay of 0.1 s necessary for the computational model stabilization (**Supplementary Fig. 20e**). In the acoustically driven case, the upper part of the air domain is harmonically actuated at $f = 100$ Hz in addition to the displacement. The wall velocity has an amplitude of 10 mm/s and a delay of 0.1 s (**Supplementary Fig. 20f**), necessary for model stabilization.

The computational domain mesh, shown in **Supplementary Fig. 21a-b**, utilizes a hybrid grid with triangular mesh elements in the bulk of the domain, complemented with a structural grid at the fluid ingress area and near nonslip boundaries. The mesh is finely resolved at the structure tip down to $d_{mesh} = 8$ µm and expands to $72 d_{mesh} = 0.57$ mm in the bulk of the domain.

The air-fluid interface aligns with the printed structure surface at the start of the simulation. The deformation of the domain propels the meniscus detachment from the structure. The displacement of the meniscus centroid $C$ is plotted in **Supplementary Fig. 21c** for DIP with the printed structure of $S_D =$ 10 mm. In the timeframe 0.1-0.96s, it translates downwards, following the boundary displacement. However, the fluid ingress along the structure causes the rebound of the meniscus and subsequent recovery. The rebound dynamics was used to evaluate the computational mesh where the half-sized mesh demonstrates practically identical dynamics of the meniscus.

**Supplementary Fig.23-24** plots the fluid velocity and velocity vectors across the approximate influx timeframe for four cases: Top-Down SLA, DIP without acoustics, DIP with acoustics driven at 40 Hz and DIP with acoustics driven at 100 Hz. These figures demonstrate an intensive flow in the interface layer adjacent to the structure. The acoustic actuation was found to induce pressure variation in the air domain with an amplitude of about 20 Pa. This pressure oscillation induces capillary-gravity waves at the liquid-air interface. **Supplementary Fig.24** demonstrates resulting fluid streaming along the acoustically actuated fluid-gas interface, ultimately accelerating the resin influx. The PEGDA ingress in turn induces recirculating flow in adjacent air domain. Acoustics actuation subsequently reduces the time required for complete wetting of the structure.

**Supplementary Fig.26a-b** plots the oscillation modes of the acoustically actuated meniscus. This visualization demonstrates the complex interactions between meniscus shape, print head size and printed structure size which governs the oscillation intensity and eventually the fluid ingress. While this study is limited to an axisymmetric case, more complex spatial modes might be observed in 3D system. These findings underline the importance of multimodal surface actuation to ensure efficient resin influx.

### S19: Particle image velocimetry (PIV)

A key component of the Dynamic Interface Printing technology is the velocity profile below and across the curved air-liquid interface. The complex nature of this system is such that the flow profile is intrinsically determined by the driving frequency, meniscus curvature, print head geometry, material properties, proximity to a solid boundary and the underlying structure of this boundary (whether printed or otherwise). The mathematical approaches discussed in previous sections aim to elucidate some of the primary driving mechanisms for acoustically driven flow. Like any model, however, they cannot capture the full extent of this behaviour.

To supplement our mathematical explanation, we used 2D particle image velocimetry (PIV) to observe the flow behaviour from both the top (**Supplementary Fig.8**) and side (**Supplementary Fig.9**) profiles of the air-liquid interface. The flow profile created by the air-liquid boundary extends in three-dimensions, therefore we aimed to capture key components of this flow profile by sectioning the interface about the $xy$ and $xz$ planes.

<u>PIV imaging, laser and material setup</u>

The PIV illumination setup consisted of a 200 mW 523nm green laser which was subsequently passed through a 45° Powell Lens (Thorlabs, LGL145) to create a uniform sheet of light. To adjust the thickness of the far-field beam, a second cylindrical lens (Thorlabs, LJ1629L2) was placed between the Powell lens and the cuvette resulting in approximately a 300 µm thick sheet at the focal plane. The focal plane of the light sheet was adjusted such that it intersected the mid-plane of the print head ($yz$) and the beam divergence was assumed to be negligible over the extent of the cuvette. For the side PIV experiments, the length of the light sheet was aligned such that it bisected the print head in the $xz$ plane. For the top-down PIV experiments, the light sheet was rotated 90° and aligned just below the maximum extent of the meniscus.

High speed images of the interface dynamics and interfacial fluid flow were captured using a Chronos 1.4 camera (Kron Technologies, Chronos 1.4 Camera) which enabled a maximum framerate of 40,413 fps. In our case, the camera framerate was varied from 200 – 2000 fps, depending on the experimental conditions. For both the side and top-down PIV experiments a microscope lens was used (Kron Technologies, Microscope lens), and the magnification of the lens was set to ensure the entire ROI was in-view.

20-50µm PMMA particles (LaVision, PMMA particles 20-50 µm) were used to seed the flow at a density of 0.1% (w/v). All PIV experiments were performed using a 25 mm diameter print head with a 20% PEGDA formulation without the photo initiator or photo-absorber to ensure adequate light-transmission through the volume. Particle detection was therefore determined by florescent excitation and scattering of the light sheet.

<u>PIV Analysis</u>

The recorded high-speed video was analyzed using PIVLab within MATLAB. For each experimental condition, a binary mask was generated to isolate the fluid region. In the top-down experiments, the mask corresponded to the diameter of the print head, and for the side-profile experiments the mask covered the print head body and maximum meniscus extent. The maximum meniscus extent was determined by edge tracking of the interface over the time series. The mask was therefore determined to be the maximum location of the interface at any point within the time series + 10%. This was to ensure that the interface movement did not contribute to the velocity profile. This approach potentially results in a lower measured maximum fluid velocity than in reality, though, as the flow close to the interface is truncated by the mask. Additionally, movement of the interface during acoustic actuation produces periodic lensing which precludes accurate PIV tracking, especially at high amplitudes. Therefore, the velocity profile was calculated approximately 90 ms after acoustic stimulation was turned off, which corresponds to the approximate interface settling time in the fluid bulk (**Supplementary Fig. 7**). Therefore, the calculated velocity in this case is lower than when acoustic stimulation is on. For each image in the video sequence, the following PIVLab settings were used:

1. ROI was determined by the binary mask.

2. Each image was pre-processed with the CLAHE filter (10px), high-pass filter (45px) and denoise filter (5px). The contrast was set to auto and the mean background was subtracted from each image.

3. The PIV analysis was performed with the default settings with an integration area of 64px and a step of 32px. A second pass integration area of 32px and a step of 16px was also performed.
4. Both velocity vector validation and image-based validation were used to remove vectors outside 2x the standard deviation of the vector-field.

The final velocity values $V_x, V_y$ and $\|V\|$ were exported from PIV lab and recreated within MATLAB. The resulting top-down and side-view PIV data is shown in **Supplementary Fig. 8** and **Supplementary Fig. 9.**

## S20: Interface release dynamics

One experimental challenge with investigating the fluid release rate close to an interface using PIV is that the fluidic displacement caused by the air-liquid boundary removes tracer particles from the region of interest. This is further exemplified by the fact that the curved oscillating boundary scatters incoming light generated by the light-sheet, resulting in image processing artefacts. As an alternative method to track the material influx rate, we doped each corresponding material with a small volume fraction of black dye and placed the cuvette on top of a bright backlight to produce a uniform light source beneath the cuvette. When the interface is in contact with the bottom of the cuvette, the dyed material is evacuated below the interface resulting in a clear circular contact region, **Supplementary Fig. 11a**. As the print head begins to move up in the z-direction, material flows in to fill available space resulting in a high contrast circle that reduces in diameter as a function of time, **Supplementary Fig. 11b**. By fitting an ellipsoid around the high-contrast region, the proportion of dry area (white areas) to wet areas (black areas) can be plotted as a function of time for acoustic and non-acoustic stimulation, **Supplementary Fig. 11c-d**. The average fluid velocity $V_{\text{fill}}$, for a period of time is given by:

$$V_{\text{fill}} = \sqrt{\frac{a_i - a_f}{\Delta t_{i \to f}} + \frac{b_i - b_f}{\Delta t_{i \to f}}}, \qquad (77)$$

where, $a, b$ denote the principal axes of the ellipsoid and $i, f$ denote the initial and final values over the time point $\Delta t_{i \to f}$.

## S21: Particle settling in dynamic interface printing

A significant obstacle in employing low-viscosity materials for cell-laden biofabrication arises from the dependency of cellular sedimentation rates on the relative densities of the materials involved and the viscosity of the bath material. The settling velocity of a spherical particle is proportional to $v \propto \Delta \rho \mu^{-1}$. In practice this causes a density-driven migration of particles towards the base of the structure prior to printing. To investigate this effect, we suspended 30 μm PMMA particles which have a density of $\sim 1.2 \text{ g/cm}^2$, into a solution of 20% PEGDA with a density approximately that of water $\sim 1.2 \text{ g/cm}^2$, and a viscosity of $\sim 3 \text{ mPa} \cdot \text{s}$. The PEGDA-particle mixture was then left for approximately 30 minutes to ensure that the majority of particles had settled to the base of the container prior to printing. To evaluate particle distribution in the final construct, a 10 mm diameter, 20 mm tall rod was printed with and without acoustic stimulation. After printing, the rod was sectioned in the $xz$ plane and imaged using an inverted microscope (**Supplementary Fig. 15a-c**). The location and distribution of particles in the sectioned plane was determined by thresholding, segmenting the image into a binary array and summating the intensity in the *x* and *y* directions of the image (**Supplementary Fig. 15d-e**).

In both cases, DIP mitigated the effect of particle settling even without acoustic stimulation. Particles were distributed throughout the rods, though in the case without acoustic stimulation, the density of particles decreased sharply above the bottom layers. We postulate that the size of the print head relative to the diameter of the cuvette (approximately ½) causes secondary flows due to the print head retraction from the volume, which aid in resuspending particles into the bulk material. Under acoustic excitation, this resuspension is further improved by circulating flows generated across the extent of the print head. In addition to the improved resuspension, the greater material influx rate under acoustic stimulation not only improves the distribution of particles in the z-direction, but also increases the total number of encapsulated particles. This is highly useful for biofabrication, wherein a greater number of cells could be encapsulated within the construct for the same cell density.

### S22: Optical power measurements

Optical power measurements were taken at the focal plane of the projection module using a commercial optical power meter (PM100A, Thorlabs) with a 200 -1100 nm Si photodiode (S120VC, Thorlabs). To determine the optical power density, a 900 µm aperture was placed in front of the photodiode.

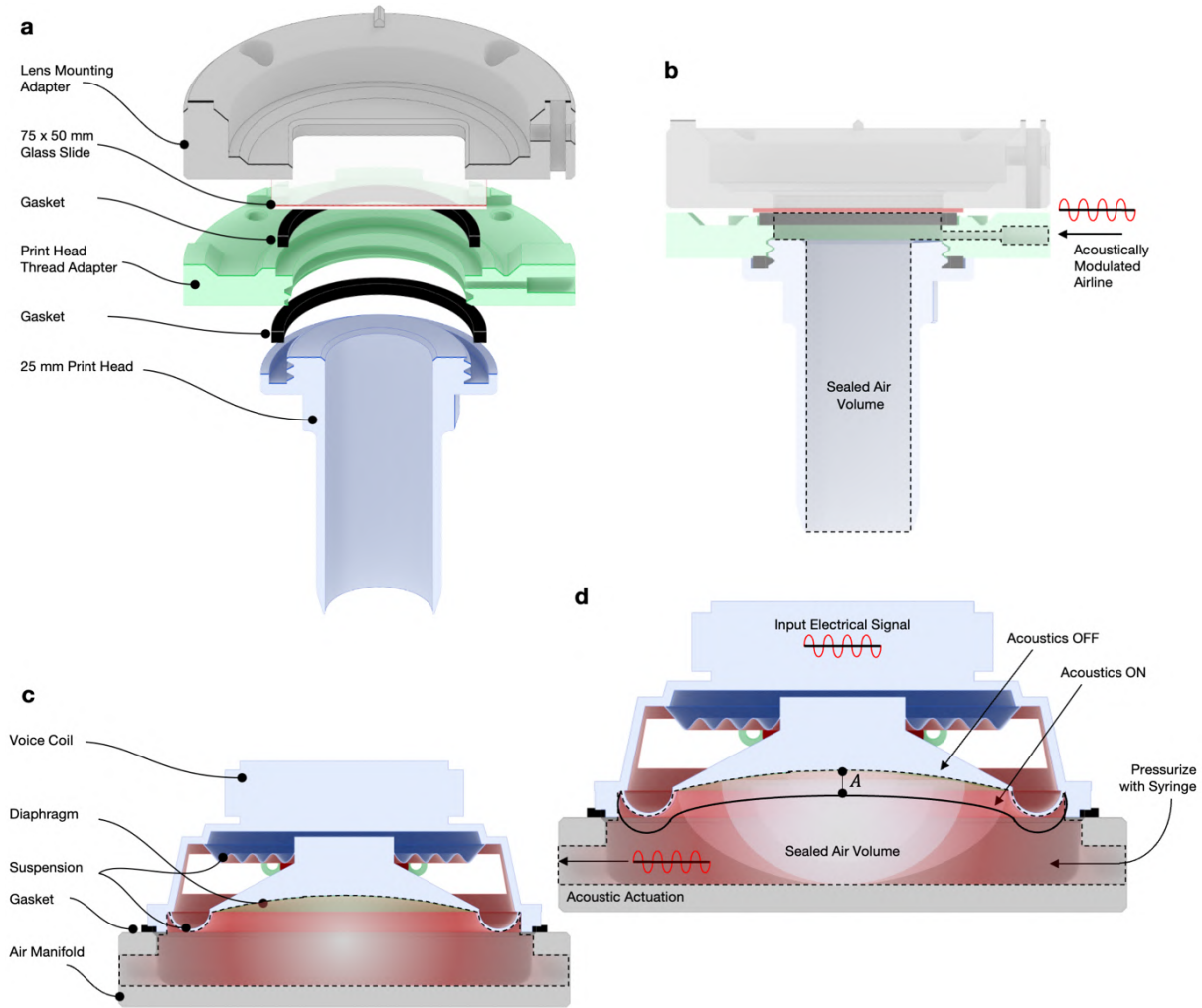

**Supplementary Fig.1 | Illustrated CAD model of the print head assembly and acoustic air-line modulation. a**, expanded half section view of the print head assembly. **b**, collapsed half section view of the print head assembly, highlighting that a sealed air-volume is formed with a transparent glass window at the top. **c**, half section view of the air-line modulation system, wherein a speaker diaphragm forms one side of an enclosed box. **d**, electrical signal applied to the voice coil causes excitation of the diaphragm which modulates the volume around a set point pressure within the air-manifold and in turn the print head.

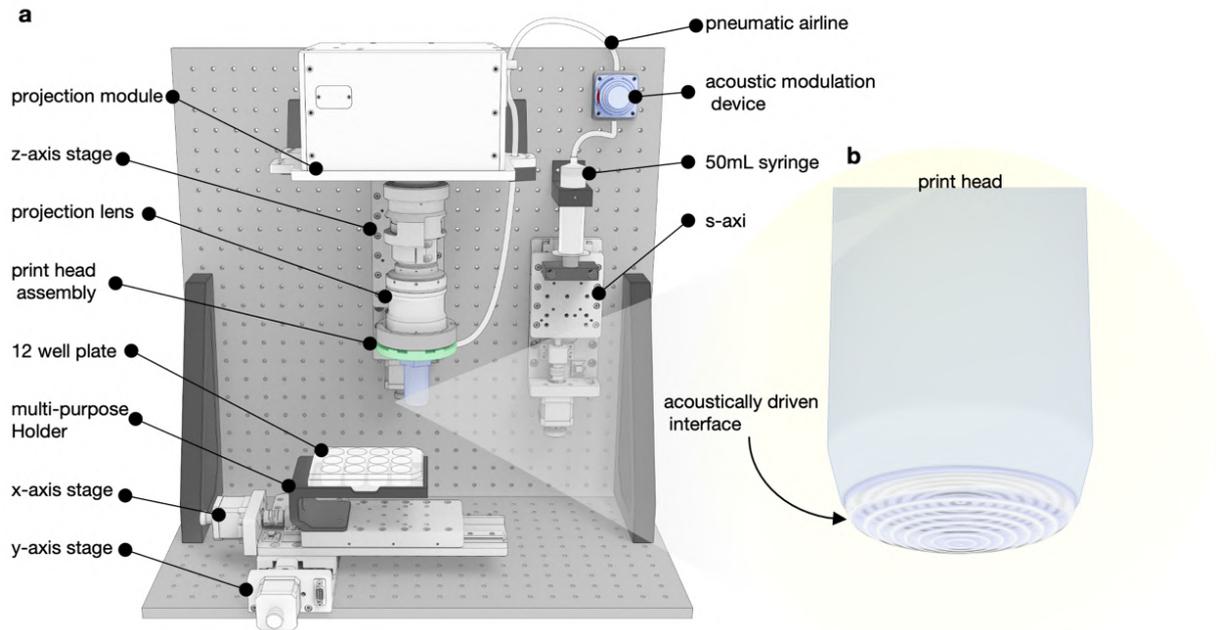

**Supplementary Fig.2| Illustrated CAD model of the mechanical components of the DIP printing system version 1 (V1). a**, labelled components of the prototype DIP system. **b,** inset view of an air-liquid boundary formed at the tip of the print head under acoustic excitation.

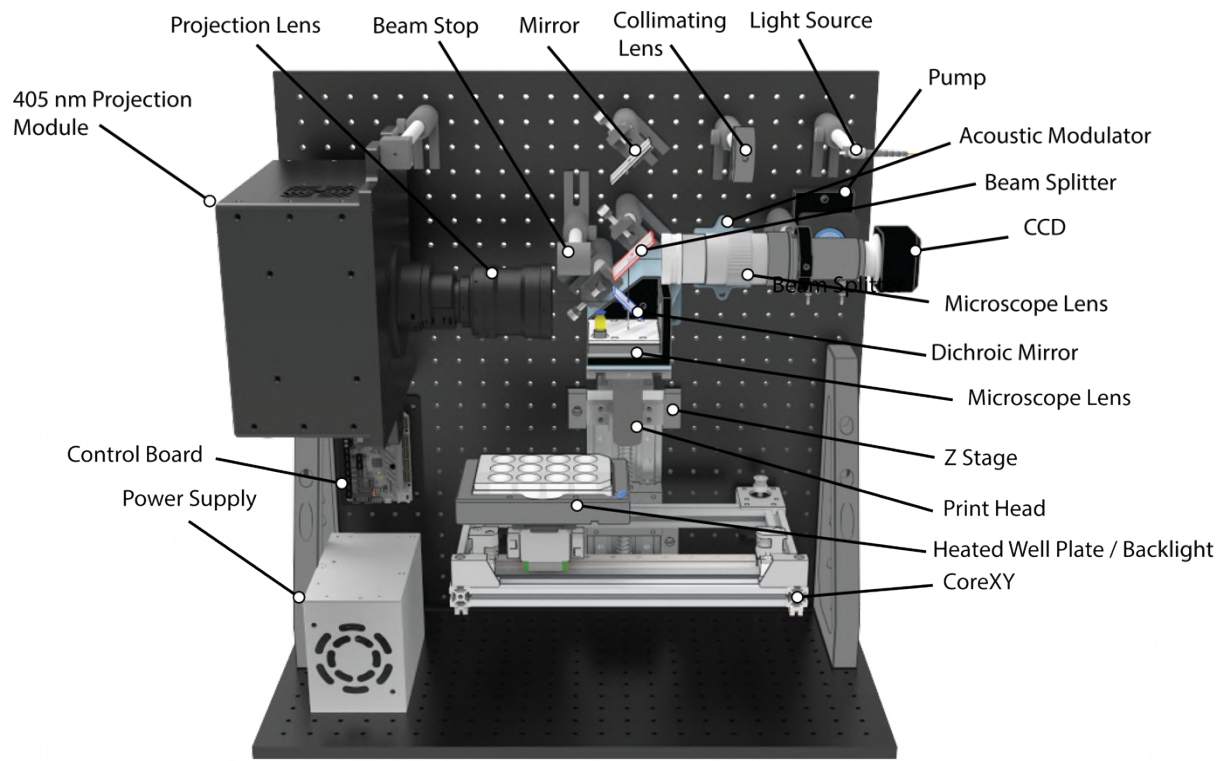

**Supplementary Fig.3 | Illustrated CAD model of the mechanical components of the DIP printing system version 2 (V2).** Version 2 of the DIP system builds upon the same principles as V1, however adds additional capabilities for in-situ imaging and a larger XY printing area.

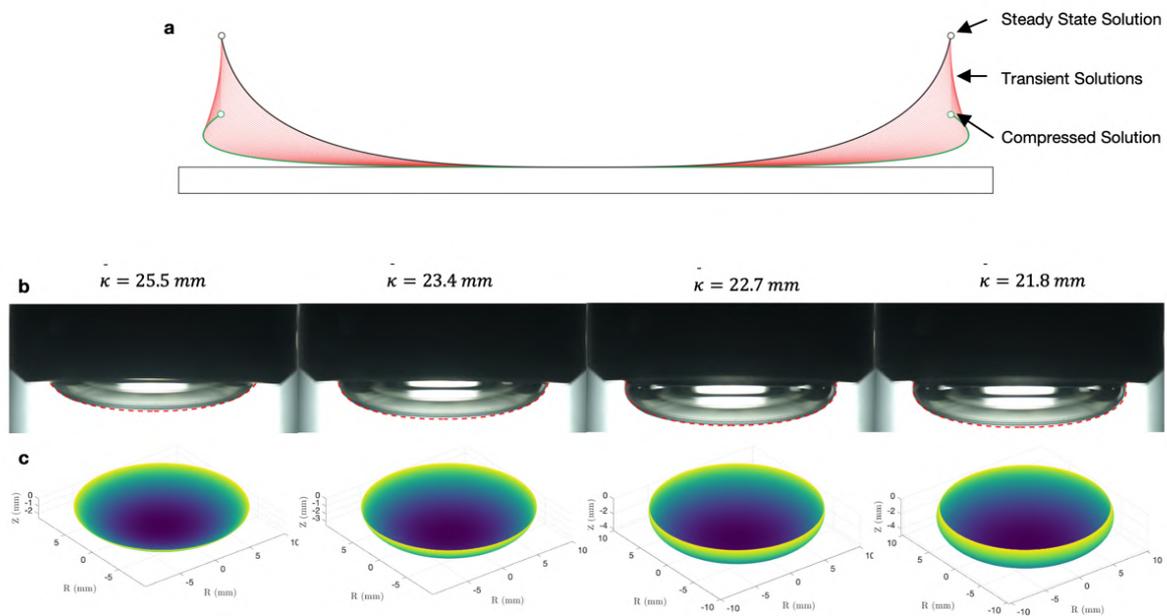

**Supplementary Fig.4 | Interface shape reconstruction based on Bezier Young-Laplace model. a**, solutions of the approximated Young-Laplace equation using Bezier curves as the interface transitions from a steady-state solution to the compressed solution, for a given pseudo-flat diameter. **b**, shape of the air-liquid meniscus formed at the tip of a 20 mm print head. Red dashed line indicates the solution from the Bezier Young-Laplace model overlayed with the real-world curvature. $\bar{\kappa}$ indicates the average radius of curvature for each interface, where for increasing pressure difference $\Delta P$ the interface curvature $\bar{\kappa}$ decreases. **c**, three-dimensional reconstruction of the interface profile assuming symmetry about the z-axis.

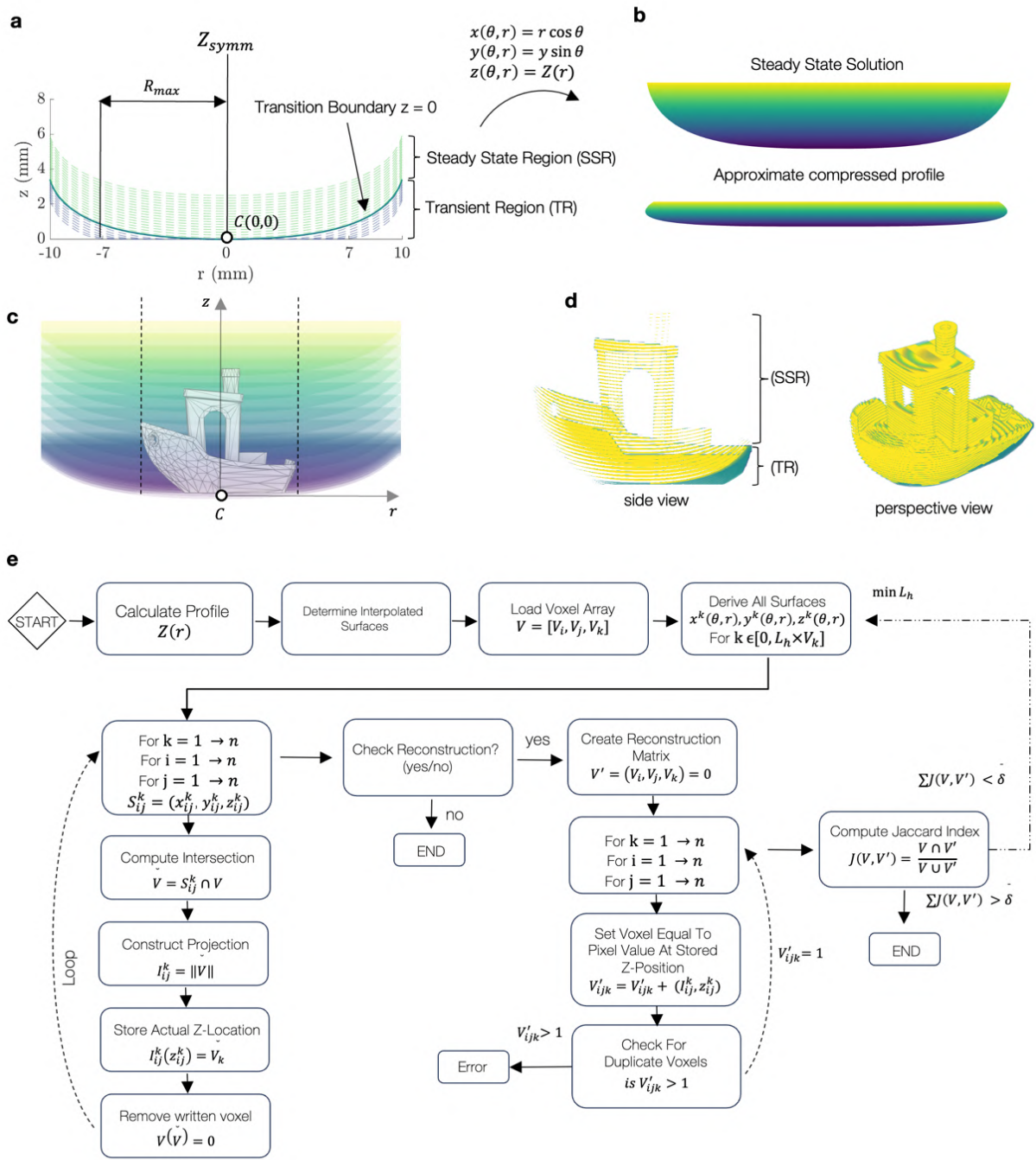

**Supplementary Fig.5| Convex slicing algorithm process flow**. **a**, examples of interface profiles produced by the Young-Laplace interface model demonstrating the steady-state solution and transient region of the interface shape corresponding to compression about the contact point C. **b**, three-dimensional surface constructed by revolving the steady-state region and transient regions about the symmetry line $Z_{symm}$. **c**, overlay of the 'Benchy' model, which is a standard test geometry, with an array of meniscus profiles corresponding to the steady-state and transient solutions over the object's height. Note the number of surfaces shown is not representative of the total number normally used when printing. **d**, reconstruction of the Benchy model based on the projections from the convex slicing algorithm, low reconstruction quality is indicative of the limited number of slice planes. **e**, process-flow diagram of the slicing algorithm illustrating key steps in both the determination of convex projections $I_{ij}^k$ and reconstruction validation via Jaccard Index $J(V, V')$.

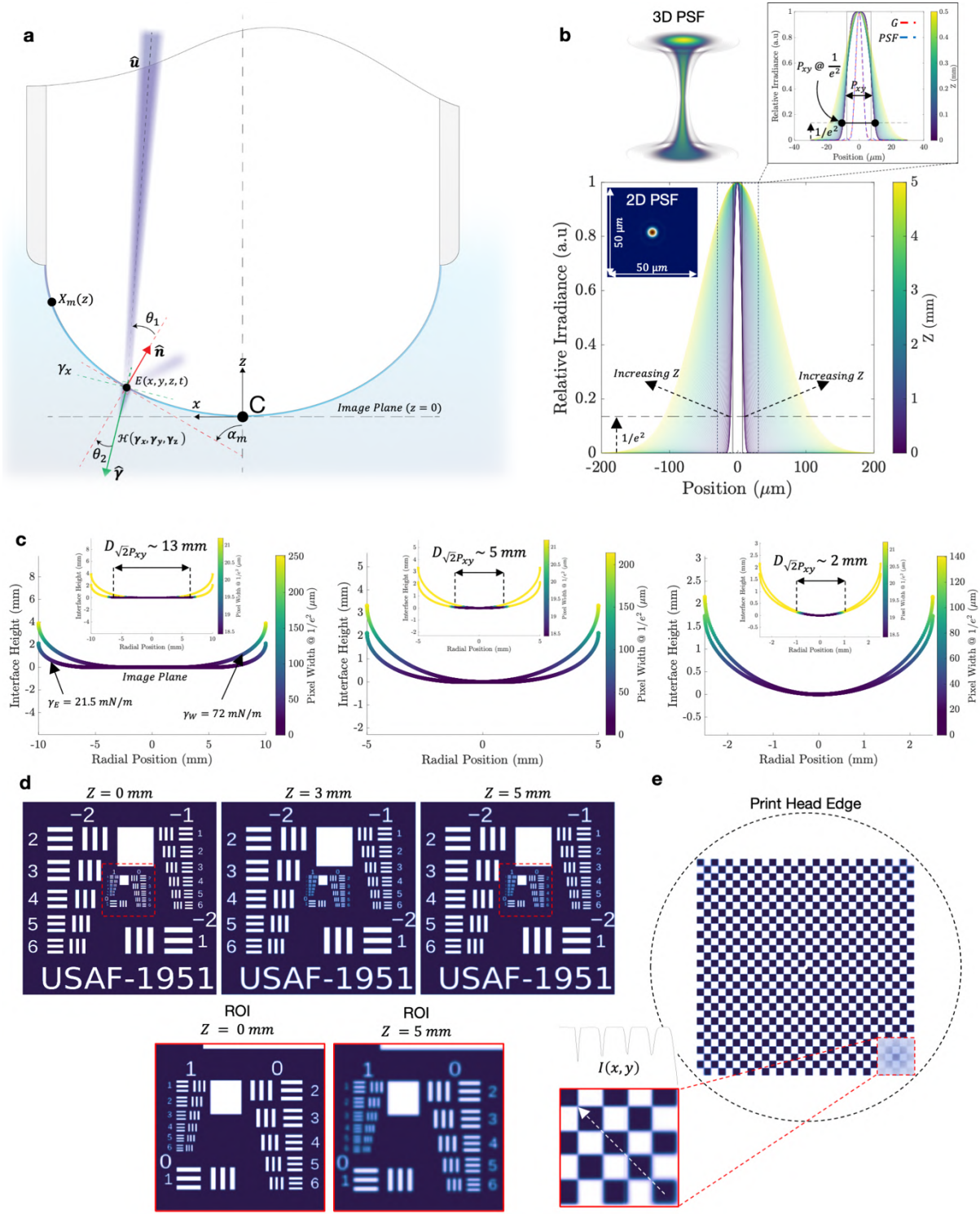

**Supplementary Fig.6| Effect of interface curvature on in-plane resolution**. **a**, optical coordinate system description denoting the incoming ray $\hat{u}$ and transmissive ray $\hat{\gamma}$ into the material. **b**, diverging effective pixel size $P_{xy}$ across a z-depth of 0 – 5 mm. Insets show the effective pixel spread close to the focal plane, the corresponding 2D PSF at $z = 0$ and the approximate thresholded PSF in the range of $z = \pm 200\ \mu m$. **c**, effective pixel resolution across the meniscus boundary corresponding to D = 20 mm, D = 10 mm and D = 5 mm respectively, for a 90-degree contact angle ($\theta_Y = 90°$). Insets show the available printing area under the assumption of an allowable pixel divergence of $\sqrt{2}P_{xy}$. **d,** resultant USAF test pattern imaged at increasing planes above $z = 0$. Inset in red shows ROI at the centre of the test pattern at $z = 0$ and $z = 5$ respectively. **e**, total image distortion across the meniscus surface corresponding to the meniscus intersection with the 3D PSF.

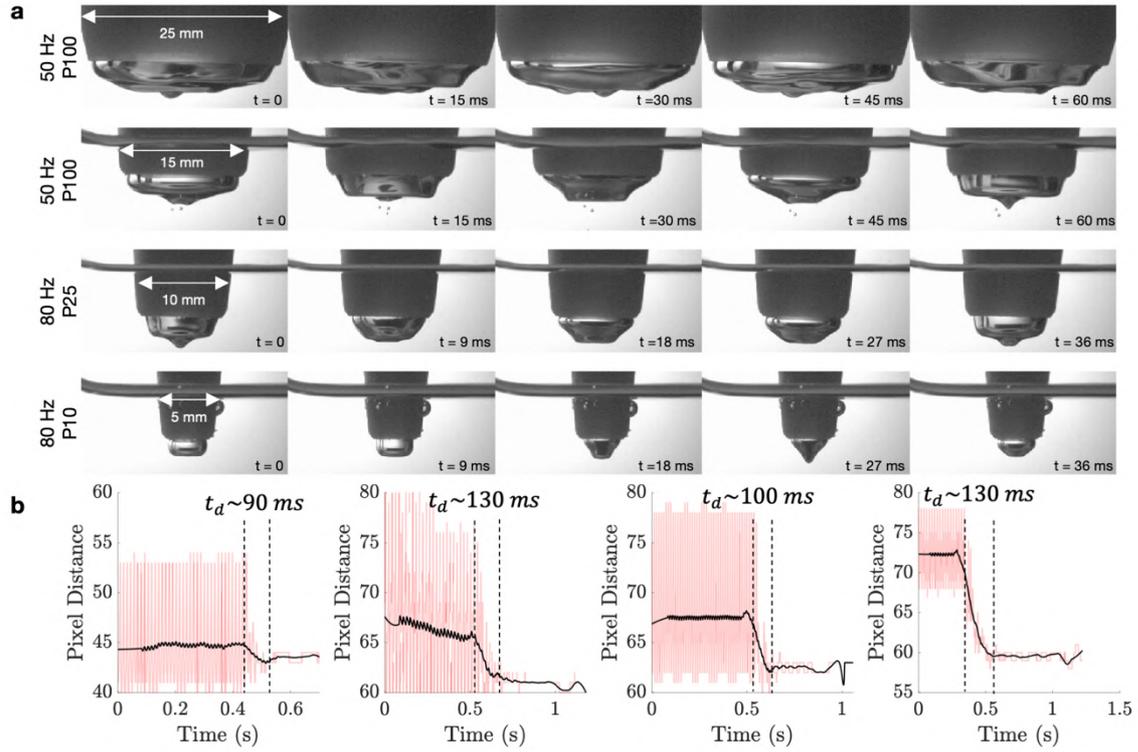

**Supplementary Fig.7| Interface re-stabilization time as a function of print head diameter for PEGDA 20%. a**, high contrast time-series images of the interface shape under high amplitude acoustic excitation. Each time series corresponds to approximately a single period of excitation. **b**, interface stabilization tracking for each of the print head configurations with decreasing print head size from left to right.

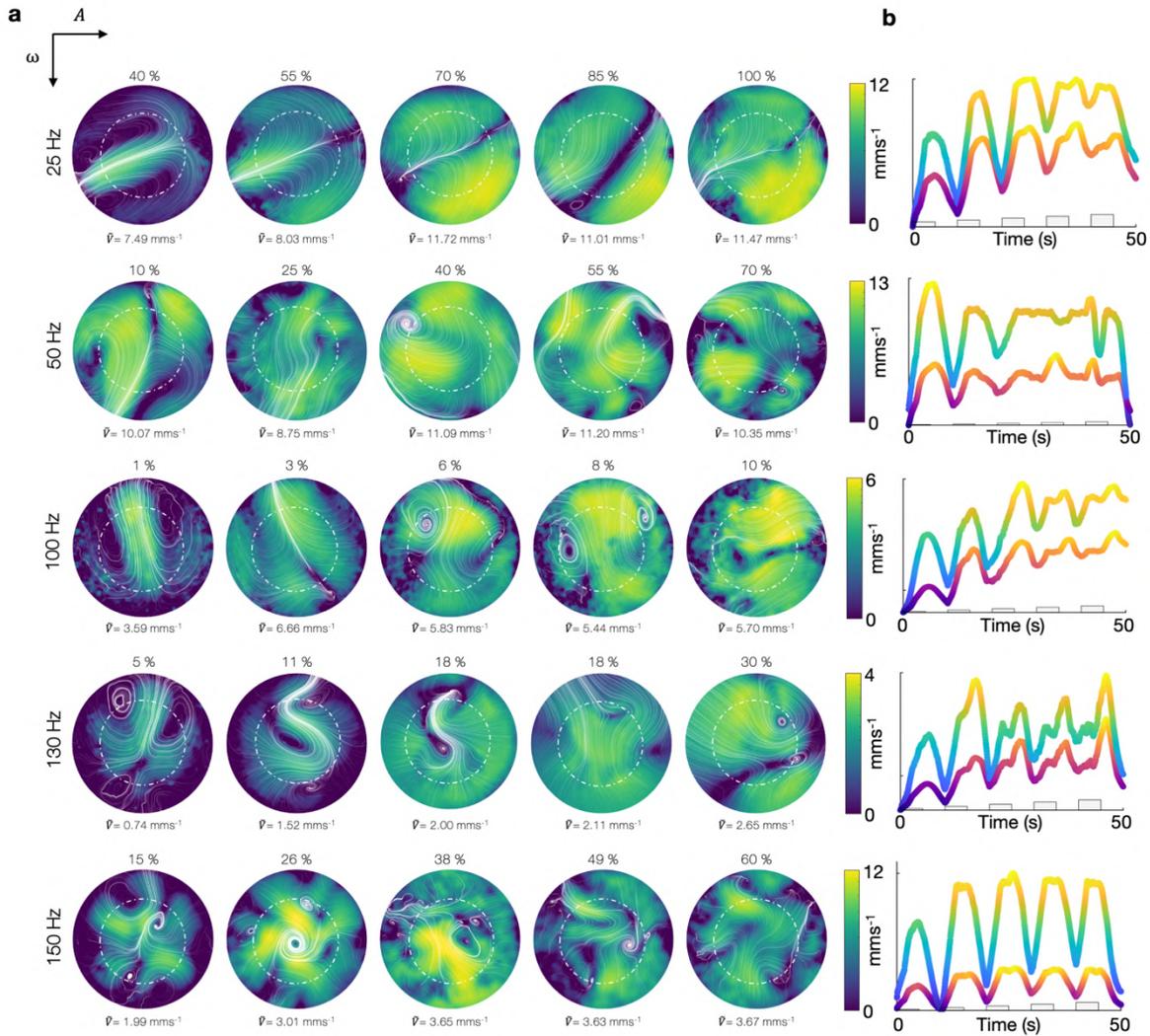

**Supplementary Fig.8| Particle image velocimetry of the printing interface as imaged from above**. **a**, time averaged velocity for 25 mm diameter print head with increasing driving amplitude $A$ and frequency $\omega$. Each image corresponds the average velocity across 20 frames, directly after acoustic actuation was turned off. $\tilde{V}$ denotes the average velocity within the centre 15 mm diameter region, highlighted by the white circular ring. **b**, average velocity and maximum velocity across the entire print field for each frequency and amplitude state. Grey bars indicate the time points and unit amplitude ($0 \rightarrow 1$) of the acoustic stimulation, where each test corresponds to 5s of stimulation followed by 5s of no stimulation as shown.

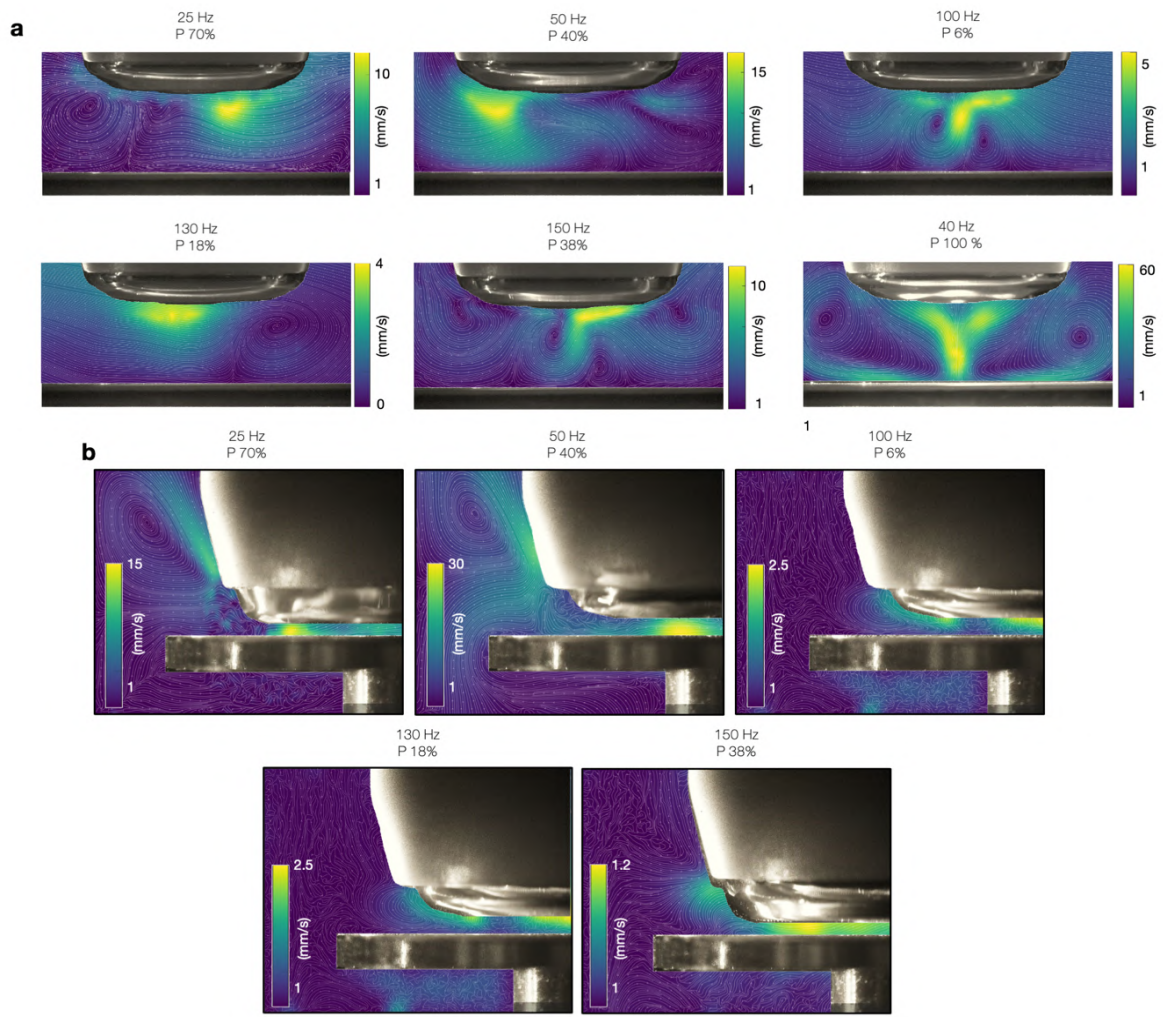

**Supplementary Fig.9| Particle image velocimetry of the printing interface as imaged from the side**. **a**, time-averaged velocity for 25 mm diameter print head for key driving frequency and amplitude pairs. Each image corresponds to the average velocity across 100 frames, with acoustic driving on for the entire test duration. The 40Hz P100 configuration highlights the 'jetting' ability of the interface under high amplitude, whereby a single high velocity jet can be formed at the centre of the print container. **b**, time averaged velocity near a solid boundary demonstrating that increased fluid velocity beyond bulk flow, can be seen under specific frequency conditions.

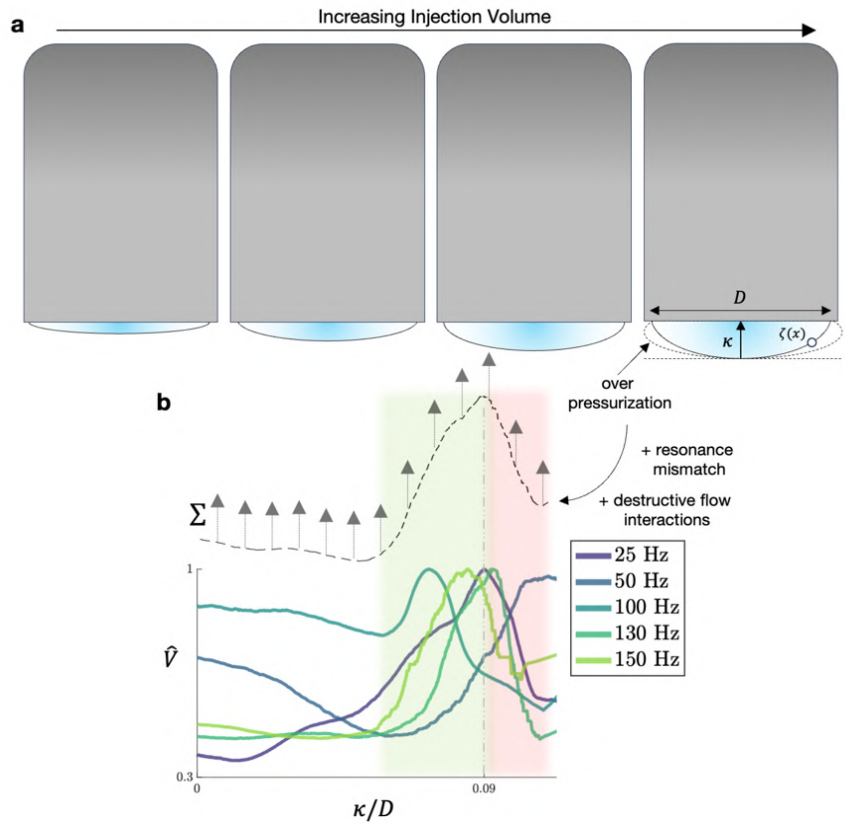

**Supplementary Fig.10| Normalized fluid velocity for increasing meniscus curvature and extension $\kappa$. a**, illustration of the parameter definitions for $\zeta(x)$, $D$ and $\kappa$. **b**, normalized maximum velocity for $\kappa/D$ over a range of driving frequencies for a fixed amplitude. Here $D = 25$ mm. Dotted line indicates the summation of all frequency response curves, emphasising a cumulative onset and peak velocity pairing, followed by a sharp decline in velocity due to over-pressurization, resonance mismatch and destructive flow interactions.

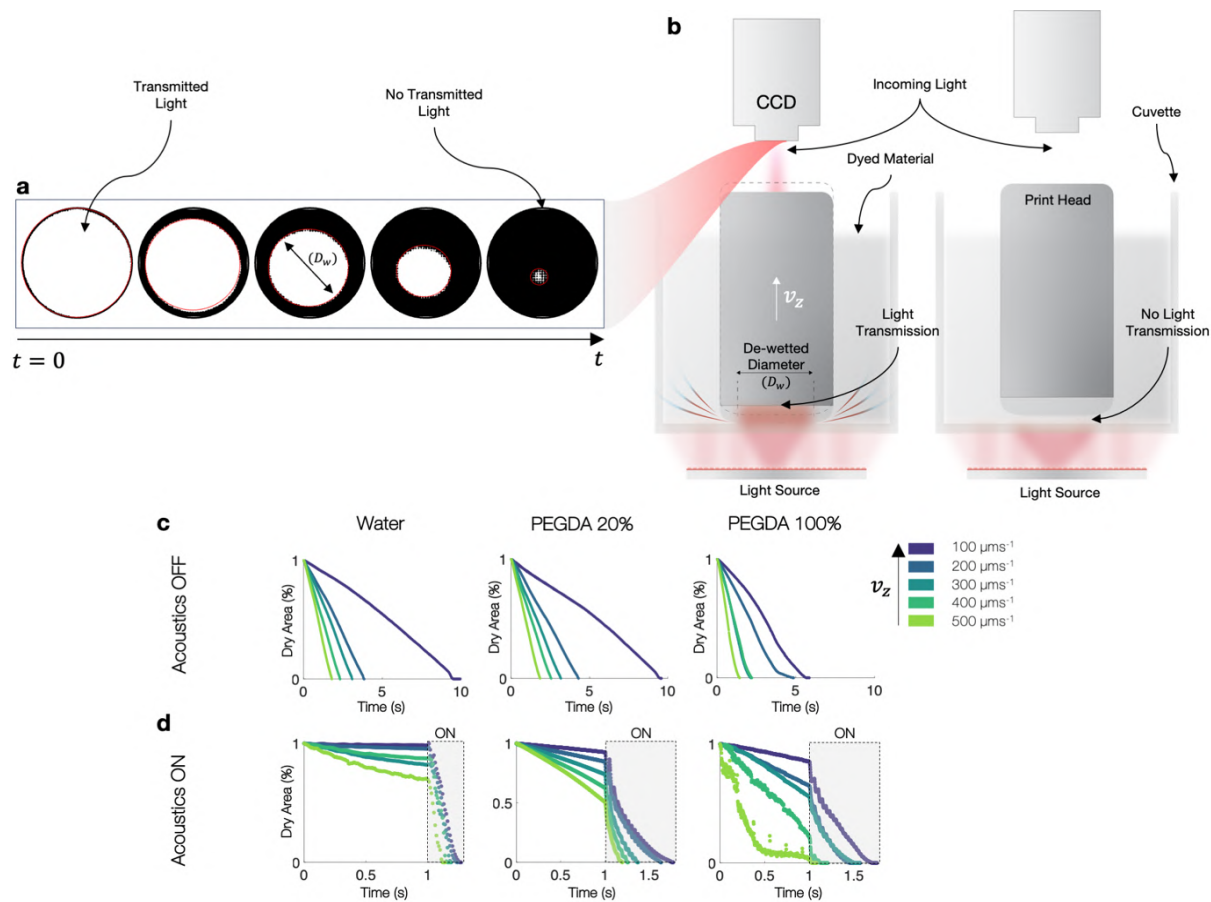

**Supplementary Fig.11| Image based analysis of the material influx rate for the 25 mm print head with varying viscosity and surface tension values. a**, time point images of the de-wetting of the meniscus as imaged from above. Dark regions of the image indicate interface release from the base of the cuvette, whereas light regions indicate interface contact resulting in light transmission. **b**, Schematic illustration of the optical setup used to capture the interface release dynamics. Each material was doped with a high concentration of black dye to attenuate the incoming light from the backlight. **c,** interface release dynamics without acoustic stimulation plotted for increasing rates of print head velocity $v_z$ and PEGDA wt.% concentration. Dry area percentage corresponds to the area of the ellipsoid that encloses the transmitted light as a function of time, normalized against the initial area of the transmitted ellipsoid. **d**, interface release dynamics with acoustic stimulation, where $\omega = 50\ Hz$ and $A = 0.40$. In each case the print container was translated initially at the same rate as in **c** for 1 second, followed by enabling acoustic stimulation.

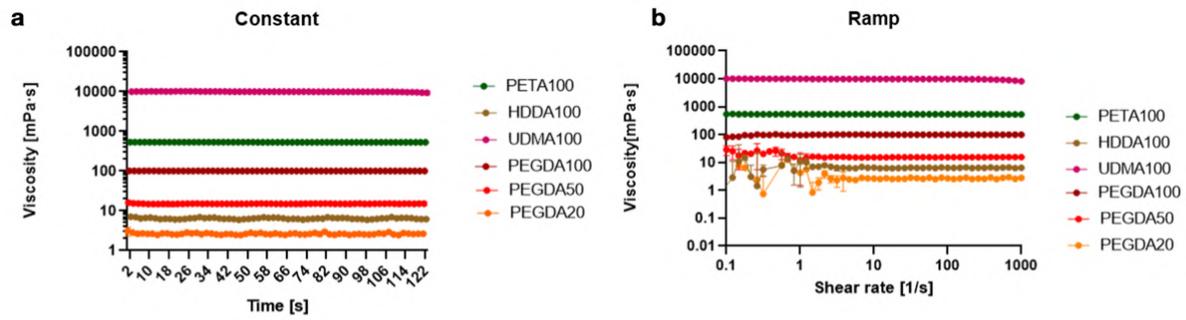

**Supplementary Fig.12| Viscosity measurements for key materials investigated within this study. a**, viscosity response under constant shear rate 10 (1/s). **b**, viscosity response under ramping shear rate from 0.01 to 1000 (1/s).

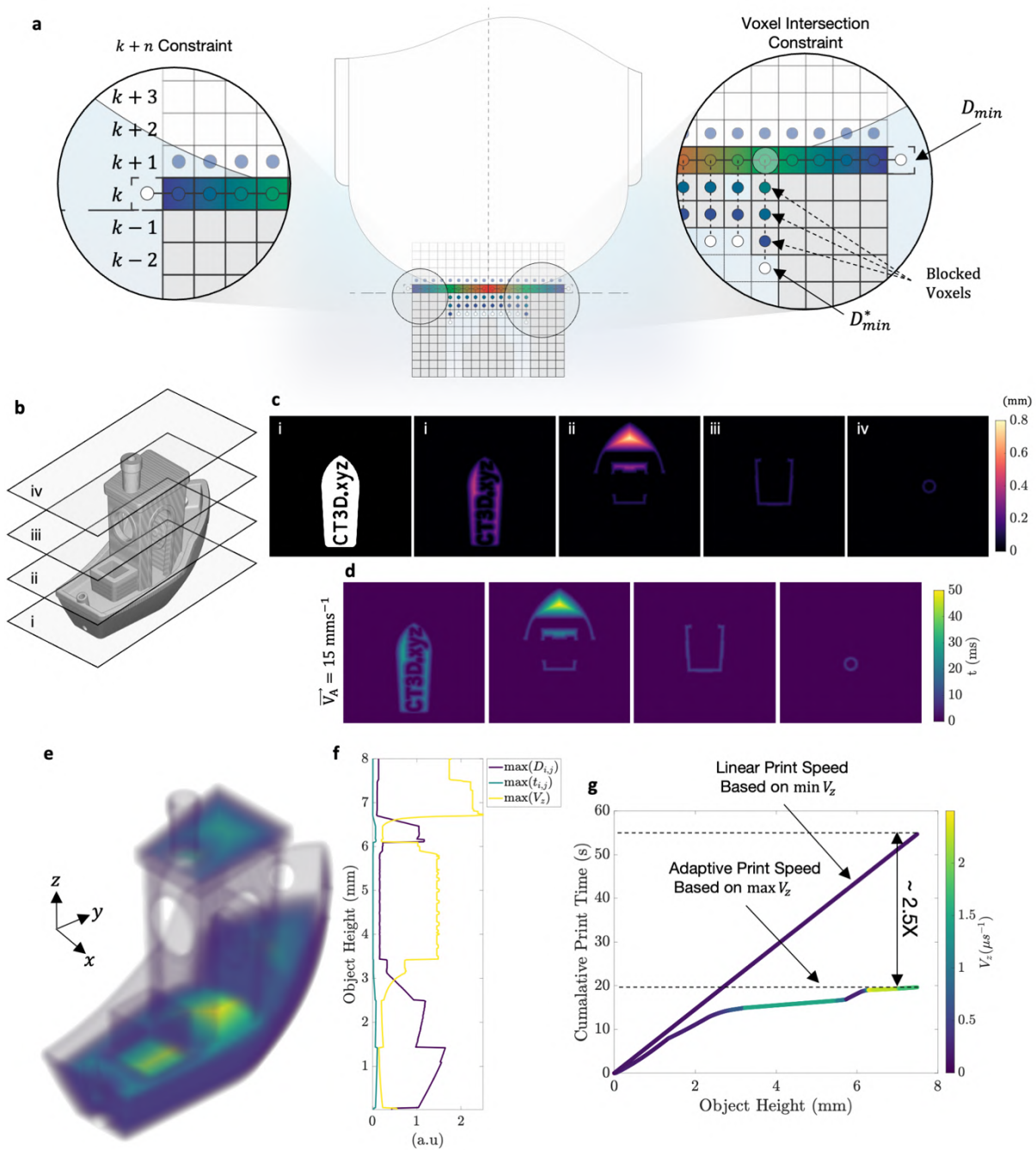

**Supplementary Fig.13| Computational prediction of interface wetting time**. **a**, illustration of algorithmic constraints applied to the 3D Euclidean distance calculation, where out of plane voxels above the calculation plane are not considered. Additionally, voxels whose distance minimizes the Euclidean distance function, but whose vector intersects with the voxel grid are also invalid solutions. **b**, voxelized Benchy model and corresponding computational target planes (i, ii, iii, iv). **c**, Euclidean distance transform applied to the corresponding target planes and normalized against the pixel size ($p_{x,y} = 15$ μm). **d**, interface wetting time in milliseconds for an acoustic flow velocity of $\vec{V}_A = 15 \text{ mms}^{-1}$. **e**, 3D distance transform applied to the entire voxelised Benchy model, colormap represents the log magnitude of the minimum source distance in 3D with imposed constraints. **f**, maximum source distance $D_{i,j}$, infill time $t_{i,j}$ and achievable print velocity $V_z$, as a function of object height. **g**, comparison between linear and adaptive print speeds on the cumulative print time. Linear print speed was taken as the minimum calculated print speed over the entire object, whereas the adaptive print speed was determined instantaneously across the object height.

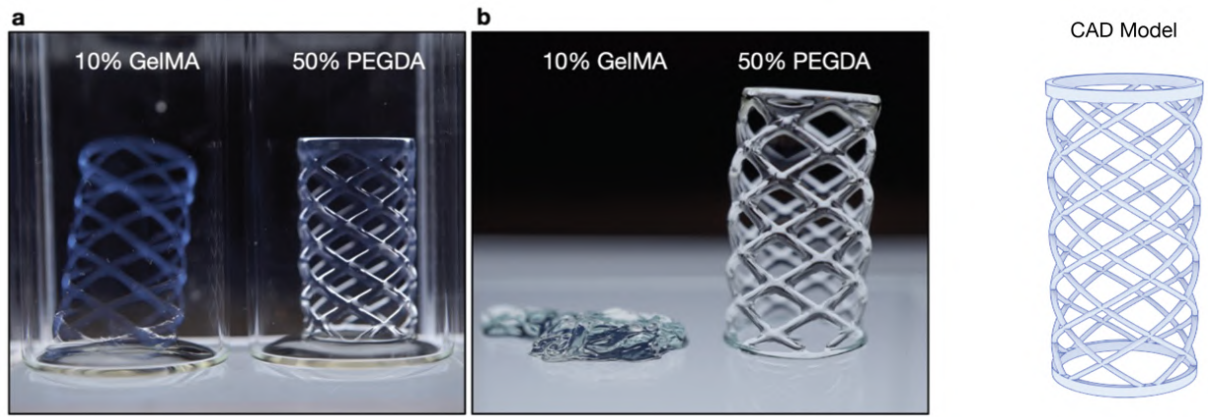

**Supplementary Fig.14| Capacity for fabricating structures with low stiffness via dynamic interface printing**. **a,** model of a helical stent geometry, constructed using a 10% GelMA hydrogel composition (left), juxtaposed with a counterpart fabricated from a 50% PEGDA composition (right). These structures, post-wash, are depicted in a liquid suspension. **b,** the GelMA and PEGDA structures after extraction from their supportive fluid, underscoring the remarkable range of stiffness that can be accomplished through the employment of DIP.

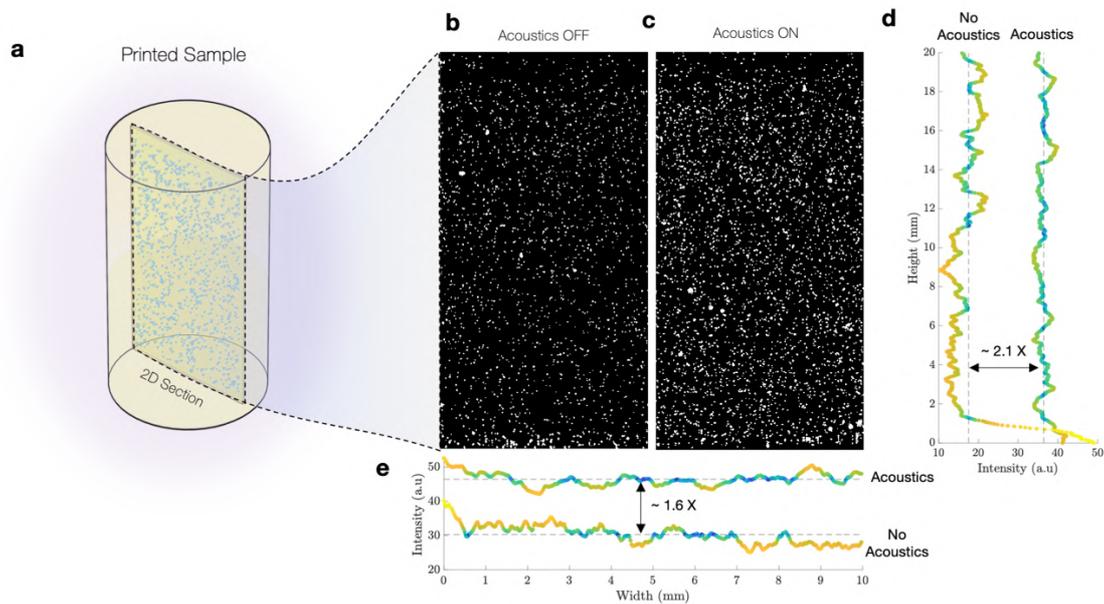

**Supplementary Fig.15| Settling of granular suspensions during printing**. **a**, illustration of the sample geometry used for this investigation, whereby a circular geometry with a diameter of 10 mm was printed. **b**, 2D cross-section of the printed sample without acoustic stimulation. **c**, 2D cross-section of the printed sample with acoustic stimulation on with a frequency of 50Hz. **d**, summed thresholded intensity of the 2D cross-sectional image in the x-direction, highlighting that approximately 2.1X the number of particles were captured with acoustic actuation on, with a more consistent distribution of particles in the y-direction. **e**, summed thresholded intensity of the 2D cross-sectional image in the y-direction.

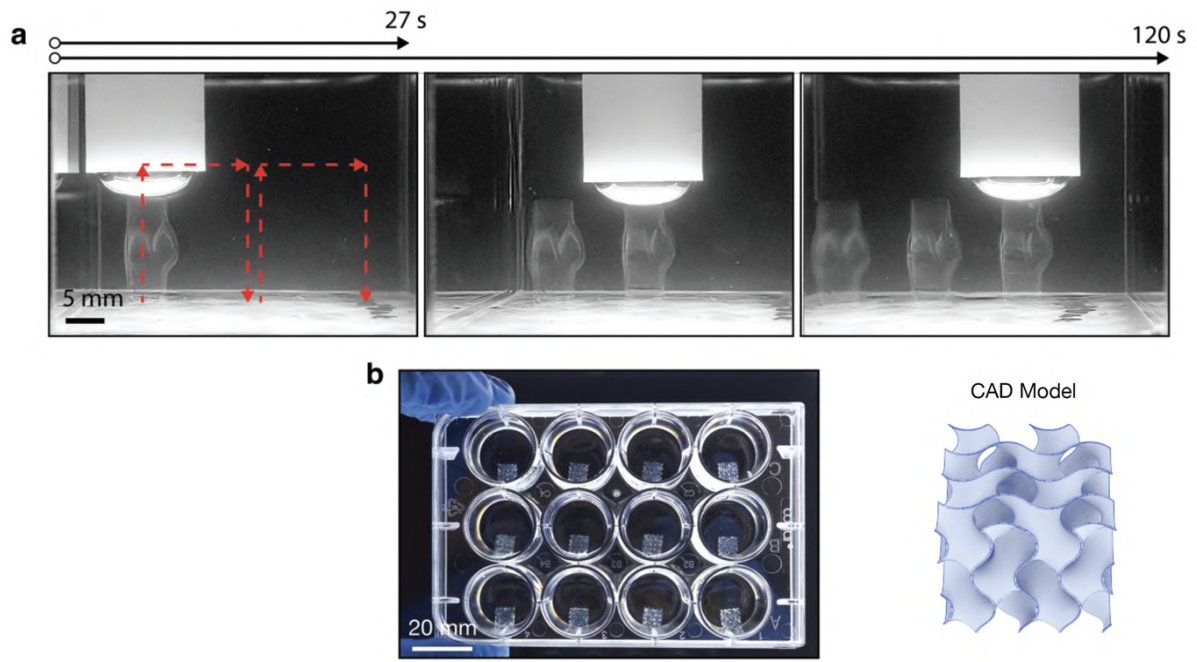

**Supplementary Fig.16| Multi-step printing and direct in well printing**. **a**, Time lapse images of three tricuspid valves printed in just under 120 seconds via three-dimensional placement of the print head. **b**, Direct in well printing of 12 gyroid lattices, created in just under 8 minutes.

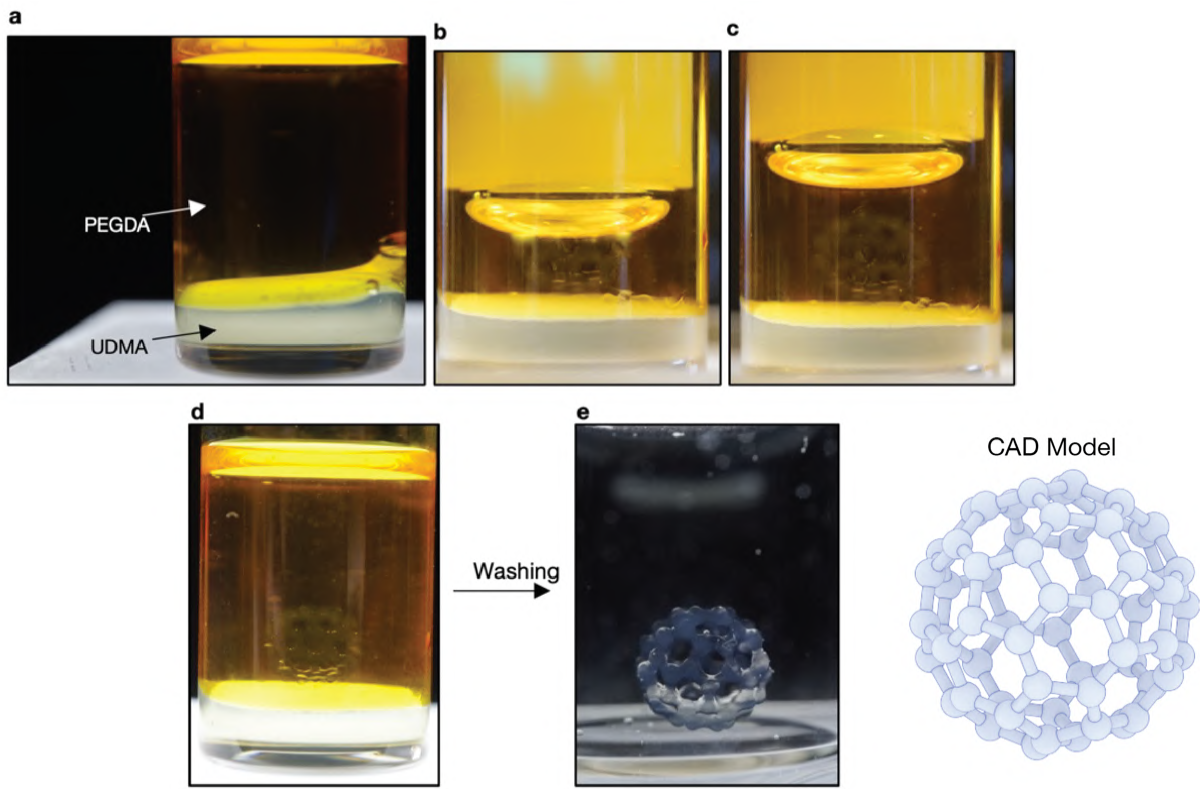

**Supplementary Fig.17| Printing of free-floating structures by utilizing high viscosity dense materials. a**, UDMA and PEGDA loaded into a 15 mm diameter vial. **b,** Bucky-ball structure during printing whereby the high viscosity UDMA acts as a fluidic supporting medium. **c-d**, Bucky-ball directly after printing. **e**, Bucky-ball (C60) after removal of UDMA and PEGDA.

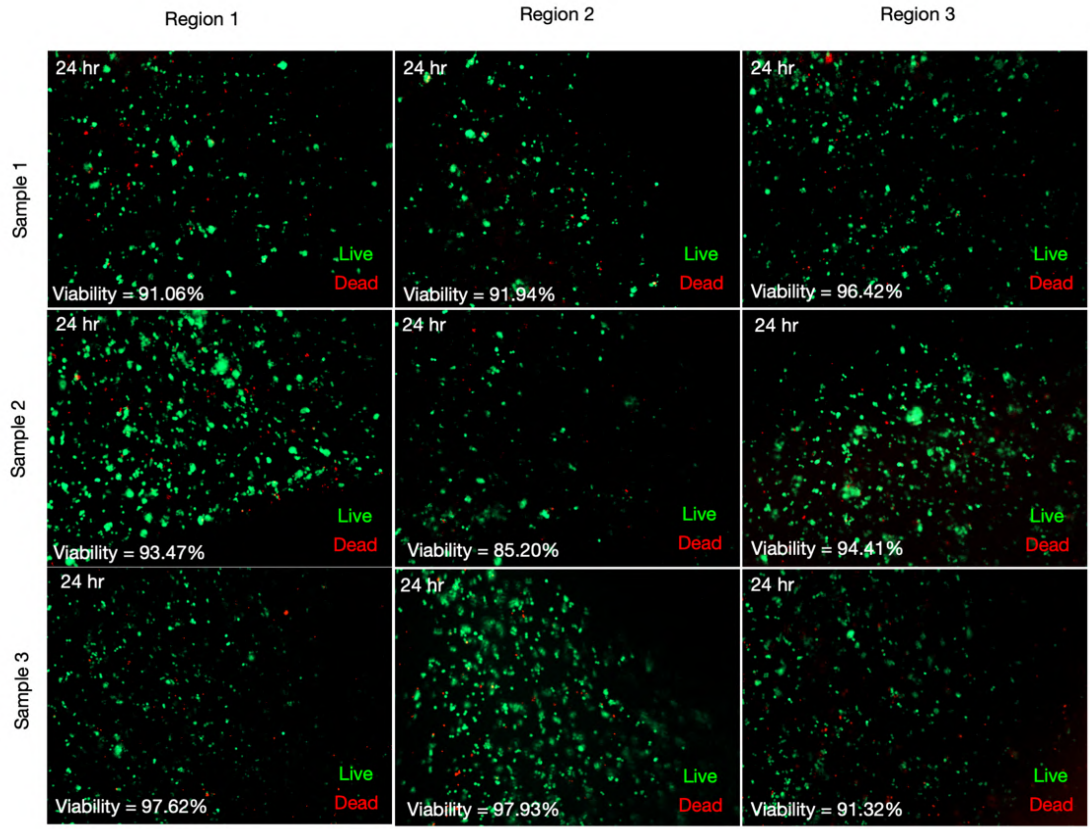

**Supplementary Fig.18|** Triplicate florescence images of the wall test structure for three representative regions at 24h after printing. High cell viability demonstrates no immediate cytotoxicity due to DIP printing. Average cell viability across the samples is 93.26% after 24h.

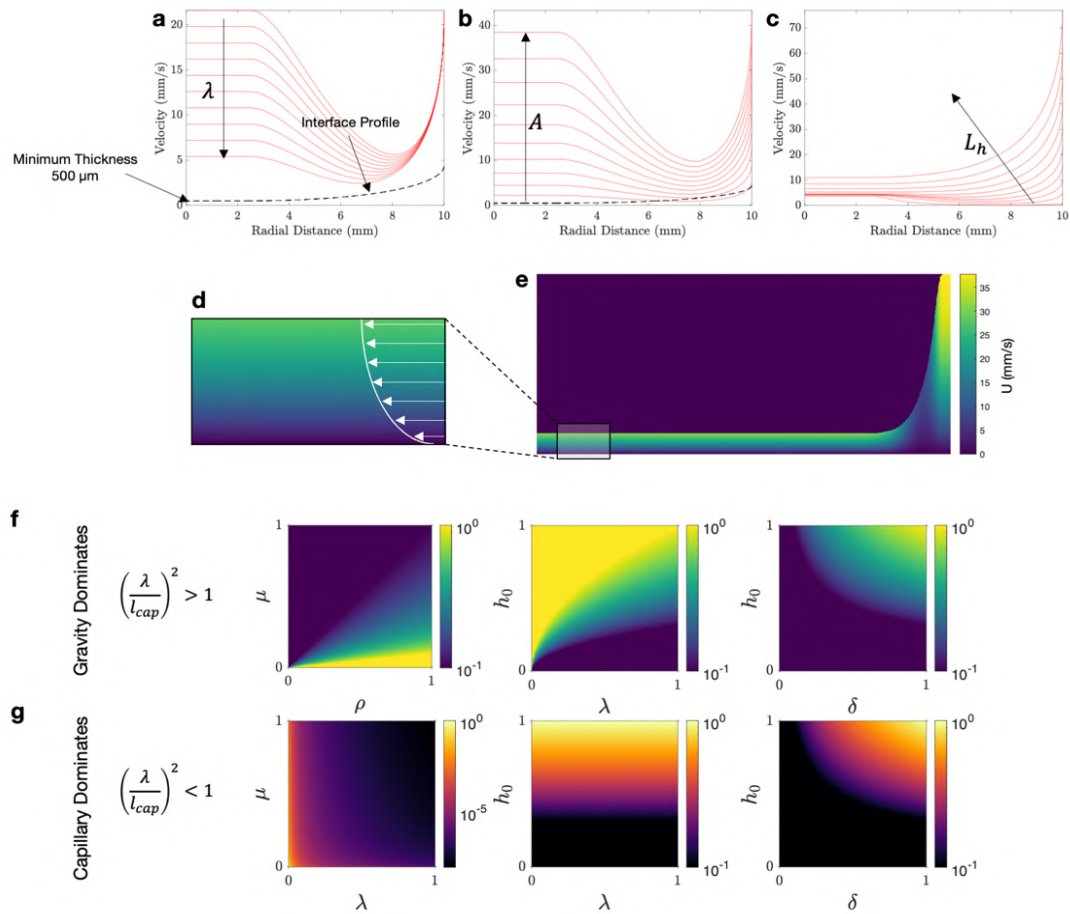

**Supplementary Fig.19| Effect of material and interface properties on acoustically driven flow profiles**. **a**, Effect of decreasing acoustic wavelength on maximum flow velocity in the x-direction as a function of the profile height. The chosen interface profile was derived from the Young-Laplace solution for PEGDA 20%. **b**, Effect of increasing the driving amplitude on the maximum flow velocity in the x-direction as a function of the interface height. **c**, Effect of increasing the layer height (gap between the interface and the lower surface) on the maximum flow velocity in the x-direction. **d**, 2D flow profile below the interface with a layer height thickness of 500 µm. **d**, 2D flow profile of the interface entire interface. **f**, dimensionless exploration of how material properties affect the velocity in the x-direction for a gravity dominated system $(\lambda/l_{cap})^2 > 1$. Velocity scaling is shown on a log scale. **g**, dimensionless exploration of how material properties affect the velocity in the x-direction for a capillary dominated system $(\lambda/l_{cap})^2 < 1$. Velocity scaling is shown on a log scale.

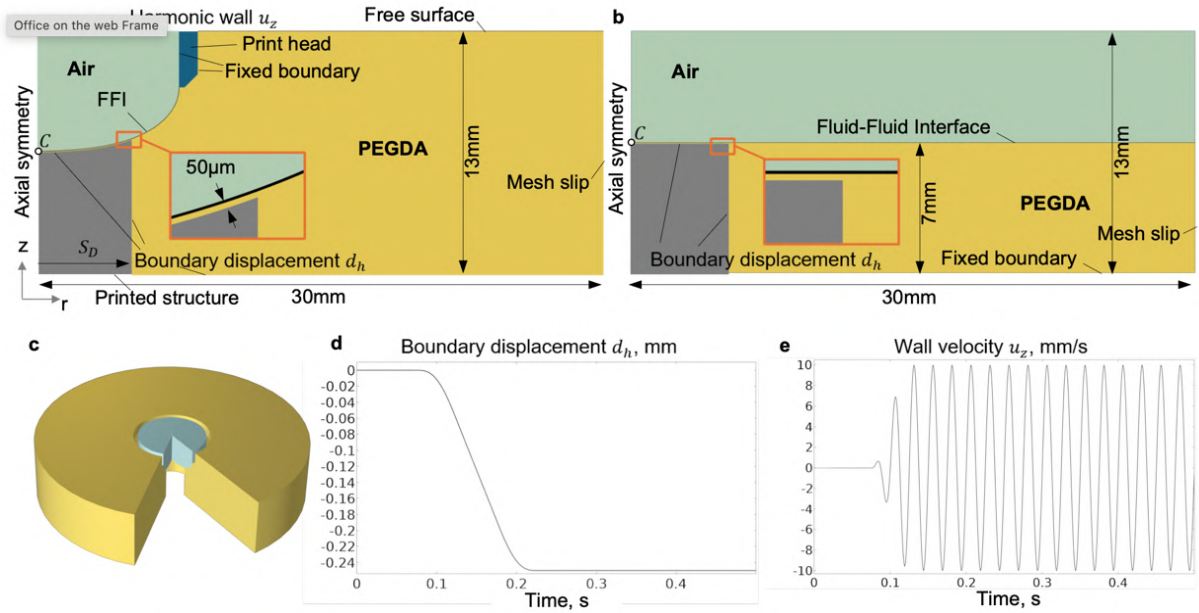

**Supplementary Fig.20| Finite element analysis (FEA) and setup of resin influx across printed structure**. **a**, Graphical view of the axisymmetric computational domain for DIP. **b**, 3D representation of the DIP simulation domain revolved around the z-axis. **c**, The view of the DIP computational domain. **d**, Modelling domain boundary displacement. **e**, Temporal evolution of the wall velocity for the acoustically actuated DIP.

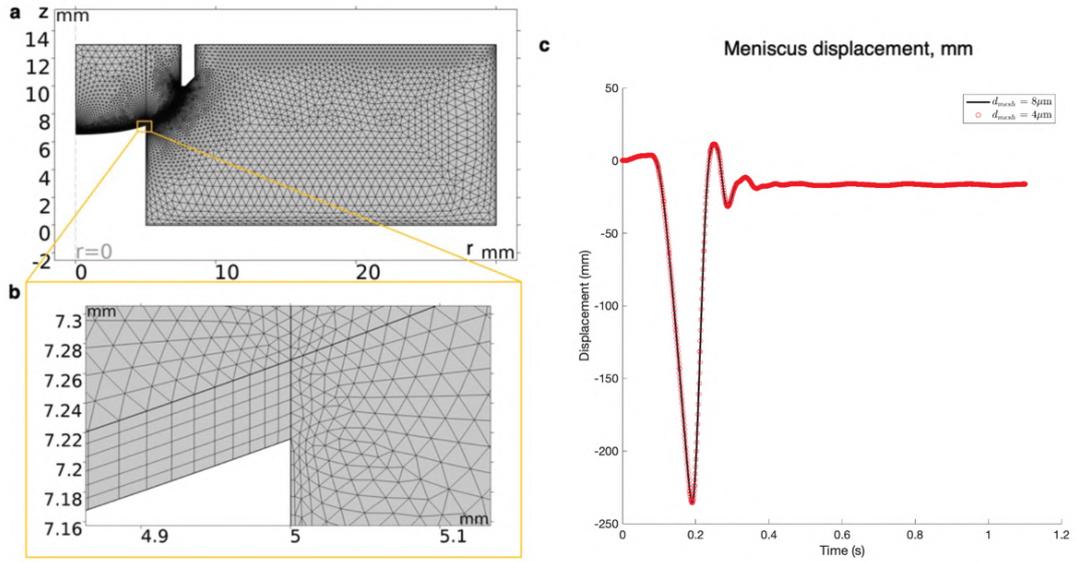

**Supplementary Fig.21| Computational mesh and mesh refinement study used for the numerical modelling. a**, Computational mesh with element size $d_{mesh} = 8$ μm. **b**, A magnified view of the mesh. **c**, The displacement of the meniscus over time yields comparable results if a refined mesh is utilized.

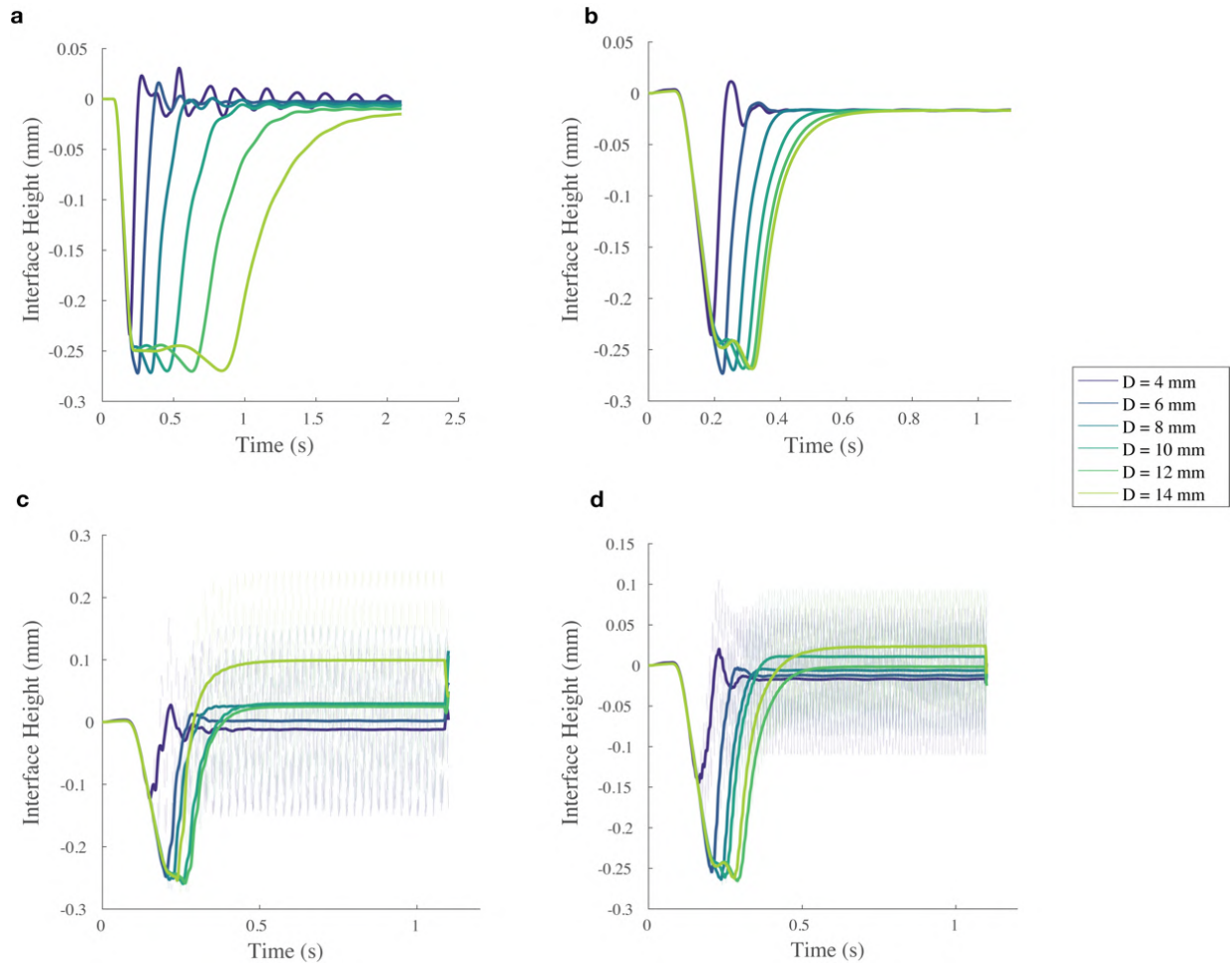

**Supplementary Fig.22| Numerical prediction of the interface release dynamics for a 15 mm diameter print head with varying circular printed structures from 4 to 14 mm in diameter. a,** Location of the central node of the interface as a function of time for Top-Down SLA. **b,** Location of the central node of the interface as a function of time for Dynamic Interface Printing (DIP) without acoustic excitation. **c,** Location of the central node of the interface as a function of time for Dynamic Interface Printing (DIP) with acoustic excitation at a frequency of 40 Hz. **d,** Location of the central node of the interface as a function of time for Dynamic Interface Printing (DIP) with acoustic excitation at a frequency of 100 Hz. For (**c-d**) the transparent plots denote the oscillatory interface height, with the solid lines representing the moving average across a single excitation period.

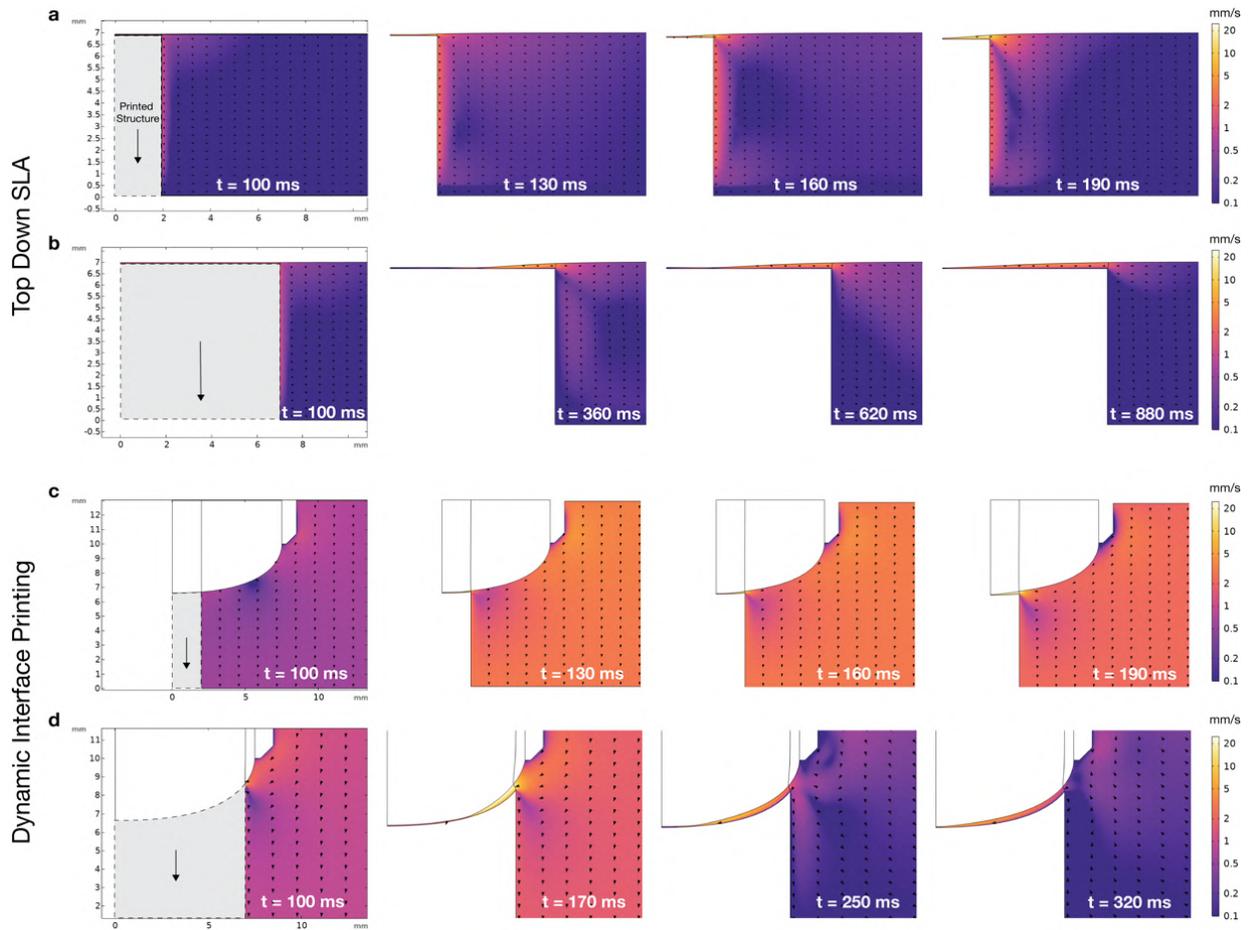

**Supplementary Fig.23| Numerical prediction of the velocity magnitude for a top-down SLA and dynamic interface printing with a 15 mm diameter print head. a,** Time sequence velocity field for top-down SLA from the time of initial displacement (t = 100ms) to central interface release (t = 190 ms) for a 4 mm diameter printed structure. **b,** Time sequence velocity field for top-down SLA from the time of initial displacement (t = 100ms) to central interface release (t = 880 ms) for a 14 mm diameter printed structure. **c,** Time sequence velocity field for dynamic interface printing without acoustics from the time of displacement (t = 100ms) to central interface release (t = 190 ms) for a 4 mm diameter printed structure. **d,** Time sequence velocity field for dynamic interface printing without acoustics from the time of displacement (t = 100ms) to central interface release (t = 320 ms) for a 14 mm diameter printed structure. Velocity magnitude for each contour plot is indicated on the right-hand side of each row. Logarithmic colormap was chosen to indicate the global flow field more effectively.

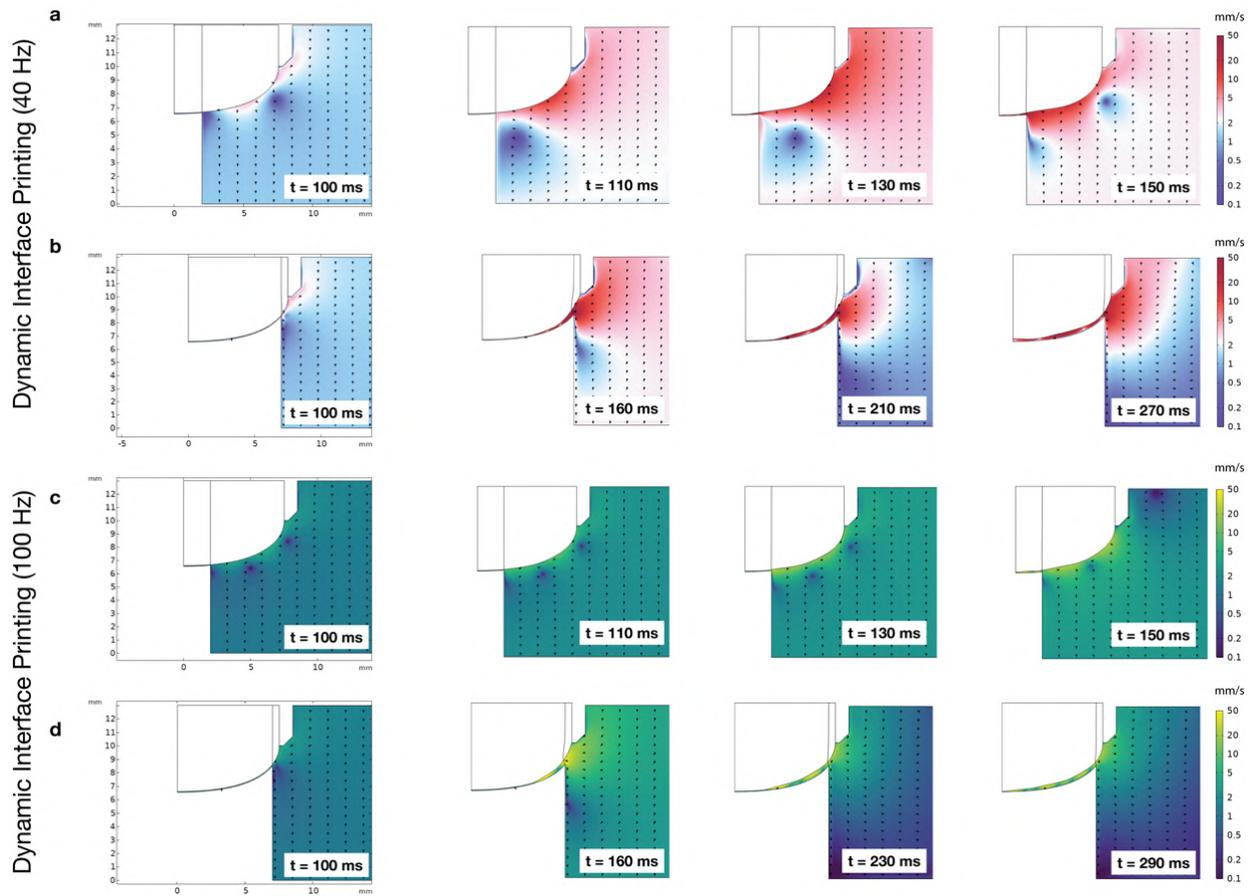

**Supplementary Fig. 24 | Numerical prediction of the velocity magnitude for acoustically driven dynamic interface printing using a 15 mm diameter print head. a,** Time-sequence velocity field for acoustically driven dynamic interface printing at a frequency of 40 Hz and a structural diameter of 4 mm. The time sequence spans from the initial displacement (t = 100 ms) to the central interface release (t = 150 ms). **b,** Time-sequence velocity field for acoustically driven dynamic interface printing at a frequency of 40 Hz and a structural diameter of 14 mm. The time sequence spans from the initial displacement (t = 100 ms) to the central interface release (t = 270 ms). **c,** Time-sequence velocity field for acoustically driven dynamic interface printing at a frequency of 100 Hz and a structural diameter of 4 mm. The time sequence spans from the initial displacement (t = 100 ms) to the central interface release (t = 150 ms). **d,** Time-sequence velocity field for acoustically driven dynamic interface printing at a frequency of 100 Hz and a structural diameter of 14 mm. The time sequence spans from the initial displacement (t = 100 ms) to the central interface release (t = 290 ms). Velocity magnitude for each contour plot is indicated on the right-hand side of each row. Logarithmic colormap was chosen to indicate the global flow field more effectively.

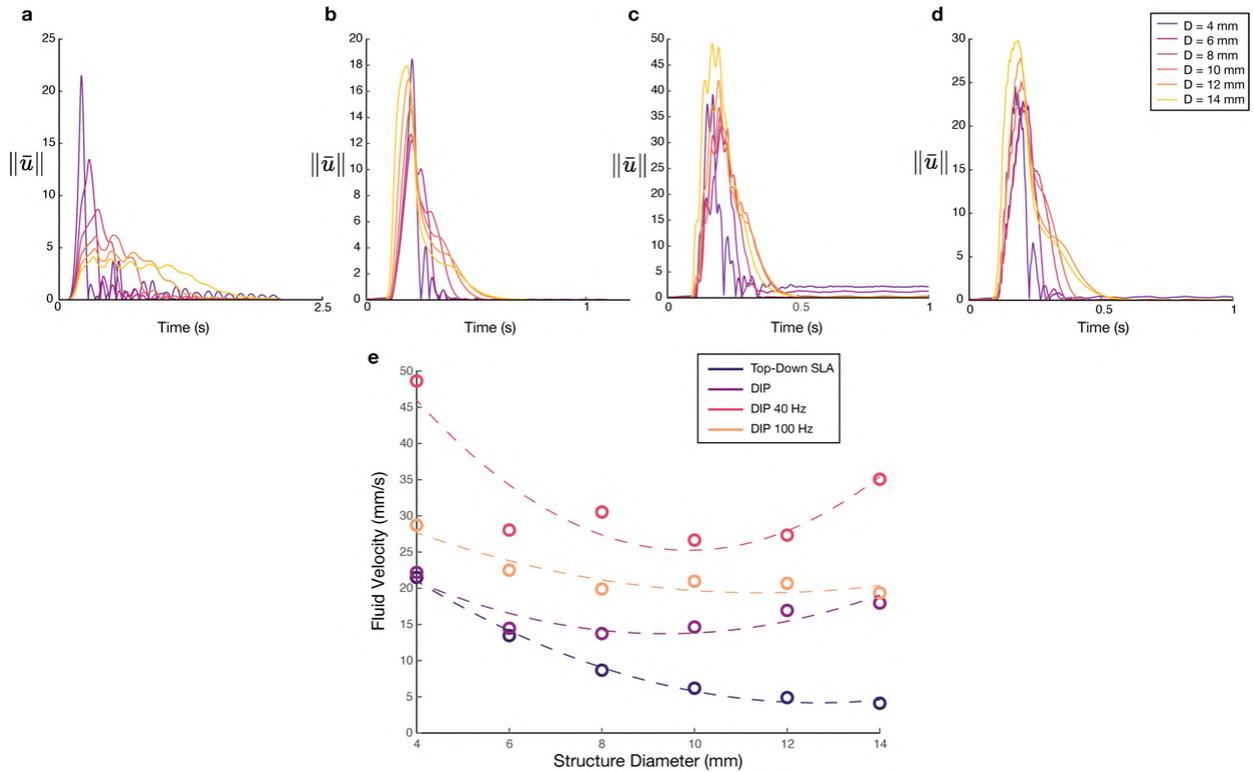

**Supplementary Fig.25| Numerical prediction of the average inflow fluid velocity for a 15 mm diameter print head with varying circular printed structures ranging from 4 to 14 mm in diameter. a,** Radial magnitude of the average fluid velocity ($\|\bar{u}\|$) for increasing structural diameter in top-down SLA. **b,** Radial magnitude of the average fluid velocity ($\|\bar{u}\|$) for increasing structural diameter in DIP without acoustics. **c,** Radial magnitude of the average fluid velocity ($\|\bar{u}\|$) for increasing structural diameter in DIP with 40 Hz acoustic driving. **d,** Radial magnitude of the average fluid velocity ($\|\bar{u}\|$) for increasing structural diameter in DIP with 100 Hz acoustic driving. **e,** Peak average fluid velocity for each printing technique as a function of structural diameter.

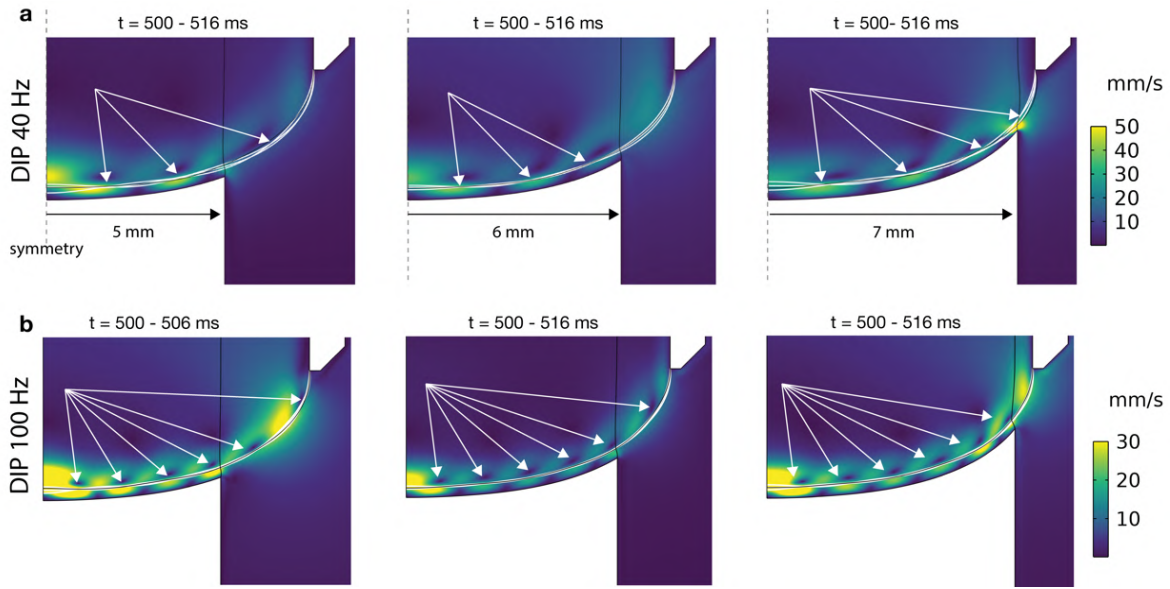

**Supplementary Fig. 26| Numerical prediction of structural-modal interaction. a,** Meniscus resonance mode shapes (indicated in white solid lines) over a single period at 40 Hz acoustic driving for structures with diameters of 10, 12, and 14 mm. **b,** Meniscus resonance mode shapes (indicated in white solid lines) over a single period at 100 Hz acoustic driving for structures with diameters of 10, 12, and 14 mm. White arrows indicate the locations of the nodal locations of the induced capillary wave. Linear colormap indicates the magnitude of the velocity field both within the fluid and within the enclosed air volume above the meniscus.

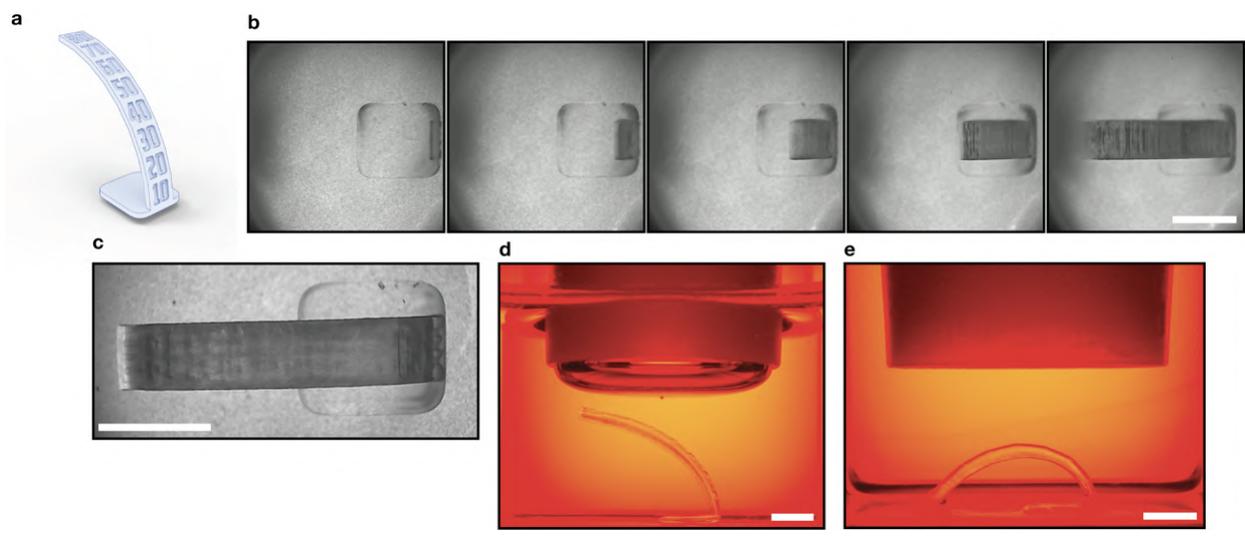

**Supplementary Fig.27| DIP printing of an overhang test structure with angles ranging from 10°- 80° in PEGDA 20%. a**, CAD model of the overhang test structure. **b**, Timelapse imaging of the test structure fabrication. **c**, Closeup image from above after fabrication. **d**, Side view after fabrication. **e**, Collapse of the test structure after the removal of the surrounding material. All scale bars are 5 mm.

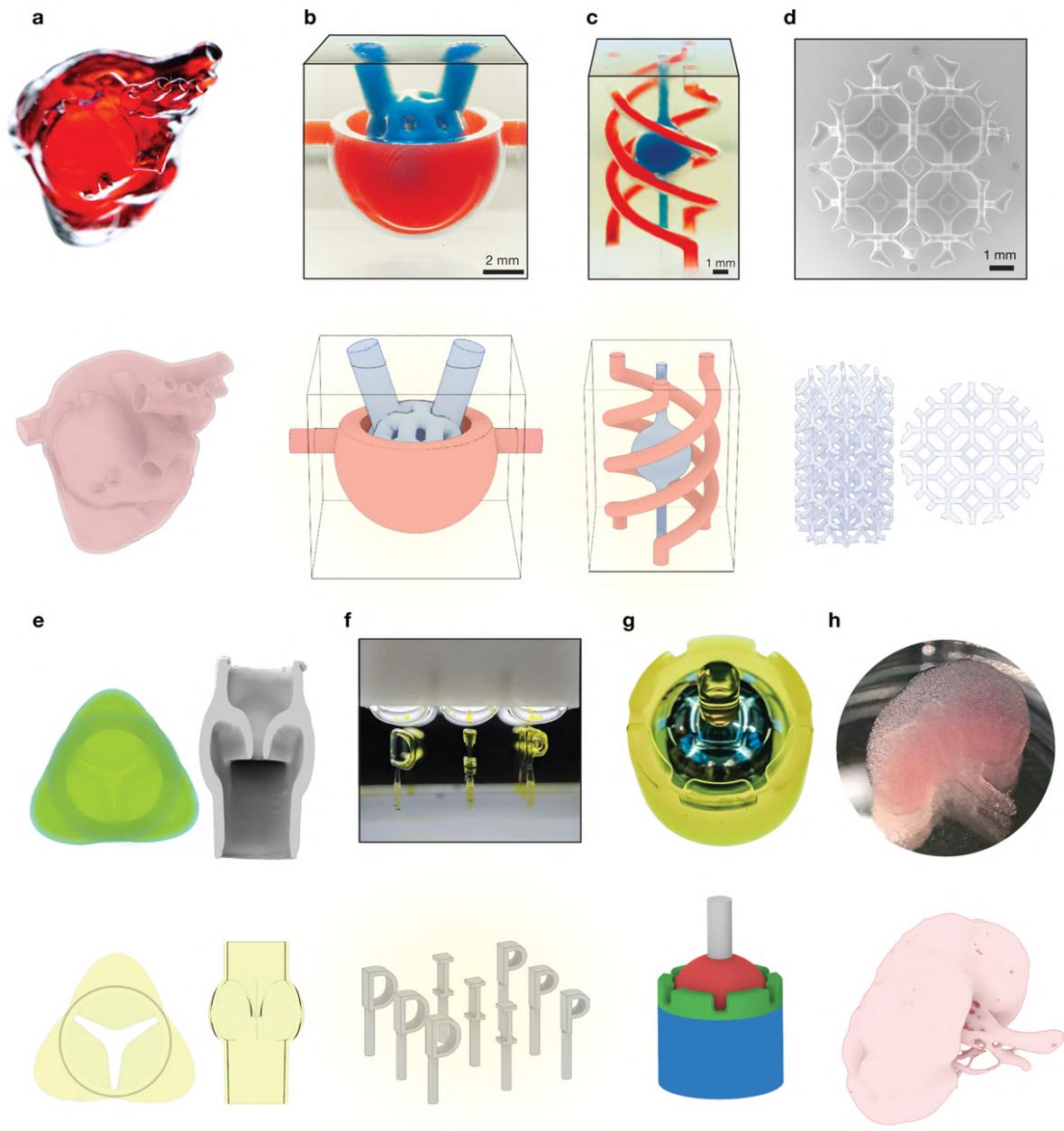

**Supplementary Fig.28| Accompanying CAD models of printed structures in main text. a**, Printed heart model. **b**, Bowmans Capsule. **c**, Tri-helix. **d**, Kelvin cell lattice. **e**, Tricuspid valve. **f**, Letters 'DIP'. **g**, Ball and socket joint. **h**, Anatomical kidney.

| Material | Viscosity (mPas) | Surface Tension (mN/m) |
| --- | --- | --- |
| PEGDA 20 | 2.75 | 64.82 (estimated) |
| PEGDA 50 | 15.23 | 52.85 (estimated) |
| PEGDA 100 | 100.50 | 32.90 |
| HDDA | 6.45 | 33.20 |
| UDMA | 9736 | 37.60 |

**Table 1|** Viscosity and surface tension values for key materials explored in this study.

**Movie 1: Dynamic interface printing of a heart model**

Real time printing of a 15 mm tall heart model within a glass cuvette. The video highlights the insertion of the print head, formation of the air-liquid boundary and rapid formation of the model in-situ.

**Movie 2: Top-down imaging of the air-liquid boundary with particles suspended particles**

PIV imaging from within the printhead at low frequency (25Hz) under increasing driving amplitude.

**Movie 3: High amplitude acoustic driving of the air-liquid interface**

High speed imaging (500 fps) of the air-liquid interface driven at 40 Hz. Overlayed flow field represents the mean fluid velocity across all frames.

**Movie 4: Direct in-well printing of a gyroid lattice.**

Printing of six identical gyroid lattices (shown in top right) across a standard 6 well plate. Left video shows real time printing of the gyroid from within the print head. Right video is approximately 6X speed, showing the print head printing and moving between wells.

**Movie 5: Printing of the letters 'DIP'**

Real time fabrication of the letters 'DIP' using a 3x3 interface print head.

References:

1. Butt, H. J. et al. Characterization of super liquid-repellent surfaces. Current Opinion in Colloid and Interface Science vol. 19 Preprint at https://doi.org/10.1016/j.cocis.2014.04.009 (2014).
2. Statics: including Hydrostatics and the Elements of the Theory of Elasticity. Nature 116, (1925).
3. Zhang, Z. & Joshi, S. An improved slicing algorithm with efficient contour construction using STL files. International Journal of Advanced Manufacturing Technology 80, (2015).
4. Lewis, K. & Matsuura, T. Bézier Curve Method to Compute Various Meniscus Shapes. ACS Omega 8, 15371–15383 (2023).
5. Lewis, K. & Matsuura, T. Calculation of the Meniscus Shape Formed under Gravitational Force by Solving the Young–Laplace Differential Equation Using the Bézier Curve Method. ACS Omega 7, 36510–36518 (2022).
6. Andrzejewska, E. Photopolymerization kinetics of multifunctional monomers. Progress in Polymer Science (Oxford) vol. 26 Preprint at https://doi.org/10.1016/S0079-6700(01)00004-1 (2001).
7. Eelbode, T. et al. Optimization for Medical Image Segmentation: Theory and Practice When Evaluating With Dice Score or Jaccard Index. IEEE Trans Med Imaging 39, (2020).
8. Behroodi, E., Latifi, H. & Najafi, F. A compact LED-based projection microstereolithography for producing 3D microstructures. Sci Rep 9, 19692 (2019).
9. Jacobs, P. F. Fundamentals of stereolithography. in 1992 International Solid Freeform Fabrication Symposium (1992).
10. Hecht, E. Hecht optics. Addison Wesley vol. 997 Preprint at (1998).
11. Born, M., Wolf, E. & Hecht, E. Principles of Optics: Electromagnetic Theory of Propagation, Interference and Diffraction of Light . Phys Today 53, (2000).
12. Hsiao, K. et al. Single-digit-micrometer-resolution continuous liquid interface production. Sci Adv 8, (2022).
13. Tumbleston, J. R. et al. Continuous liquid interface production of 3D objects. Science (1979) 347, 1349–1352 (2015).
14. Basic lubrication theory. Wear 84, (1983).
15. Porte, E., Cann, P. & Masen, M. A lubrication replenishment theory for hydrogels. Soft Matter 16, (2020).
16. Oron, A., Davis, S. H. & Bankoff, S. G. Long-scale evolution of thin liquid films. Rev Mod Phys 69, (1997).
17. Zhang, X. Observations on waveforms of capillary and gravity-capillary waves. European Journal of Mechanics, B/Fluids 18, (1999).
18. Périnet, N., Gutiérrez, P., Urra, H., Mujica, N. & Gordillo, L. Streaming patterns in Faraday waves. J Fluid Mech 819, (2017).
19. Huang, Y., Wolfe, C. L. P., Zhang, J. & Zhong, J. Q. Streaming controlled by meniscus shape. J Fluid Mech 895, (2020).
20. Dendukuri, D. et al. Modeling of oxygen-inhibited free radical photopolymerization in a PDMS microfluidic device. Macromolecules 41, (2008).
21. Elizondo-Leal, J. C. et al. Parallel raster scan for euclidean distance transform. Symmetry (Basel) 12, (2020).
22. Sramek, M. & Kaufman, A. Fast ray-tracing of rectilinear volume data using distance transforms. IEEE Trans Vis Comput Graph 6, 236–252 (2000).
23. Amanatides, J. & Woo, A. A Fast Voxel Traversal Algorithm for Ray Tracing. Eurographics 87, (1987).